\documentclass[12pt,british]{report}
\usepackage[T1]{fontenc}
\usepackage[latin9]{inputenc}
\usepackage{geometry}
\usepackage{fancyhdr}
\usepackage{color}
\usepackage{babel}
\usepackage{wrapfig}
\usepackage{amsmath}
\usepackage{amssymb}
\usepackage{graphicx}
\usepackage{esint}
\usepackage[unicode=true,pdfusetitle,
 bookmarks=true,bookmarksnumbered=false,bookmarksopen=false,
 breaklinks=false,pdfborder={0 0 1},backref=false,colorlinks=true]
 {hyperref}
\hypersetup{
 linkcolor=red,citecolor=blue,urlcolor=blue}

\makeatletter
\numberwithin{equation}{section}

\usepackage{tensind}
\tensordelimiter{?}
\usepackage{dsfont}
\usepackage[numbers,sort&compress]{natbib}

\renewcommand{\chaptermark}[1]{%
\markboth{\@chapapp\ \thechapter.  #1}{\chaptername\ \thechapter.  #1}%
}

\fancypagestyle{plain}{
\fancyhf{}
\fancyfoot[LE,RO]{\thepage}

}

\usepackage{tocloft}
\makeatother

\begin{document}
\global\long\def\m#1{\mathcal{#1}}
\global\long\def\dd{\mathrm{d}}
 \global\long\def\t#1{{#1}}
 \global\long\def\bC#1#2{?{\t{\bar{C}}}^{#1}{}_{#2}?}
\global\long\def\tr{\mathrm{tr}}
\global\long\def\id{\mathds{1}}

\title{I've Got the World on a Brane}

\author{John Omotani}

\date{{\normalsize Thesis submitted to the University of Nottingham}\\
{\normalsize for the degree of Doctor of Philosophy\bigskip{}
}\\
{\normalsize July 2012}}
\maketitle
\begin{abstract}
This thesis treats several topics in the study of extra-dimensional
models of the world, concerning Heterotic M-Theory and the dynamics
of branes.

We describe a reduction to five dimensions, over a Calabi-Yau manifold,
of an improved version of Heterotic M-Theory, which is valid to all
orders in the gravitational coupling. This provides a starting point
for considering the consequences of the improved theory for the very
fruitful phenomenology of the original.

We investigate the singularities formed by the collision of gravitating
branes in scalar field theory. By considering the asymptotic structure
of the spacetime, the properties of the horizons formed and the growth
of the curvature we argue that the singularity is not a black brane,
as one might have expected, but rather a big crunch.

Finally, we construct a restricted class of multi-galileon theories
as braneworld models with codimension greater than one, developing
in the process some of the formalism needed for the general construction.
\end{abstract}
\renewcommand\abstractname{Acknowledgements}
\begin{abstract}
Firstly I would like to express my gratitude to my parents for their
love and support throughout my life. Thanks are due of course to my
supervisor Paul Saffin for his guidance and his invaluable help and
also to Ian Moss for his assistance and insight during our collaboration.
Finally I would like to thank everyone in the Particle Theory and
Astronomy groups for making Nottingham such a friendly place to work
and especially Rob Chuter and Duncan Buck, my office mates for much
of the journey. My PhD Studentship was funded by STFC.\end{abstract}

\tableofcontents{}

\chapter{Introduction}

\section{Supergravity}

The story of particle physics in the 20th century was one of finding
symmetries and exploiting them to construct quantum field theories,
culminating in the construction of the standard model of particle
physics with (exact) colour gauge symmetry and (broken) electroweak
gauge symmetry. However, the theory of spacetime symmetry, that is
the other great pillar of 20th century physics, general relativity,
stubbornly refuses formulation as a quantum theory, which our experience
of particle physics teaches us that it would have to be at short enough
distances or high enough energies.

In the spirit of the search in particle physics for ever more symmetries
to exploit, it is natural to ask whether there can be further symmetries
of spacetime, in the hope that they might help to reconcile the incompatibility
of gravity and quantum theory. The initial answer to this question
is that there can be no further symmetries due to the theorem of Coleman
and Mandula \cite{Coleman:1967ad} which states that, under reasonable
assumptions, the most general Lie algebra of symmetry operators consists
of the Poincaré algebra plus internal symmetry algebras. However,
this is not the end of the story as the Coleman-Mandula theorem can
be evaded by relaxing its restrictions.

In particular, supersymmetry is a spacetime symmetry which circumvents
the theorem by including anti-commutators as well as commutators in
its algebra: it has a \emph{graded} Lie algebra rather than just a
Lie algebra. In addition to the bosonic generators of Poincaré symmetry
there are fermionic generators of the supersymmetry. These fermionic
generators transform bosonic fields to fermionic ones and vice versa.
We can then group particles into `supermultiplets' of `superpartners'
which form a closed cycle under the supersymmetry. If supersymmetry
were unbroken, particles in a supermultiplet would all have the same
mass and so it is clear that if the world is supersymmetric then that
supersymmetry is broken somehow. Nevertheless, supersymmetry as a
theory of physics beyond the standard model predicts that the observed
particles have superpartners, albeit ones made heavy enough by the
breaking of supersymmetry that they are unobservable. Indeed it is
possible to construct phenomenological extensions to the standard
model such as the `minimal supersymmetric standard model' (MSSM) which
reproduce the same results for experiments to date but predict superpartners
observable by the Large Hadron Collider at CERN\@. One motivation
for such models is that they may solve the hierarchy problem, which
is the puzzle of why the scale of the electroweak theory (i.e.\ the
mass of the Higgs boson) is so much less than the natural scale of
some more fundamental theory: $\sim10^{2}$GeV as opposed to $\sim10^{16}-10^{17}$GeV
for a grand unified theory or perhaps even $\sim10^{19}$GeV (the
Planck scale). Spontaneously broken supersymmetry provides a mechanism
for a Higgs mass (which would be zero with unbroken supersymmetry)
which is exponentially suppressed, thus solving the problem (if one
has such a supersymmetry breaking mechanism in hand).

As local Poincaré symmetry gives us general relativity, so local supersymmetry
gives us supergravity. It is an interesting fact that there is an
upper limit to the number of supersymmetries in a given number of
spacetime dimensions, and this limit decreases as the number of dimensions
increases. Indeed there is a maximum number of spacetime dimensions
in which it is possible to construct supersymmetric theories. The
reason is as follows: For the graviton to be the unique spin two particle,
it must be the highest spin member of some supermultiplet. The supersymmetry
generators are the components of $\m N$ spinors, where $\m N$ is
the number of supersymmetries. In $2n$ or $2n+1$ dimensions a spinor
has $2^{n}$ components, so there are $\m N\cdot2^{n}$ supersymmetry
generators. Consider their actions on a single particle state. For
a massless particle we cannot choose the zero momentum state, but
for a single particle state it suffices to consider only states with
momentum in a particular direction, call it the 1 direction. States
then have $p_{1}=\pm p_{0}$ and we can choose the basis of the supersymmetry
generators so that half annihilate the states with $p_{1}=+p_{0}$
and the other half annihilate the states with $p_{1}=-p_{0}$. Furthermore,
when a generator acts on a state with some helicity $h$, it changes
$h$ by $\frac{1}{2}$ with half of the generators raising, and the
other half lowering, $h$. Since the generators are fermionic and
anti-commute with themselves, if we act twice with any one generator
we destroy the state. So for states with say $p_{1}=+p_{0}$ we have
$\m N\cdot2^{n-1}$ generators which do not annihilate the states.
The lowest helicity state of a supermultiplet is given by acting once
on the highest helicity state with each of the $\m N\cdot2^{n-2}$
of these generators which also lower helicity. If the graviton, which
has spin two, is the highest spin state then we start from helicity
$+2$ and we can only have at most 8 generators, each taking us down
by $\frac{1}{2}$, or the lowest helicity state would be below $-2$
(which would contradict our assertion that the graviton is the highest
spin state). Thus we have a limit on the number of spacetime dimensions
in which we can have a supergravity, for we must have $\m N\cdot2^{n-2}\leq8$
in order for there to be a supermultiplet without any state having
spin higher than the graviton. The highest dimension possible is eleven,
with also $\m N=1$. This maximal eleven dimensional supergravity,
which was first described by Cremmer, Julia and Scherk \cite{Cremmer:1978km},
is unique because if there is only one graviton then there can be
only one supermultiplet, whose structure is entirely dictated by supersymmetry.

\section{Quantum Anomalies}

It may happen that a theory which is classically invariant under some
symmetry is not invariant quantum mechanically, for instance it may
be that it is impossible to construct a regularization scheme for
divergent Feynman diagrams which respects that symmetry. This quantum
variation is called the anomaly. In the case of a global symmetry
this may simply indicate that certain processes which are forbidden
classically are allowed quantum mechanically, which is not a problem.
For a local symmetry however, states related by a gauge transformation
are supposed to be identified with each other and the presence of
a quantum anomaly would render the theory inconsistent. Therefore
one must ensure that if there are sources of such local anomalies,
then their contributions cancel each other exactly.

The quantum anomaly is the variation under gauge transformations or
diffeomorphisms of the effective action describing the gauge fields
or the metric which is given by integrating out the fermions in the
path integral. Anomalies in $d$ dimensions are associated to certain
$(d+2)$-forms. The anomaly must obey what is known as the Wess-Zumino
consistency condition. Let us call the effective action $\Gamma$.
Then the anomaly, $G(\epsilon)$, is the variation by an amount $\epsilon$
of $\Gamma$: $G(\epsilon)=\delta_{\epsilon}\Gamma$. If we consider
another variation, this time by $\eta$, then $\delta_{\eta}G(\epsilon)-\delta_{\epsilon}G(\eta)=\delta_{\eta}\delta_{\epsilon}\Gamma-\delta_{\epsilon}\delta_{\eta}\Gamma$,
but this is just the Lie bracket acting on $\Gamma$, giving $G([\eta,\epsilon])$,
so we have the consistency condition $\delta_{\eta}G(\epsilon)-\delta_{\epsilon}G(\eta)=G([\eta,\epsilon])$.
This condition can be fulfilled straightforwardly if we can describe
the anomaly, in a $d$-dimensional theory, by starting from a (formal)
$(d+2)$-form $I_{(d+2)}$ which is an invariant polynomial of the
2-form (gauge or gravitational) curvature. We only need even rank
forms because quantum anomalies only occur in even dimensions. According
to the Chern-Weil theorem invariant polynomials of the curvature are
closed so $\dd I_{(d+2)}=0$. Therefore it can be written as the derivative
of a $(d+1)$-form, $I_{(d+2)}=\dd I_{(d+1)}$, at least locally.
Since $I_{(d+2)}$ is gauge invariant, the variation of $I_{(d+1)}$
must be a total derivative $\delta_{\epsilon}I_{(d+1)}=\dd I_{d}(\epsilon)$.
Now if the quantum anomaly is given by the integral of such a $d$-form,
$G(\epsilon)=\int_{\m M_{d}}I_{d}(\epsilon)$ and if we regard $\m M_{d}$
(formally) as the boundary of $\m M_{(d+1)}$ then $G(\epsilon)=\int_{\m M_{(d+1)}}\dd I_{d}(\epsilon)=\int_{\m M_{(d+1)}}\delta_{\epsilon}I_{(d+1)}$
and so it is easy to see that the Wess-Zumino consistency condition
is obeyed: $\delta_{\eta}G(\epsilon)-\delta_{\epsilon}G(\eta)=\int_{\m M_{(d+1)}}\left(\delta_{\eta}\delta_{\epsilon}I_{(d+1)}-\delta_{\epsilon}\delta_{\eta}I_{(d+1)}\right)=\int_{\m M_{(d+1)}}\delta_{[\eta,\epsilon]}I_{(d+1)}=G([\eta,\epsilon])$.
This illustrates why $(d+2)$-form polynomials might be useful for
calculating quantum anomalies. In fact the connection is deeper, all
the gauge and gravitational anomalies can be so calculated and the
coefficients of the polynomial can be determined. The reason is that
the anomaly can be related to the index of the Dirac operator in $(d+2)$
dimensions, which can in turn be calculated using the Atiyah-Singer
index theorem \cite{AlvarezGaume:1984dr}. A careful discussion useful
to our purposes later is \cite{Bilal:2003es}. We can ensure anomaly
cancellation just by requiring that the sum of the polynomials from
all the sources of variation cancel; this is helpful because there
is an ambiguity in the definition of the anomalies themselves since
their explicit form depends upon a choice of gauge. Anomalies can
occur where we have chiral fermions (the index of the Dirac operator
is the difference in the number of zero modes with positive and negative
chirality): in particular Yang-Mills gauge theory has an anomaly in
any even dimension and gravity has an anomaly in dimension $d=4k+2$.
In ten dimensions, where there are both gauge and gravitational anomalies,
enforcing anomaly cancellation is a strong constraint on the possible
theories, such as those discussed below, which are formulated either
in ten dimensions (superstring theories) or in spaces with ten dimensional
boundaries (Heterotic M-Theory).

\section{M-Theory}

String theory promises to be a quantum theory of gravity. Since the
fundamental objects have a finite size, the traditional problem for
quantum gravity of how to deal with ultraviolet divergences is ameliorated
since the fundamental objects have finite size rather than being point
particles. The questions are how to formulate string theory without
inconsistencies such as tachyons and quantum anomalies and, that being
accomplished, whether and how such a theory can describe the real
world. Tachyons can be banished by incorporating supersymmetry and
it is possible to construct such superstring theories which are free
of anomalies. However, this puts rather stringent constraints on the
theories, in particular they are all ten dimensional, but still there
are five consistent superstring theories known. 

\begin{wrapfigure}{O}{0.5\columnwidth}%
\begin{centering}
\includegraphics[bb=60mm 97mm 156mm 181mm,clip,scale=0.7]{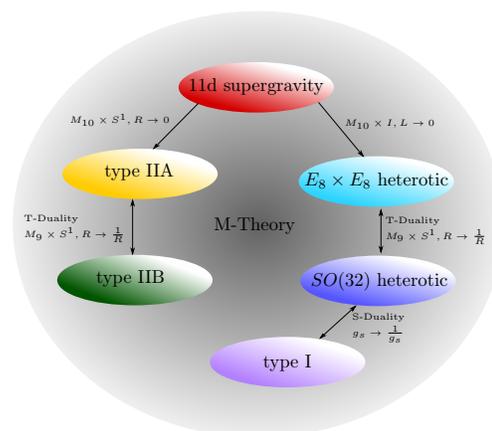}
\par\end{centering}

\caption{M-Theory Dualities}
\end{wrapfigure}%
From the point of view of discovering the correct description of quantum
gravity for the real world, this might be viewed as an embarrassment
of riches, for how are we to decide which of these theories ought
to apply to us? However, it seems that the various superstring theories
are not in fact independent. They are related by duality transformations
which imply that they are merely different limits of a single underlying
theory, which is known as M-Theory. Though the complete structure
of M-Theory remains unknown, still we can say some things about it:
in particular its low energy limit is eleven dimensional. This was
first observed \cite{Townsend:1995kk,Witten:1995ex} in the type IIA
superstring theory, which in the limit of large string coupling `decompactifies':
its spacetime is seen to be $R^{10}\times S^{1}$ rather than $R^{10}$
and the size of the circle is set by the string coupling, so that
both tend to infinity in this limit. We have then, in the low energy
limit of M-Theory, an eleven dimensional, supersymmetric theory of
gravity, so in light of the uniqueness of the eleven dimensional supergravity
mentioned above it should not be surprising that this is, in fact,
eleven dimensional supergravity. The strong coupling limit of the
$E_{8}\times E_{8}$ heterotic superstring is also eleven dimensional
supergravity, as discussed below. The other superstring theories,
the type IIB, $SO(32)$ heterotic and the type I do not have clear
eleven dimensional limits themselves but are continuously related
to the type IIA and $E_{8}\times E_{8}$ heterotic theories and so
must be part of the same picture.

With the unification of the various superstring theories into M-Theory,
we then have progress of a sort towards contact with the real world.
Though we have lost (at least for now) a complete description of our
theory of the world, at least there is only the one and we know its
low energy limit; which limit would seem to be necessarily an important
part of any phenomenological model, even if not absolutely the whole
story. Clearly however further effort is needed to describe the real
world. We have an eleven dimensional theory (or ten dimensional in
some limits) and a four dimensional world. One way to progress is
to adopt a Kaluza-Klein picture: we compactify the extra dimensions
so that we have four visible, extended dimensions and a tower of massive
excited states which are too massive for us to have yet observed due
to the small size of the extra dimensions. A particularly attractive
option for a supersymmetric theory is to compactify on a Calabi-Yau
manifold. This is a six dimensional space (though they can also be
defined in other even dimensions) whose essential property for this
purpose is that it has a single covariantly constant spinor. This
means that the single supersymmetry parameter of the eleven (or ten)
dimensional theory gives only a single supersymmetry parameter in
five (or four) dimensions and so we are left after compactification
with an $\m N=1$ supersymmetric theory. If we started in eleven dimensions
we still have one yet to compactify in order to describe the real
world, which might be useful phenomenologically; alternatively we
could compactify the eleven dimensional model directly over a $G_{2}$
manifold, which is a seven dimensional analogue of a Calabi-Yau manifold
sharing the property of having a single covariantly constant spinor.
The demand of supersymmetry after compactification allows us to make
contact with the phenomenology of supersymmetric field theory and
thus gives us a route to physically plausible scenarios, which with
no such restriction on the possible compactifications would be a challenge
of daunting complexity. It also allows us to retain attractive features
of supersymmetric extensions to the standard model such as resolving
the hierarchy problem with spontaneous breaking of supersymmetry.

\section{Ho\v{r}ava-Witten Theory}

Heterotic M-Theory, which was invented by Ho\v{r}ava and Witten \cite{Horava:1995qa,Horava:1996ma},
is eleven dimensional supergravity with two boundaries, one at each
end of the eleventh dimension. It represents the limit of the heterotic
$E_{8}\times E_{8}$ string theory which is both low energy and strongly
coupled; dimensional reduction of the interval gives the supergravity
which is the low energy limit of that theory. The construction of
the theory depends intimately on anomaly cancellation. The supergravity
in the bulk induces gravity on the boundaries and the bulk gravitino
becomes chiral on the boundaries. Thus the boundaries are ten dimensional
spacetimes with gravity and chiral fermions, and so they have a gravitational
quantum anomaly. In order to cancel this anomaly, one is forced to
introduce further chiral fermions on the boundary (as well as some
Green-Schwarz terms in the bulk). This can be done by adding an $\m N=1$
Yang-Mills supermultiplet (which in ten dimensions is the maximal
supersymmetric theory that is not a supergravity) with the gauge group
chosen to give the right number of fermions to cancel the gravity
anomaly. This requirement dictates that the gauge group be $E_{8}$,
and so we have two independent $E_{8}$ gauge theories, one on each
boundary, which gives us the $E_{8}\times E_{8}$ gauge group of the
string theory on compactification of the eleventh dimension. The Yang-Mills
theory also has a gauge anomaly in ten dimensions which must be cancelled
by coupling the three-form of the bulk supergravity to the boundary
so that a classical variation of the three-form can cancel the gauge
quantum anomaly on the boundary. This structure rigidly determines
the theory since we have the maximal supersymmetric theory in the
bulk and the maximal (non-gravitational) supersymmetric theory on
the boundary with the gauge group and couplings between the two dictated
by anomaly cancellation.

\subsection{Phenomenology}

The heterotic theory is an attractive corner of M-Theory for phenomenological
model building. Since it contains $E_{8}$ gauge theory as an intrinsic
part, it is a natural place to build particle physics models by breaking
the $E_{8}$ gauge symmetry to give a grand unified theory or even
a theory with the gauge group of the standard model itself. This is
of course true of both ($E_{8}\times E_{8}$) heterotic string theory
and Heterotic M-Theory and much phenomenological work was done on
the string theory before the advent of M-Theory (as well as after).
Since models of the real world (at least the real world today) are
necessarily low energy ones though, the limit taken in Heterotic M-Theory
is more or less forced upon us and so this is the natural place to
build particle physics models as we do not need then to assume weak
string coupling. On the other hand in cosmological contexts it might
well prove not to be the case that the low energy limit is sufficient,
and since the full M-Theory is not known the string theory is the
only way to relax this limit and gain insight into such situations.

To turn an eleven dimensional theory into a model of the real world,
we need to hide seven dimensions somehow. As mentioned above one way
of doing this is to take those seven dimensions to be small, and to
get an $\m N=1$ supersymmetric theory in four dimensions the extra
dimensions must be a $G_{2}$ manifold. In the context of Heterotic
M-Theory one of those seven must be the interval since the standard
model particles are supposed to come from the Yang-Mills fields on
one of the boundaries, meaning that the boundaries must span the extended
dimensions. The other six dimensions are curled up into some small
manifold, for which an attractive choice is a Calabi-Yau manifold.
For a given topology of the Calabi-Yau manifold, we can describe the
geometry of the compact space by some (geometrical) `moduli' which
measure the size of the eleventh dimension (the interval), the overall
volume of the Calabi-Yau and the size of any topologically non-trivial
cycles of the Calabi-Yau. A particular model will be given by specifying
a background with a particular compactification manifold having some
values of these moduli and possibly also non-zero configurations of
the fields (which might give some additional moduli) which solve the
eleven dimensional equations of motion. The effective four dimensional
theory is then the theory of the fluctuations around this background.
For a realistic model we obviously need ultimately to have an effective
theory which resembles the standard model; models close to the MSSM
are a useful intermediate step in string or M-Theory phenomenology.
We also need a stable background, which means that we need some mechanism
for making sure that the moduli which specify the particular model
we are looking at are fixed.

A substantial amount of work has been done building particle physics
models with increasing degrees of similarity to the standard model.
For example, fairly soon after the invention of Heterotic M-Theory,
models were constructed with the $E_{8}$ on the visible brane broken
to the standard model gauge group \cite{Donagi:1999ez} by choosing
a particular compactification background. Recently techniques have
been developed for computational searches of heterotic models \cite{Anderson:2009mh,He:2011rs}
and a large class of `heterotic standard models' with `the precise
matter spectrum of the MSSM, at least one pair of Higgs doublets,
the standard model gauge group and no exotics charged under the standard
model of any kind' have been identified \cite{Anderson:2011ns}. There
has also been recent progress on the problem of moduli stabilization
in the context of particle physics model building where \cite{Anderson:2011cz,Anderson:2011ty}
it has been shown that it is possible to stabilize all the geometric
moduli using the gauge bundle plus non-perturbative effects, for some
choices of gauge bundle.

Given the successes of Heterotic M-Theory for particle physics model
building, it is desirable also to have cosmological models constructed
in the same framework. This motivated the first cosmological solutions
\cite{Lukas:1998qs} and continues to motivate, for example, models
of assisted inflation driven by M5-brane dynamics \cite{Becker:2005sg,Moniz:2009ax}.
Heterotic M-Theory is also a natural home for braneworld cosmological
models, about which more below, the famous example being the ekpyrotic
model \cite{Khoury:2001wf} in which the big bang is generated by
a collision of branes in a five dimensional bulk, which is of continued
interest \cite{Lehners:2006pu,Lehners:2010ug} as an alternative to
inflation.

\section{Gravity with Boundaries}

The question of how to construct consistently a theory of gravity
which includes boundaries is one of direct relevance for Heterotic
M-Theory and for braneworld models in general. For a theory of gravity
without boundaries, the variation of the Einstein-Hilbert action vanishes
for metrics that solve the Einstein equations. However in the presence
of a boundary this is no longer true. Even when the variation of the
metric is constrained to vanish on the boundary, there is still a
non-vanishing boundary term involving the normal derivative of the
variation. In order to make the variational principle consistent,
it is necessary to introduce boundary terms into the action. The Einstein-Hilbert
action in the bulk is made consistent by the inclusion of the Gibbons-Hawking
action on the boundary \cite{Gibbons:1976ue}. More complex gravity
theories in the bulk also require additional surface terms. Of relevance
for us will be supergravity, where a gravitino boundary contribution
is required by similar considerations \cite{Luckock:1989jr}, and
Lovelock gravity, where the higher curvature terms require corresponding
Myers terms on the boundary \cite{Myers:1987yn}.

\section{Improved Heterotic M-Theory}

The original Ho\v{r}ava-Witten formulation of Heterotic M-Theory has
a serious problem. It is only entirely well defined up to first order
in the eleven-dimensional gravitational coupling, $\m O(\kappa^{2/3})$.
It was described as an orbifold $M_{10}\times S^{1}/\mathbb{Z}_{2}$,
which is equivalent to an interval with boundaries at the fixed points
of the $\mathbb{Z}_{2}$ symmetry and which was considered to be technically
convenient. In this description the coupling of the bulk three-form
to the boundary is accomplished with $\delta$-function sources in
the Bianchi identity of its field strength. Imposition of supersymmetry
of the action at the orbifold fixed points then brought terms proportional
to $\delta(0)$ into the action at $\m O(\kappa^{4/3})$. As Heterotic
M-Theory is only supposed to be an effective theory of a particular
low-energy limit of a complete (albeit still unknown) quantum M-Theory
this inconsistency need not necessarily be considered problematic
as long as one is content to work within the approximations it entails.
In practice however, this may in fact cause difficulties: taking reasonable
values for the GUT scale and Newton's constant, it has been found
\cite{Banks:1996ss} that the expansion parameter of the theory, $\epsilon=\kappa^{2/3}\rho V^{-2/3}$
is of order one (where $\rho$ is the size of $S^{1}/\mathbb{Z}_{2}$
and $V$ is the volume of the Calabi-Yau manifold used in the compactification
to a four dimensional theory). This calls into question the validity
of the first order expansion in $\kappa^{2/3}$. More recently, terms
second order in $\kappa^{2/3}$ have been found to be important in
calculating the back-reaction of anti-branes \cite{Gray:2007zza}
and the contribution of gaugino condensation in the presence of anti-branes
\cite{Gray:2007qy}, using such $\m O(\kappa^{4/3})$ terms involving
gauge matter fields as were considered able to be computed reliably
\cite{Lukas:1997fg}.

These phenomenological considerations give one motivation for relaxing
the restriction of Heterotic M-Theory to first order in $\kappa^{2/3}$.
On the other hand, if one considers Heterotic M-Theory as supergravity
on a manifold with boundary it is an attractive theory: eleven-dimensional
supergravity is the maximal supergravity and is unique, and Heterotic
M-Theory is the only way to include boundaries without introducing
gravitational anomalies. However, considered just as a supergravity
theory it is highly unsatisfactory that it should be well-defined
only up to first order in the gravitational coupling, providing another
motivation for improvement.

The problem has been addressed by Ian Moss who has constructed an
improved version of Heterotic M-Theory \cite{Moss:2003bk,Moss:2004ck,Moss:2005zw,Moss:2008ng}
which is consistent and supersymmetric to all orders in $\kappa^{2/3}$.
This construction is performed taking the view of the eleventh dimension
as an interval with boundaries, with careful attention paid to the
boundary conditions which must be satisfied by the bulk fields. The
structure of the eleven-dimensional improved theory is described in
Section \ref{cha:The-11d-Theory}. It is constructed as an expansion
in the curvature up to $R^{2}$ terms and at each order is rigidly
constrained by the requirements of anomaly cancellation at the boundaries
and supersymmetry, leaving only one free parameter: the gravitational
coupling, $\kappa$.

The problems in the Ho\v{r}ava-Witten version of the theory seem to
arise principally from its failure to fully account for the effect
of the energy-momentum density localized on the orbifold fixed planes.
Its effect as a source for the flux $G$ was carefully taken into
account by modifying the Bianchi identity for $G$, but it must also
be considered as a source for the spacetime curvature of the bulk,
and possibly also for the gravitino, modifying the junction conditions
across the brane. In particular, given some energy-momentum localized
on a codimension one surface, the Israel junction conditions \cite{Israel:1966rt},
which relate the extrinsic curvature of the surface to its energy-momentum,
must apply. In order for these to be imposed by the action, the Gibbons-Hawking
term must be included on the boundary. One might wonder whether it
could be possible to impose the junction conditions in the upstairs
picture by modifying the Bianchi identity for the Riemann curvature
to $\nabla_{[I|}?R^{J}{}_{K|LM]}?=\delta\left(x^{11}\right)\ldots$
in analogy with the modification of the Bianchi identity for $G$.
However, the boundary action approach seems more natural to us and
we work here in the manifold with boundary picture everywhere. This
picture also does not require carrying around a redundant, and physically
irrelevant, copy of the bulk as the orbifold picture does. As for
the gravitino, it will turn out (Section \ref{sec:11d-Boundary-Conditions})
that supersymmetry of the boundary forces us to include the Yang-Mills
fields in its boundary condition as well.

\section{Dimensional Reduction of the Improved Theory}

We will describe in Chapter \ref{cha:Heterotic-Reduction} the reduction
to five dimensions of the improved Heterotic M-Theory over a Calabi-Yau
manifold, to find the model analogous to that found in \cite{Lukas:1998tt}
for the original Ho\v{r}ava-Witten theory but with the inclusion here
of terms with up to two fermions, which have not been presented either
for the original or for the improved theory before our work \cite{Moss:2011pi}.
The five dimensional reduction is a natural intermediate step to phenomenological
models since the scales in the problem suggest \cite{Banks:1996ss}
that the size of interval (the eleventh dimension) is about ten times
the length scale of the Calabi-Yau, and the latter can therefore be
integrated out before the former. It is also a useful starting point
for the introduction of five-branes \cite{Lukas:1999kt}, which seem
to be vital for many phenomenological applications, and anti-five-branes
\cite{Gray:2007zza}, and also for the study of gaugino condensation
\cite{Gray:2007qy}. The five dimensional reduction of the improved
version of Heterotic M-Theory is then a first step in assessing the
impact on phenomenology of a theory consistent to all orders in $\kappa^{2/3}$.

The topic of gaugino condensation has received some attention already,
as discussed in Section \ref{sec:Gaugino-Condensation}. It was considered
in \cite{Ahmed:2008jz,Ahmed:2009ty} in a much simpler reduction,
in which attention was focused on the gravitino and the Calabi-Yau
volume modulus. It was found there that the condensate gives a contribution
to the flux through the boundary condition on the supergravity three-form
and also induces a twist in the chirality condition on the gravitino
between the two boundaries. The latter is a particular effect of the
improved theory where the gauginos appear in the boundary condition
of the gravitino. The twisted boundary conditions break supersymmetry
and give a Casimir contribution to the vacuum energy which can lift
the cosmological constant to give de Sitter vacua. Since in this scenario
both moduli stabilization and uplift depend on the gaugino condensate
they naturally have similar scales, though fine tuning of a parameter
in the superpotential is still required to give a small four dimensional
cosmological constant. Since the improved theory is valid to all orders
in the gravitational coupling warping of the bulk metric which is
not small can be consistently accommodated when considering the gaugino
condensate, though for the Casimir energy calculations in \cite{Ahmed:2009ty}
small warping was still assumed.

The other motivation for this work, apart from any possible consequences
for the phenomenology of Heterotic M-Theory, is the investigation
of supergravity on a manifold with boundary per se. Such supergravities
with boundary matter have recently been constructed; the most detailed
are in three dimensions \cite{Belyaev:2007bg,Belyaev:2010as}, where
an off-shell formulation is available, but there are also models in
five \cite{Belyaev:2005rt} and seven \cite{Pugh:2010ii} dimensions.
Five dimensional theories are obviously of particular interest since
their four dimensional boundaries might correspond to the physical
universe. The contribution of the present work in this context is
that by dimensionally reducing a consistent eleven-dimensional supergravity
with boundaries we find an explicit example of a consistent five-dimensional
supergravity on a manifold with boundary, which includes boundary
matter, and we do so without having to include distributions in the
theory. In the absence of a general formalism in five dimensions,
hopefully such an example is of some interest.

\section{Braneworlds}

In order to hide extra dimensions, rather than postulating that the
extra dimensions are just too small to be observed à la Kaluza-Klein,
we could allow them to be arbitrarily large if we suppose that we
are constrained to live on a four dimensional brane and so cannot
see them. This braneworld scenario is further motivated by the possibility
that something like Heterotic M-Theory might be the fundamental theory
of the universe, since in Heterotic M-Theory the fields which give
rise to standard model matter are perforce confined to branes.

The simplest model of a braneworld is a scalar field model, which
can be arranged to have topological defects representing the branes
and the fluctuations of the scalar field (the `matter' in the theory)
can be confined to the defect. Even if such models do not give a realistic
picture of the universe, their simplicity makes them amenable to analysis,
in particular to numerical simulation, and one can hope that studying
such models may give insight into some generic features of braneworld
scenarios.

Generically, one would expect gravity to propagate away from the brane
into the bulk spacetime, and so the effects of large extra dimensions
would be visible in gravitational effects even if they are hidden
from standard model type processes. However, if the bulk spacetime
is appropriately warped then gravity can be confined to the brane
as well, as was demonstrated by the Randall-Sundrum model \cite{Randall:1999vf},
which fact makes it plausible that the universe today might be described
as a braneworld with large extra dimensions even though gravity appears
in experiments to date to be just as four dimensional as the other
forces of nature (at least up to solar system scales).

The braneworld picture also provides an alternative to inflation as
a description of the very early universe. In the ekpyrotic scenario
\cite{Khoury:2001wf} the big bang is caused by a collision of branes
in a five dimensional bulk. The horizon and flatness problems are
solved not by superluminal expansion, as in inflationary theories,
but rather by the dynamics of the branes, by letting the branes approach
each other rather slowly before the collision so that areas widely
separated relative to the Hubble scale after collision were in causal
contact before it.

In light of this ekpyrotic cosmological scenario, it is interesting
to study the topic of brane collision in scalar field theory models
of braneworlds, with an eye on learning about the generic features
of brane collisions \cite{Gibbons:2006ge,Saffin:2007ja,Saffin:2007qa,Takamizu:2004rq,Takamizu:2006gm,Takamizu:2007ks}.
A contribution to this effort is described in Chapter \ref{cha:Colliding-Branes},
concerning the nature of the singularities which may be formed by
the collision of scalar field theory domain walls.

Braneworld models can also provide a means of modifying (four dimensional)
gravity on large scales, as in the DGP model \cite{Dvali:2000hr}.
Here the bulk spacetime is flat, rather than warped as it is in the
Randall-Sundrum case, and four dimensional gravity is provided by
having an Einstein-Hilbert term on the brane, built from the induced
metric, as well as one in the bulk built from the full metric. In
this theory gravity appears four dimensional at short distances but
five dimensional at long distances. There is an extra scalar degree
of freedom, but this has only second order equations of motion even
though the Lagrangian contains higher derivative terms. This property
allows the construction of solutions in which the scalar field is
strongly coupled at small scales and its fluctuations are suppressed
by that strong coupling. This sort of hiding of scalar fields in modified
gravity by strong coupling at small scales is known generically as
Vainshtein screening. The importance of this phenomenologically is
that the scalar field can be hidden from observation on solar system
scales, which have been well tested, while being active on larger
scales to affect the expansion of the universe.

The attractive properties of this scalar when considered in the four
dimensional effective theory inspired a generalization to find a broader
class of theories having a scalar with similar couplings to gravity
and to itself as the one in the DGP model and also retaining second
order equations of motion \cite{Nicolis:2008in}. The outcome was
a set of theories of gravity coupled to a scalar having a `Galilean'
symmetry, leading the scalar field to be christened the `galileon'.
The restriction of Galilean symmetry and second order equations of
motion resulted in an action with just five free parameters. As the
DGP model is a particular example of these galileon theories, it is
natural to ask whether there is a braneworld description of the whole
class. Indeed there is, as first described in \cite{deRham:2010eu}.
The picture is similar to DGP, with a four dimensional probe brane
in a five dimensional spacetime and the scalar field being just the
displacement of the brane in the extra dimension. Now, however, we
include in the brane action all the curvature terms that give second
order equations of motion, that is the Lovelock terms in the intrinsic
curvature \cite{Lovelock:1971yv} and the Myers terms in the extrinsic
curvature \cite{Myers:1987yn}. In the number of dimensions here,
in addition to the Einstein-Hilbert term there is a cosmological constant;
the first two Myers terms; and a tadpole term. One can also consider
such theories with de Sitter or anti-de Sitter rather than Minkowski
spaces, as was done in \cite{Goon:2011qf}. On the other hand from
the four dimensional field theory perspective it is also natural to
consider how one might generalize to more than one galileon field
\cite{Padilla:2010de,Padilla:2010ir}, though in such theories one
encounters a rapidly proliferating number of terms in the action as
more galileon fields are added. To describe such theories in a braneworld
picture we need a higher codimension bulk, as each galileon field
corresponds to the fluctuations in one of the extra dimensions. This
is more challenging than in the codimension one case as we can no
longer describe the branes as boundaries and the toolkit for building
the actions is less well developed. A particular model has been found
for the most symmetric case with a Minkowski brane in a Minkowski
bulk \cite{Hinterbichler:2010xn}. In Chapter \ref{cha:Galileons}
we develop the higher codimension construction allowing for other
maximally symmetric spaces. In order to do so we develop some machinery
that may be useful for a further generalization to the full class
of models which can be constructed in the field theory approach.

\subsection*{Statement of Original Research}

Chapter \ref{cha:Heterotic-Reduction} describes work previously reported
in \cite{Moss:2011pi} done by the author in collaboration with Ian
Moss and Paul Saffin. The original research is described in Sections
\ref{sec:Reduction-Ansatz} to \ref{sec:The-Boundaries} and \ref{sec:Reduction_Conclusions},
with the results summarized in Appendix \ref{cha:Summary}; the work
relating to the bosonic fields obviously has strong parallels with
\cite{Lukas:1998tt} which started from Ho\v{r}ava and Witten's eleven
dimensional theory \cite{Horava:1995qa,Horava:1996ma}, whereas the
work here starts from Ian Moss's improved version \cite{Moss:2003bk,Moss:2004ck,Moss:2005zw,Moss:2008ng}.
Section \ref{sec:Gaugino-Condensation} describes some work by Ahmed
and Moss \cite{Ahmed:2008jz,Ahmed:2009ty}, and not by the author,
to illustrate a consequence of the reduction of the improved theory.

Chapter \ref{cha:Colliding-Branes} describes work previously reported
in \cite{Omotani:2011un} done by the author in collaboration with
Jorma Louko and Paul Saffin. The original results are described in
Sections \ref{sec:Asymptotic-Structure}, \ref{sec:Dynamical-Solutions}
and \ref{sec:Simulation-Results} to \ref{sec:Colliding_Branes_Conclusions}.

Chapter \ref{cha:Galileons} describes previously unpublished work
done by the author in collaboration with Ian Moss, Antonio Padilla
and Paul Saffin. The original research is described in Sections \ref{sec:galileons-formalism}
to \ref{sec:Galileons_Conclusions} with further details of the calculations
given in Appendix \ref{cha:galileon-appendix}.

\chapter{The Reduction to Five Dimensions of Heterotic M-Theory\label{cha:Heterotic-Reduction}}

The improved version introduced by Ian Moss \cite{Moss:2003bk,Moss:2004ck,Moss:2005zw,Moss:2008ng}
cured Heterotic M-Theory of the inconsistency of the original formulation
of Ho\v{r}ava and Witten \cite{Horava:1995qa,Horava:1996ma} beyond
leading order in the gravitational coupling. Here we give a self-contained
description of the improved theory and its reduction to five dimensions.
The aim of the reduction is to provide a starting point for exploring
the consequences for the phenomenology of the theory, which has been
extensively investigated in the original version.

We begin in Section \ref{cha:The-11d-Theory} with a derivation of
the improved version of the theory. The purpose of this is to show
how its structure is entirely determined by the requirements of anomaly
cancellation and supersymmetry and also to make explicit the form
of the eleven dimensional theory from which our reduction begins.
We will perform the reduction restricted to terms with two fermions
or less and so for simplicity we neglect higher terms in the eleven
dimensional description as well, though they were given in the original
presentation of the improved theory \cite{Moss:2004ck}. The difference
between the Ho\v{r}ava-Witten theory and the improved version is that
in the manifold-with-boundaries picture, which we use everywhere,
the coupling of the bulk three-form of the eleven dimensional theory
to the boundaries can be simply given by a boundary condition, in
which distributions do not appear, rather than by modifying its Bianchi
identity, as was done by Ho\v{r}ava and Witten working on the orbifold
$S^{1}/\mathbb{Z}_{2}$, where distributions do appear. The importance
of avoiding distributions is that in the Ho\v{r}ava-Witten theory
they contrive to appear in the boundary action as $\delta(0)$ terms,
which can be swept under the carpet at leading order in the gravitational
coupling but make the theory inconsistent at higher orders. In the
improved theory the other bulk fields, the metric and the gravitino,
are given a consistent description by the inclusion of boundary terms
in the action and have non-trivial boundary conditions which include
contributions from the gauge fields on the boundary.

We perform a reduction to find the equivalent, for the improved version,
of the five dimensional theory found by Lukas et al.\ from the Ho\v{r}ava-Witten
version of Heterotic M-Theory \cite{Lukas:1998tt}. Here we include
everywhere the fermion sector (up to terms with two fermions), but
the major difference of course is in the treatment of the boundaries.
Thus the main results here are the boundary action and the boundary
conditions on the fields derived from the eleven dimensional three-form.
These include fermion bilinear terms, which are particularly important
if there is a gaugino condensate on the boundary as they give the
coupling of the condensate to the bulk fields. Like \cite{Lukas:1998tt}
we allow the Hodge number $h^{1,1}\geq1$ to be arbitrary but take
$h^{2,1}=0$ ($h^{p,q}$ is the number of independent, covariantly
constant $(p,q)$-forms). A full reduction to include $h^{2,1}>0$
would be a logical future extension of the work described here. The
reason we include the $h^{1,1}>1$ sector first is that there is a
`non-zero mode' induced by boundary sources whose charges are related
to the $h^{1,1}$ basis elements of (1,1)-forms on the Calabi-Yau
space and so to fully describe this effect we must include the whole
$h^{1,1}>1$ sector. One might also hope that the complex structure
moduli, which are associated with the $h^{2,1}$ (2,1)-cycles of the
Calabi-Yau, are stabilized at a higher scale than the Kähler moduli
associated with the $h^{1,1}$ (2,2)-cycles. In that case, which seems
to be plausible both in the IIB case \cite{Kachru:2003aw} and in
Heterotic M-Theory \cite{Anderson:2011cz}, the low energy theory
without a (2,1) sector and the stabilization of the complex structure
moduli could be treated independently.

We perform the reduction entirely directly (apart from the potential
of the chiral multiplets on the boundary, where we use the structure
of the general four dimensional supersymmetric theory to help us:
Section \ref{sub:Chiral-Multiplets}) rather than by making use of
the known five dimensional supergravity theories and four dimensional
super-Yang-Mills theories to construct the reduced theory once the
field content has been ascertained, as in \cite{Lukas:1998tt}. The
reason for this is that although the general forms of the five dimensional
supergravity and the four dimensional Yang-Mills theory are known
the supersymmetric coupled theory with boundary matter and bulk, five
dimensional supergravity is not known. Therefore we must derive the
coupling from that in the eleven dimensional theory, which we do know.
As we will see in Section \ref{cha:The-11d-Theory} the eleven dimensional
theory is rigidly constructed: there is only one way of coupling the
bulk supergravity to the boundary matter and the gauge coupling constant
is fixed in terms of the gravitational coupling. If we want to make
sure that the reduced theory is coupled according to the dictates
of the eleven dimensional theory then we need to know not just the
form of the five and four dimensional theories but also their relative
normalizations. In addition to the explicit boundary conditions, there
are bulk fields in the boundary action which we would not be able
to find in the reduced theory other than directly. The direct reduction
will give us, by virtue of the eleven dimensional supersymmetry and
the choice of a Calabi-Yau as the reduction manifold, a supersymmetric
supergravity-with-boundary-matter theory whose bulk-boundary couplings
we could not find exactly by comparison with a `bottom-up' approach,
because the general coupled theory is not known in five dimensions.
Thus performing a direct reduction of the bulk and boundary actions
and of the boundary conditions from eleven to five dimensions is necessary
to give us a reduced theory which respects the rigid constraints which
anomaly cancellation and supersymmetry place upon the bulk-boundary
coupling in eleven dimensions.

\begin{wrapfigure}{O}{0.5\columnwidth}%
\noindent \begin{centering}
\def\svgwidth{200pt}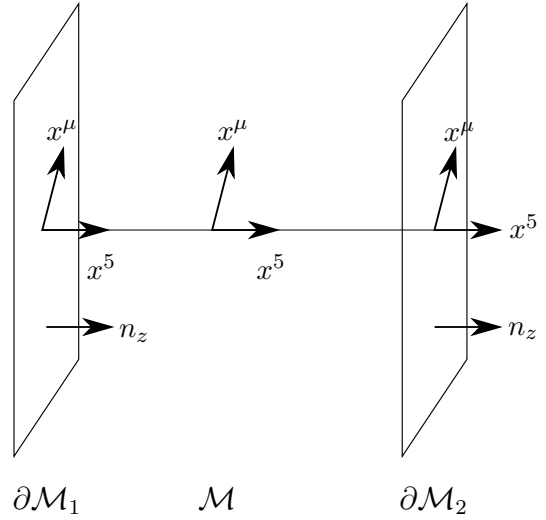
\par\end{centering}

\caption{The five dimensional spacetime}
\end{wrapfigure}%
We perform the reduction with the standard embedding of the spin connection
into the $SU(3)$ subgroup of the $E_{8}$ gauge group on $\partial\m M_{1}$,
which breaks it to $E_{6}$. If both gauge groups are left unbroken
then the sources for the non-zero mode on either boundary have opposite
signs and so we cannot match both with a constant background flux
(as would occur in Section \ref{sub:Spin_embedding} if we set $\tr F^{2}=0$
on $\partial\m M_{1}$ instead of using the standard embedding). Therefore
some measure needs to be taken to give us equal sources on either
boundary (unless perhaps one has five-branes in the bulk to provide
extra sources, but we do not consider these here). The standard embedding
gives us an explicit and fairly simple way of achieving this which
is familiar from the literature, thus facilitating comparison and
possible extension to non-standard embeddings that have been considered
before (in \cite{Lukas:1998hk} for example).

The reduction splits an eleven dimensional spacetime, with boundaries,
into the product of a five dimensional spacetime with boundaries and
a Calabi-Yau threefold and we perform a Kaluza-Klein reduction to
five dimensions. The reduction proceeds in several stages. First,
in order to obtain the five dimensional theory we need to find a suitable
ansatz for the split of the eleven dimensional fields into their five
dimensional spacetime and their Calabi-Yau components. The ansatz
we use and the reasons for choosing it are described in Section \ref{sec:Reduction-Ansatz}.
Having found a suitable ansatz, we substitute it into the eleven dimensional
action, boundary conditions and supersymmetry transformations, and
then turn the expressions we thus find into products of spacetime
and Calabi-Yau parts. We integrate out the Calabi-Yau space in the
action and identify the parts of boundary conditions and supersymmetry
transformations which share identical Calabi-Yau components, in order
to identify the boundary conditions and supersymmetry transformations
of the five dimensional fields. We omit the details of this set of
rather lengthy calculations, just giving the results through Sections
\ref{sec:The-Bulk} and \ref{sec:The-Boundaries}, to spare the reader
the large number of terms resulting from the product of combinations
of spacetime and Calabi-Yau components of tensors with the various
combinations of fermions. Finally we organize the results we have
found to show the structure of the five dimensional supergravity in
the bulk and the four dimensional super-Yang-Mills on the boundaries.
We find the supergravity structure by dualizing the four-form field
strength $G_{\alpha\beta\gamma\delta}$ resulting from the reduction
to a scalar $\sigma$ (Section \ref{sec:Dualization}) and identifying
a hypermultiplet with quaternionic structure (Section \ref{sub:Hypermultiplet-Quaternionic-Structure}).
The identification, on the boundary $\partial\m M_{1}$, of the scalars
and fermions in the fundamental representation of $E_{6}$ as chiral
multiplets is described in Section \ref{sub:Chiral-Multiplets}.

A comprehensive list of the conventions used in this chapter for the
metric, gamma matrices, naming of indices, etc.\ is given in Appendix
\ref{cha:Conventions}. The model we find as a result of the reduction
is summarized in Appendix \ref{cha:Summary} where the full action,
boundary conditions and supersymmetry transformations are gathered
together.

\section{The Eleven Dimensional Theory\label{cha:The-11d-Theory}}

\subsection{Supergravity With Boundaries}

If we introduce boundaries into a theory described by an action we
must take care that the variational principle that gives us the equations
of motion from the action remains well defined. Let us first consider
the simple example of a free scalar field on a manifold $\m M$. In
the absence of a boundary the actions $S=-\int_{\m M}dv\nabla_{I}\phi\nabla^{I}\phi$
and $S^{\prime}=\int_{\m M}dv\phi\nabla^{2}\phi$ are equivalent:
they are related by the total derivative $-\int_{\m M}dv\nabla_{I}\left(\phi\nabla^{I}\phi\right)$
which vanishes if the boundary is empty. In either case $\delta S=\delta S^{\prime}=2\int dv\delta\phi\nabla^{2}\phi$
giving us the equation of motion $\nabla^{2}\phi=0$. However, consider
now the variation of $S^{\prime}$ \emph{with} a boundary, $\partial\m M$.
Now $\delta S^{\prime}=\int_{\m M}dv\delta\phi\nabla^{2}\phi+\int_{\partial\m M}dv\left(\phi\nabla_{N}\delta\phi-\delta\phi\nabla_{N}\phi\right)$
where $\nabla_{N}=n^{I}\nabla_{I}$ is the component of the derivative
parallel to the outward unit normal vector to the boundary, $n^{I}$.
This is not consistent since given an arbitrary variation $\delta\phi$,
the derivatives of the variation are determined, and so $\nabla_{N}\delta\phi$
is not independent. Since $\nabla_{N}$ is perpendicular to the boundary
$\partial\m M$, we cannot get rid of it by integrating by parts.
Instead we must introduce a boundary term $S_{b}^{\prime}=-\int_{\partial\m M}dv\phi\nabla_{N}\phi$
whose variation, $\delta S_{b}^{\prime}=-\int_{\partial\m M}dv\left(\delta\phi\nabla_{N}\phi+\phi\nabla_{N}\delta\phi\right)$
cancels the $\nabla_{N}\delta\phi$ part of the variation of $S^{\prime}$
and leaves us with a consistent variational principle for $S^{\prime}+S_{b}^{\prime}$.
In this simple case we can see that in fact $S_{b}^{\prime}$ is just
the total derivative $S_{b}^{\prime}=-\int_{\m M}dv\nabla^{I}\left(\phi\nabla_{I}\phi\right)$
which is the difference between $S$ and $S^{\prime}$. Thus $S=-\int_{\m M}dv\nabla_{I}\phi\nabla^{I}\phi$
is in this case a `better' choice of action in the presence of a boundary.
In a less simple case the boundary term will not generally be a total
derivative in this way and we will have to augment the bulk action
with a boundary term, as we will now see for gravity.

Gravity is described by the Einstein-Hilbert action $S_{EH}=\frac{1}{2\kappa^{2}}\int_{\m M}dvR$.
The variational principle gives us the equation of motion 
\[
0=\delta S_{EH}=\frac{1}{2\kappa^{2}}\int_{\m M}dv\left(\frac{1}{2}g^{IJ}R-R^{IJ}\right)\delta g_{IJ}
\]
\[
\Rightarrow R_{IJ}-\frac{1}{2}g_{IJ}R=0
\]
 which is the Einstein equation in vacuum. Once again this variation
includes a total derivative. If we introduce a boundary, $\partial\m M$
to the manifold $\m M$ the boundary terms arising from the total
derivative mean that the variational problem is no longer well posed,
in a similar way to that which we just saw for the scalar field.
\begin{align*}
\delta S_{EH} & =\frac{1}{2\kappa^{2}}\int_{\m M}dv\left(\frac{1}{2}g^{IJ}R-R^{IJ}\right)\delta g_{IJ}\\
 & \quad+\frac{1}{2\kappa^{2}}\int_{\partial\m M}dv\left(-K^{IJ}\delta h_{IJ}-n^{I}h^{JK}\nabla_{I}\delta h_{JK}+2Kn^{I}\delta n_{I}\right)
\end{align*}
where $h_{IJ}$ is the induced metric on the boundary, $n^{I}$ is
the outward pointing unit normal vector on the boundary and $K_{IJ}=?h_{I}^{K}??h_{J}^{L}?\nabla_{K}n_{L}$
is the extrinsic curvature. We only have freedom to vary independently
$\delta g_{IJ}$ and $\delta h_{IJ}$, not also $n^{I}\nabla_{I}\delta h_{JK}$
or $\delta n_{I}$. The resolution is to introduce the Gibbons-Hawking-York
boundary term, that is the extrinsic curvature scalar, $K$, into
the boundary action. $\delta K=\frac{1}{2}h^{IJ}n^{K}\nabla_{K}\delta h_{IJ}-Kn^{K}\delta n_{K}$
and so the variation of the action $\frac{1}{2\kappa^{2}}\int_{\m M}dvR+\frac{1}{2\kappa^{2}}\int_{\partial\m M}dv\left(2K\right)$
depends only on $\delta g_{IJ}$ in the bulk and $\delta h_{IJ}$
on the boundary and we can consistently apply the variational principle.
Here $\int_{\partial\m M}dvK$ is not equivalent to a total derivative
term in the bulk, so we necessarily have an action with both a bulk
and a boundary term.

Heterotic M-Theory is eleven dimensional supergravity on a manifold
with boundary. The requirements of supersymmetry and anomaly cancellation
dictate the structure of the theory completely. We have an eleven
dimensional bulk with two disjoint, spatially separated, ten dimensional
boundaries. In the bulk we have eleven dimensional supergravity \cite{Cremmer:1978km}.
In the presence of a boundary the bulk Einstein-Hilbert action must,
as we have just seen, be supplemented by an extrinsic curvature term
\cite{York:1972sj,Gibbons:1976ue} $\frac{1}{2\kappa_{11}^{2}}\int dv\left(2K\right)$.
The bulk Rarita-Schwinger action also requires a boundary term \cite{Luckock:1989jr}
$\int dv\left(\frac{1}{2}\bar{\Psi}_{A}\Gamma^{AB}\Psi_{B}\right)$.
Finally, in the supergravity action there is torsion whose variation
gives a boundary term and so to cancel this we need an additional
piece in the boundary action \cite{Moss:2004ck} $\int dv\left(\bar{\Psi}_{A}\Gamma^{A}\Psi_{N}\right)$.
So we have as a starting point the action
\begin{align}
S & =\frac{1}{2\kappa_{11}^{2}}\int_{\m{\mathcal{M}}_{11}}dv\m L\left(\m M_{11}\right)+\frac{1}{2\kappa_{11}^{2}}\int_{\partial\m M_{10,1}}dv\m L\left(\partial\m M_{10,1}\right)+\frac{1}{2\kappa_{11}^{2}}\int_{\partial\m M_{10,2}}dv\m L\left(\partial\m M_{10,2}\right)
\end{align}
with
\begin{align}
\m L\left(\m M_{11}\right)=\; & R\left(\Omega\right)+\bar{\Psi}_{I}\Gamma^{IJK}D_{J}\left(\Omega^{*}\right)\Psi_{K}\nonumber \\
 & -\frac{1}{96}\left(\bar{\Psi}_{M}\Gamma^{IJKLMO}\Psi_{O}+12\bar{\Psi}^{I}\Gamma^{JK}\Psi^{L}\right)G_{IJKL}^{*}\nonumber \\
 & -\frac{1}{48}G_{IJKL}G^{IJKL}+\frac{1}{12^{4}}\epsilon^{I_{1}\ldots I_{11}}G_{I_{1}I_{2}I_{3}I_{4}}G_{I_{5}I_{6}I_{7}I_{8}}C_{I_{9}I_{10}I_{11}}\\
\m L\left(\partial\m M_{10,1}\right)=\; & -2K+\frac{1}{2}\bar{\Psi}_{A}\Gamma^{AB}\Psi_{B}-\bar{\Psi}_{A}\Gamma^{A}\Psi_{N}\\
\m L\left(\partial\m M_{10,2}\right)=\; & 2K-\frac{1}{2}\bar{\Psi}_{A}\Gamma^{AB}\Psi_{B}+\bar{\Psi}_{A}\Gamma^{A}\Psi_{N}
\end{align}
where $I,J,K,\ldots$ are eleven dimensional spacetime indices; $A,B,C,\ldots$
are ten dimensional spacetime indices; $K=g^{AB}K_{AB}$ is the trace
of the extrinsic curvature $K_{AB}=?h_{A}^{I}??h_{B}^{J}?\nabla_{I}n_{J}$;
$n^{I}$ is the unit normal vector to the boundary which is inward
pointing on $\partial\m M_{10,1}$ and outward pointing on $\partial\m M_{10,2}$
(so that its definition can be consistently extended across the bulk);
$\Omega_{IJK}=\omega_{IJK}-\frac{1}{4}\left(\bar{\Psi}_{I}\Gamma_{J}\Psi_{K}-\bar{\Psi}_{I}\Gamma_{K}\Psi_{J}+\bar{\Psi}_{J}\Gamma_{K}\right)-\frac{1}{8}\bar{\Psi}^{L}\Gamma_{IJKLM}\Psi^{M}$
where $\omega_{IJK}$ is the Levi-Civita spin connection, the supercovariant
connection is $\hat{\Omega}_{IJK}=\omega_{IJK}-\frac{1}{4}\left(\bar{\Psi}_{I}\Gamma_{J}\Psi_{K}-\bar{\Psi}_{I}\Gamma_{K}\Psi_{J}+\bar{\Psi}_{J}\Gamma_{K}\right)$
and $\Omega^{*}=\frac{1}{2}\left(\Omega+\hat{\Omega}\right)$; the
supercovariant field strength is $\hat{G}_{IJKL}=G_{IJKL}+3\bar{\Psi}_{[I}\Gamma_{JK}\Psi_{L]}$
and $G_{IJKL}^{*}=\frac{1}{2}\left(G_{IJKL}+\hat{G}_{IJKL}\right)$.
The supersymmetry transformations are (with supersymmetry parameter
$S$)
\begin{align}
\delta_{S}?E^{\hat{I}}{}_{J}? & =-\frac{1}{2}\bar{S}\Gamma^{\hat{I}}\Psi_{J}\\
\delta_{S}\Psi_{I} & =D_{I}S-\frac{1}{288}\left(?\Gamma_{I}^{JKLM}?-8\delta_{I}^{J}\Gamma^{KLM}\right)G_{JKLM}S\label{eq:Psi-susy}\\
\delta_{S}C_{IJK} & =-\frac{3}{2}\bar{S}\Gamma_{[IJ}\Psi_{K]}
\end{align}
where $?E^{\hat{I}}{}_{J}?$ is the eleven dimensional vielbein. This
action gives boundary conditions (with the upper signs on $\partial\m M_{10,1}$
and the lower signs on $\partial\m M_{10,2}$)
\begin{align}
K_{AB} & =\mp\kappa_{11}^{2}T_{AB}=0\label{eq:provisional_K_bc}\\
P_{-}\Psi_{A} & =0\label{eq:provisional_Psi_bc}
\end{align}
while the boundary condition
\begin{align}
C_{ABC} & =0\label{eq:provisional_C_bc}
\end{align}
must be imposed in addition to the action, similarly to a Bianchi
identity which is not imposed by variation of an action. The supersymmetry
parameter is chiral ($P_{-}S=0$) on the boundary, with the chiral
projection operator being $P_{\pm}=\frac{1}{2}\left(\id\pm\Gamma_{N}\right)$.

This action is supersymmetric but it has a quantum anomaly since the
gravitino is a chiral fermion on the ten dimensional boundary \cite{Horava:1995qa}.

\subsection{Gravitational Anomaly Cancellation}

The anomaly in ten dimensional may be found using a twelve dimensional
index theorem \cite{AlvarezGaume:1984dr} (I have found \cite{Bilal:2003es,Metzger:2003ud}
helpful expositions of anomaly cancellation and I follow here the
notation of \cite{Bilal:2003es}). The anomalous variation of the
Euclidean quantum effective action under diffeomorphisms, $\delta\Gamma_{E}$
may be calculated from a twelve-form polynomial in the curvature as
\begin{align}
\delta\Gamma_{E} & =-i\int_{\m M_{10}}\hat{I}_{10}^{1}
\end{align}
where $\hat{I}_{10}^{1}$ is given by
\begin{equation}
\dd\hat{I}_{10}^{1}=\delta\hat{I}_{11},\:\dd\hat{I}_{11}=\hat{I}_{12}
\end{equation}
and
\begin{align}
\hat{I}_{12}^{gravitino}=\; & \frac{1}{4}\left(-\hat{I}_{12}^{spin\frac{3}{2}}\left(R\right)+\hat{I}_{12}^{spin\frac{1}{2}}\left(R\right)\right)\nonumber \\
=\; & -\frac{1}{4}\cdot2\pi\cdot\frac{1}{\left(4\pi\right)^{6}}\left(-\frac{495}{5670}\tr R^{6}+\frac{225}{4320}\tr R^{4}\tr R^{2}-\frac{63}{10368}\left(\tr R^{2}\right)^{3}\right)\nonumber \\
 & +\frac{1}{4}\cdot2\pi\cdot\frac{1}{\left(4\pi\right)^{6}}\left(\frac{1}{5670}\tr R^{6}+\frac{1}{4320}\tr R^{4}\tr R^{2}+\frac{1}{10368}\left(\tr R^{2}\right)^{3}\right)\nonumber \\
=\; & \frac{1}{8\left(4\pi\right)^{5}}\left(\frac{496}{5670}\tr R^{6}-\frac{224}{4320}\tr R^{4}\tr R^{2}+\frac{64}{10368}\left(\tr R^{2}\right)^{3}\right)
\end{align}
where $R$ is the curvature two-form with two indices regarded as
$SO(1,9)$ matrix indices so that, for example, $\tr R^{2}=\frac{1}{4}\tr\left(R_{AB}R_{CD}\right)\mathbf{d}x^{A}\wedge\mathbf{d}x^{B}\wedge\mathbf{d}x^{C}\wedge\mathbf{d}x^{D}=\frac{1}{4}?R_{ABE}^{F}??R_{CDF}^{E}?\mathbf{d}x^{A}\wedge\mathbf{d}x^{B}\wedge\mathbf{d}x^{C}\wedge\mathbf{d}x^{D}$.

A theory with local quantum anomalies is not well defined so we must
find a mechanism to cancel them. The factorizable part of the anomaly,
that is the $\tr R^{4}\tr R^{2}$ and $\left(\tr R^{2}\right)^{3}$
terms, can be cancelled by a Green-Schwarz type mechanism (discussed
below) but to cancel the irreducible part, $\tr R^{6}$, it is necessary
to introduce some gauge fields on the ten dimensional boundary so
that the irreducible part of the quantum anomaly of the Yang-Mills
theory can cancel the irreducible part of the anomaly from the gravitino.
The gauge theory has gauge, gravity and mixed anomalies. The pure
gravity part of the Yang-Mills anomaly polynomial is
\begin{align}
\hat{I}_{12}^{YM}\left(R\right) & =-\frac{1}{2}\hat{I}_{12}^{spin\frac{1}{2}}\left(R\right)=-\frac{1}{4\left(4\pi\right)^{5}}\left(\frac{1}{5670}\tr R^{6}+\frac{1}{4320}\tr R^{4}\tr R^{2}+\frac{1}{10368}\left(\tr R^{2}\right)^{3}\right)
\end{align}
so we can see that we need 248 gauginos on each boundary so that $248\hat{I}_{12}^{YM}\left(R\right)$
cancels the $\tr R^{6}$ part of $\hat{I}_{12}^{gravitino}$, thus
cancelling the irreducible part of the anomaly. In the ten dimensional
string theory anomaly cancellation, the possible groups (having rank
496) are $SO(32)$ and $E_{8}\times E_{8}$. Since $SO(32)$ is irreducible
it cannot be split between the two boundaries and so we can only have
$E_{8}\times E_{8}$ with an $E_{8}$ gauge theory on each boundary
giving us the 248 gauginos we need. The criterion for choosing acceptable
gauge groups is that $\tr F^{6}$ must factorize so that the Green-Schwarz
mechanism can be used to cancel all of the remaining anomalies.%
\footnote{Note that for $E_{8}$ we define $\tr=\frac{1}{30}\mathrm{Tr}$ where
$\mathrm{Tr}$ is the trace in the adjoint of $E_{8}$. For $E_{8}$
there is no fundamental representation smaller than the adjoint in
which to define $\tr$. This definition is given by analogy with $SO(n)$
where a distinct fundamental does exist and gives the correct relation
for the fundamental of the $SO(16)$ subgroup of $E_{8}$.%
} For $E_{8}$, $\tr F^{6}=\frac{1}{8}\tr\left(F^{2}\right)^{3}$ and
also $\tr F^{4}=\frac{3}{10}\tr\left(F^{2}\right)^{2}$.

The overall anomaly is
\begin{align}
\hat{I}_{12}=\; & \hat{I}_{12}^{gravitino}+\hat{I}_{12}^{YM}\left(R\right)+\hat{I}_{12}^{YM}\left(R,F\right)+\hat{I}_{12}^{YM}\left(F\right)\nonumber \\
=\; & \frac{1}{8\left(4\pi\right)^{5}}\left(\frac{496}{5670}\tr R^{6}-\frac{224}{4320}\tr R^{4}\tr R^{2}+\frac{64}{10368}\left(\tr R^{2}\right)^{3}\right)\nonumber \\
 & -\frac{1}{4\left(4\pi\right)^{5}}\tr\left(\id_{248}\right)\left(\frac{1}{5670}\tr R^{6}+\frac{1}{4320}\tr R^{4}\tr R^{2}+\frac{1}{10368}\left(\tr R^{2}\right)^{3}\right)\nonumber \\
 & -\frac{2\pi}{2}\left(\frac{-30}{2\left(2\pi\right)^{2}}\right)\tr F^{2}\frac{1}{\left(4\pi\right)^{4}}\left(\frac{1}{360}\tr R^{4}+\frac{1}{288}\left(\tr R^{2}\right)^{2}\right)\nonumber \\
 & -\frac{2\pi}{2}\cdot\frac{30}{24\left(2\pi\right)^{4}}\tr F^{4}\frac{1}{\left(4\pi\right)^{2}}\frac{1}{12}\tr R^{2}\nonumber \\
 & -\frac{2\pi}{2}\cdot\left(\frac{-30}{720\left(2\pi\right)^{6}}\tr F^{6}\right)\nonumber \\
=\; & \frac{1}{\left(4\pi\right)^{5}}\left(-\frac{1}{48}\tr R^{2}\tr R^{4}+\frac{1}{24}\tr F^{2}\tr R^{4}-\frac{1}{192}\left(\tr R^{2}\right)^{3}+\frac{5}{96}\tr F^{2}\left(\tr R^{2}\right)^{2}\right.\nonumber \\
 & \qquad\quad-\frac{1}{8}\left(\tr F^{2}\right)^{2}\tr R^{2}+\frac{1}{12}\left(\tr F^{2}\right)^{3}\nonumber \\
=\; & \frac{1}{24\left(4\pi\right)^{5}}\left(\tr F^{2}-\frac{1}{2}\tr R^{2}\right)\left(\tr R^{4}+\frac{1}{4}\left(\tr R^{2}\right)^{2}-2\tr F^{2}\tr R^{2}+2\left(\tr F^{2}\right)^{2}\right)\nonumber \\
=\; & \frac{1}{12\left(4\pi\right)^{5}}I_{4}\left(I_{4}^{2}+\frac{1}{2}\tr R^{4}-\frac{1}{8}\left(\tr R^{2}\right)^{2}\right)\nonumber \\
=\; & \frac{1}{12\left(4\pi\right)^{5}}\left(I_{4}^{3}+I_{4}X_{8}\right)
\end{align}
where $I_{4}=\tr F^{2}-\frac{1}{2}\tr R^{2}$ and $X_{8}=\frac{1}{2}\tr R^{4}-\frac{1}{8}\left(\tr R^{2}\right)^{2}$

We have cancelled the irreducible part of the gravity anomaly, so
that the anomaly is now factorizable. Therefore it can be cancelled
by classical inflow from the bulk fields, but first we need to find
the boundary conditions consistent with supersymmetry now that we
have a Yang-Mills gauge theory on the boundary and thus a non-zero
energy-momentum localized on the boundary.

\subsection{Supersymmetric Boundary Conditions\label{sec:11d-Boundary-Conditions}}

Since there is a gauge theory on the boundary, the stress-energy tensor
is non-vanishing and so the boundary has extrinsic curvature, from
\eqref{eq:provisional_K_bc}. This makes the boundary condition on
$\Psi_{A}$ and $G$, \eqref{eq:provisional_Psi_bc} and \ref{eq:provisional_C_bc},
inconsistent with supersymmetry since
\begin{align}
\delta_{S}\left(P_{-}\Psi_{A}\right) & =P_{-}\delta_{S}\Psi_{A}\nonumber \\
 & =P_{-}D_{A}S-\frac{1}{288}\left(?\Gamma_{A}^{JKLM}?-8\delta_{A}^{J}\Gamma^{KLM}\right)G_{JKLM}S
\end{align}
but since $P_{-}S=0$, $D_{A}\left(P_{-}S\right)=0$ and so 
\begin{align}
P_{-}D_{A}S & =\left(D_{A}P_{-}\right)S\nonumber \\
 & =\mp\frac{1}{2}K_{AB}\Gamma^{B}S\neq0
\end{align}
which does not vanish when $T^{AB}\neq0$ as \eqref{eq:provisional_K_bc}
shows. Since the remaining part of the variation does vanish if $G=0$,
we see that the non-vanishing energy-momentum now present on the boundary
breaks the supersymmetry of the theory, which we need to restore.
We accomplish this by extending the boundary conditions \eqref{eq:provisional_Psi_bc}
and \eqref{eq:provisional_C_bc}. By writing down all the terms in
the boundary fields that have the right tensor structure and requiring
the variation of the boundary conditions to vanish under the supersymmetry
transformations (\ref{eq:E_susy}-\ref{eq:chi_susy_11d}) it can be
shown that the unique way in which we may recover supersymmetry is
to have
\begin{align}
P_{-}\Psi_{A} & =\mp\frac{\kappa_{11}^{2}}{\lambda^{2}}\left(?\Gamma_{A}^{BC}?-10\delta_{A}^{B}\Gamma^{C}\right)\tr\left(F_{BC}\chi\right)\label{eq:Psi_bc_2}\\
C & =\mp\frac{\kappa_{11}^{2}}{\lambda^{2}}\left(\omega_{3}^{Y}+\omega_{3}^{\chi}\right)\label{eq:C_bc_2}
\end{align}
where $\frac{1}{\lambda^{2}}$ is the coupling constant of the Yang-Mills
theory, $\dd\omega_{3}^{Y}=\tr\left(F\wedge F\right)$ and $\left(\omega_{3}^{\chi}\right)_{ABC}=-\frac{1}{4}\tr\left(\bar{\chi}\Gamma_{ABC}\chi\right)$.

\subsection{Gauge Anomaly Cancellation}

We start by noting that the three-form field in supergravity has (in
the bulk) an Abelian symmetry $C\rightarrow C+\dd a$ since under
this transformation $G$ is invariant and $C\wedge G\wedge G$ is
a total derivative. If we associate this transformation of $C$ to
the gauge and diffeomorphism transformations of the boundary we can
arrange classical inflow from that total derivative in the bulk to
cancel the quantum anomaly on the boundary.

Consider first just the gauge part of the anomaly, i.e.\ just the
$\tr F^{2}$ parts. We choose $a$ so that on the boundary $\dd a=\mp\frac{\kappa_{11}^{2}}{\lambda^{2}}\delta\omega_{3}^{Y}$,
consistent with the boundary condition \eqref{eq:C_bc_2}. Now we
note that the $C\wedge G\wedge G$ term in the bulk action contributes
a total derivative which gives a variation $\frac{1}{12\kappa_{11}^{2}}\int_{\partial\m M_{1}}a\wedge G\wedge G-\frac{1}{12\kappa_{11}^{2}}\int_{\partial\m M_{2}}a\wedge G\wedge G$
and since on the boundary $G=\mp\frac{\kappa_{11}^{2}}{\lambda^{2}}\tr F^{2}+\ldots=\mp\frac{\kappa_{11}^{2}}{\lambda^{2}}I_{4}+\ldots$
this gives a contribution to the anomaly polynomial of $-\frac{1}{12}\frac{\kappa_{11}^{2}}{\lambda^{4}}I_{4}^{3}$.
The gauge part of the anomaly is then cancelled if the gauge coupling
$\frac{1}{\lambda^{2}}$ is given by 
\begin{equation}
\frac{\kappa_{11}^{2}}{\lambda^{2}}=\frac{1}{4\pi}\left(\frac{\kappa_{11}}{4\pi}\right)^{\frac{2}{3}}
\end{equation}

It is interesting to note that one might, as we did initially, take
the supersymmetry transformations of the Yang-Mills theory, \eqref{eq:A_susy_11d}
and \eqref{eq:chi_susy_11d}, to both have the opposite sign, as the
supersymmetry of the Yang-Mills theory itself is obviously unaffected
by the sign of the arbitrary parameter. In that case one finds that
the necessary boundary condition on $G$ has the opposite sign. Then
the gauge anomaly would be cancelled by $\frac{\kappa_{11}^{2}}{\lambda^{2}}$
having the same magnitude, but negative sign. Hence we see that the
requirements of anomaly cancellation give us extra constraints on
the supersymmetry: the relative sign of the supersymmetry parameter
in the transformations of the bulk fields and the transformations
of the boundary fields cannot be freely chosen.

To cancel the full $I_{4}^{3}$ part of the anomaly, we use the same
mechanism with the boundary condition on $C$ extended to
\begin{align}
C & =\mp\frac{\kappa_{11}^{2}}{\lambda^{2}}\left(\omega_{3}^{Y}-\frac{1}{2}\omega_{3}^{L}\right)+\text{fermion terms}\label{eq:bosonic_C_bc}
\end{align}
where $\dd\omega_{3}^{L}=\tr R^{2}$. In order to do this supersymmetrically
we must have $R^{2}$ terms in the boundary action. These can be included
\cite{Moss:2008ng} by constructing a Yang-Mills like `Lorentz multiplet',
inspired by a method introduced by Bergshoeff and de Roo \cite{Bergshoeff:1988nn,Bergshoeff:1989de}
but since we will not use the fermion terms it suffices to say that
the $\tr\left(F_{AB}F^{AB}\right)$ term in the boundary Lagrangian
is extended to $\tr\left(F_{AB}F^{AB}\right)-\frac{1}{2}\tr\left(R_{AB}R^{AB}\right)$.
Since we now have \eqref{eq:bosonic_C_bc} instead of just \eqref{eq:C_bc_2},
the inflow from the $C\wedge G\wedge G$ term completely cancels the
$I_{4}^{3}$ part of the anomaly.

We can cancel the $I_{4}X_{8}$ part of the anomaly by introducing
a Green-Schwarz term in the bulk $S_{GS}=-\frac{1}{\kappa_{11}^{2}12\left(4\pi\right)^{5}}\int_{\m M}C\wedge X_{8}$.
Since $X_{8}$ is invariant and $\dd X_{8}=0$, $\delta S_{GS}=\frac{1}{\kappa_{11}^{2}12\left(4\pi\right)^{5}}\int_{\partial\m M_{1}}a\wedge X_{8}-\frac{1}{\kappa_{11}^{2}12\left(4\pi\right)^{5}}\int_{\partial\m M_{2}}a\wedge X_{8}$.
Hence the contribution of $\delta S_{GS}$ to the anomaly polynomial
is $-\frac{1}{12\left(4\pi\right)^{5}}I_{4}\wedge X_{8}$ cancelling
the second term in the quantum anomaly. Since $S_{GS}$ is $\m O(R^{4})$
and we will be working in the low curvature regime, keeping terms
at most $\m O(R^{2})$, it will not appear henceforth.

\subsection{The Lorentz Multiplet}

We have just seen that in order to cancel the gravitational quantum
anomaly on the boundaries, we must introduce $R^{2}$ terms into the
boundary action. The question then is how to do this in a way that
respects supersymmetry. The $R^{2}$ term appears in the boundary
condition of $G$ as part of $I_{4}=\tr\left(F_{AB}F^{AB}\right)-\frac{1}{2}\tr\left(R_{AB}R^{AB}\right)$;
it appears in exactly the same way as $F^{2}$ except for a factor
of $-\frac{1}{2}$. The way to include the $R^{2}$ terms on the boundary
is, as was explained in \cite{Moss:2008ng}, to exploit this analogy
between the Yang-Mills curvature and the gravitational curvature by
considering the latter as the curvature of a gauge field of the Lorentz
symmetry group $SO(1,9)$. This is done by identifying a `Lorentz
multiplet' built to give supersymmetry transformations as close as
possible to an exact analogy of the transformations of the Yang-Mills
multiplet. Since we know already the supersymmetric action for a Yang-Mills
multiplet we can exploit the analogy to construct a supersymmetric
action for the Lorentz multiplet. The construction in eleven dimensions
is complicated, as compared to the ten dimensional case, by the components
of the spacetime curvature and the gravitino normal to the boundary.
These are necessary for the consistency of the construction, and are
included in \cite{Moss:2008ng}, but for simplicity we will ignore
them for this overview.

The supersymmetry variation of the gravitino, \eqref{eq:Psi-susy},
is the covariant derivative of the supersymmetry parameter $S$ plus
some terms depending on the flux $G$. If we use this to define a
new derivative $D_{I}^{(L)}$ so that $\delta\Psi_{I}=D_{I}^{(L)}S$
then we can construct a gravitino curvature which has two `$SO(1,9)$
gauge indices'
\begin{align}
\Psi_{AB} & =2D_{[A}^{(L)}\Psi_{B]}^{\vphantom{(}}
\end{align}
and which varies as
\begin{align}
\delta\Psi_{AB} & =\left[D_{A}^{(L)},D_{B}^{(L)}\right]S=\frac{1}{4}R_{CDAB}^{(L)}\Gamma^{CD}S
\end{align}
and so is our analogue of the gaugino $\chi$.

Naively one would expect the analogue of the gauge field to be just
the spin connection $\Omega_{ABC}$, but this does not transform in
the proper way. The field which does transform properly contains an
additional $G$-flux term: it is
\begin{align}
\Omega_{ABC}^{-} & =\Omega_{ABC}-\frac{1}{2}G_{NABC}
\end{align}

This pair of fields then transforms almost as a Yang-Mills multiplet
\begin{align}
\delta\Omega_{ABC}^{-} & =\frac{1}{2}\bar{S}\Gamma_{A}\Psi_{BC}+y_{ABC}\\
\delta\Psi_{AB} & =\frac{1}{4}R_{CDAB}^{-}\Gamma^{CD}S+y_{AB}
\end{align}
where $y_{ABC}$ and $y_{AB}$ are correction terms which depend on
the $G$-flux and are higher order in the derivative expansion used
in \cite{Moss:2008ng}. To construct the supersymmetric $R^{2}$ action
we then add this Lorentz multiplet to the action and boundary conditions
just as the Yang-Mills multiplet appears but with a coupling constant
which is $-\frac{1}{2}$ of the Yang-Mills one. The $R^{2}$ action
is
\begin{align}
S_{R^{2}} & =-\frac{1}{2\lambda^{2}}\int_{\partial\m M_{10}}\bigg(-\frac{1}{4}R_{ABCD}^{-}R^{-ABDC}+\frac{1}{2}\bar{\Psi}_{BC}\Gamma^{A}D_{A}\Psi^{CB}\nonumber \\
 & \qquad\qquad\qquad\qquad-\frac{1}{4}\bar{\Psi}_{C}\Gamma^{AB}\Gamma^{C}R_{ABDE}\Psi^{ED}\bigg)
\end{align}
with $R^{-}$ the curvature of the connection $\Omega^{-}$. The boundary
condition on $C$ is
\begin{align}
C_{ABC} & =\mp\frac{\kappa_{11}^{2}}{\lambda^{2}}\left(\omega_{ABC}^{Y}+\omega_{ABC}^{\chi}\right)\pm\frac{\kappa_{11}^{2}}{2\lambda^{2}}\left(\omega_{ABC}^{L}+\omega_{ABC}^{\Psi}\right)
\end{align}
where $\dd\omega^{L}=\tr(R^{-}\wedge R^{-})$ and $\omega_{ABC}^{\Psi}=-\frac{1}{4}\bar{\Psi}_{DE}\Gamma_{ABC}\Psi^{ED}$.

In the reduction to five dimensions of this theory described below
we will assume that the spacetime curvature is small enough that we
can neglect almost all of this structure, except for the contribution
of the Calabi-Yau components of $R^{2}$ as a boundary source of $G$-flux
(Section \ref{sub:Spin_embedding}). This approximation was not our
first thought, but as explained in Section \ref{sub:Reducing-the-Lorentz-Multiplet}
it proved impractical to include more than just this piece.

\subsection{Heterotic M-Theory}

In summary, the complete eleven dimensional theory, which is the starting
point for our reduction to five dimensions, is supergravity on a manifold
with boundary
\begin{align}
S_{SG} & =\frac{1}{2\kappa_{11}^{2}}\int_{\m M_{11}}dv\bigg(R\left(\Omega\right)+\bar{\Psi}_{I}\Gamma^{IJK}D_{J}\left(\Omega^{*}\right)\Psi_{K}\nonumber \\
 & \qquad\qquad\qquad\qquad-\frac{1}{96}\left(\bar{\Psi}_{M}\Gamma^{IJKLMO}\Psi_{O}+12\bar{\Psi}^{I}\Gamma^{JK}\Psi^{L}\right)G_{IJKL}^{*}\nonumber \\
 & \qquad\qquad\qquad\qquad-\frac{1}{48}G_{IJKL}G^{IJKL}+\frac{1}{12^{4}}\epsilon^{I_{1}\ldots I_{11}}G_{I_{1}I_{2}I_{3}I_{4}}G_{I_{5}I_{6}I_{7}I_{8}}C_{I_{9}I_{10}I_{11}}\bigg)\nonumber \\
 & \quad+\frac{1}{2\kappa_{11}^{2}}\int_{\partial\m M_{10,1}}dv\left(-2K+\frac{1}{2}\bar{\Psi}_{A}\Gamma^{AB}\Psi_{B}-\bar{\Psi}_{A}\Gamma^{A}\Psi_{N}\right)\nonumber \\
 & \quad+\frac{1}{2\kappa_{11}^{2}}\int_{\partial\m M_{10,2}}dv\left(2K-\frac{1}{2}\bar{\Psi}_{A}\Gamma^{AB}\Psi_{B}+\bar{\Psi}_{A}\Gamma^{A}\Psi_{N}\right)\label{eq:11d-bulk-action}
\end{align}
with an $E_{8}$ Yang-Mills multiplet and a Lorentz multiplet on each
boundary (though here we neglect the fermion terms of the Lorentz
multiplet)
\begin{align}
S_{g1} & =\frac{1}{\lambda^{2}}\int_{\partial\m M_{10,1}}dv\left(-\frac{1}{4}\left(\tr\left(F_{AB}F^{AB}\right)-\frac{1}{2}\tr\left(R_{AB}R^{AB}\right)\right)\right.\nonumber \\
 & \qquad\qquad\qquad\quad\left.+\frac{1}{2}\tr\left(\bar{\chi}\Gamma^{A}D_{A}\left(\Omega^{**}\right)\chi\right)-\frac{1}{4}\bar{\Psi}_{C}\Gamma^{AB}\Gamma^{C}\tr\left(F_{AB}^{*}\chi\right)\right)\label{eq:11d-boundary-action}
\end{align}
where $\Omega^{**}=\frac{1}{2}\left(\Omega+\Omega^{*}\right)$, the
supercovariant field strength is $\hat{F}_{AB}=F_{AB}-2\bar{\Psi}_{[A}\Gamma_{B]}\chi$
and $F_{AB}^{*}=\frac{1}{2}\left(F_{AB}+\hat{F}_{AB}\right)$. $S_{g2}$
is identical to $S_{g1}$ but for $\partial\m M_{10,1}\rightarrow\partial\m M_{10,2}$.
The theory is completed by the specification of the boundary condition
on $C$. In a similar way to the Bianchi identity which cannot be
imposed by an action, this boundary condition must be imposed in addition
to the action. It is
\begin{align}
C & =\begin{cases}
-\frac{\kappa_{11}^{2}}{\lambda^{2}}\left(\omega_{3}^{Y}+\frac{1}{2}\omega_{3}^{L}+\omega_{3}^{\chi}\right) & \text{on }\partial\m M_{10,1}\\
\frac{\kappa_{11}^{2}}{\lambda^{2}}\left(\omega_{3}^{Y}+\frac{1}{2}\omega_{3}^{L}+\omega_{3}^{\chi}\right) & \text{on }\partial\m M_{10,2}
\end{cases}\label{eq:bulkC_bc}
\end{align}
The remaining boundary conditions, which do come from the action but
are given here for completeness, are
\begin{align}
K_{AB}-g_{AB}K & =\begin{cases}
-\kappa_{11}^{2}T_{AB} & \text{on }\partial\m M_{10,1}\\
\kappa_{11}^{2}T_{AB} & \text{on }\partial\m M_{10,2}
\end{cases}\\
P_{-}\Psi_{A} & =\begin{cases}
-\frac{\kappa_{11}^{2}}{12\lambda^{2}}\left(?\Gamma_{A}^{BC}?\chi-10\delta_{A}^{B}\Gamma^{C}\right)\tr\left(F_{BC}\chi\right) & \text{on }\partial\m M_{10,1}\\
\frac{\kappa_{11}^{2}}{12\lambda^{2}}\left(?\Gamma_{A}^{BC}?\chi-10\delta_{A}^{B}\Gamma^{C}\right)\tr\left(F_{BC}\chi\right) & \text{on }\partial\m M_{10,2}
\end{cases}
\end{align}
Finally the supersymmetry transformations are
\begin{align}
\delta_{S}?E^{\hat{I}}{}_{J}? & =-\frac{1}{2}\bar{S}\Gamma^{\hat{I}}\Psi_{J}\label{eq:E_susy}\\
\delta_{S}\Psi_{I} & =D_{I}(\hat{\Omega})S-\frac{1}{288}\left(?\Gamma_{I}^{JKLM}?-8\delta_{I}^{J}\Gamma^{KLM}\right)\hat{G}_{JKLM}S\label{eq:Psi_susy}\\
\delta_{S}C_{IJK} & =-\frac{3}{2}\bar{S}\Gamma_{[IJ}\Psi_{K]}\label{eq:C_susy_11d}\\
\delta_{S}A_{A} & =\frac{1}{2}\bar{S}\Gamma_{A}\chi\label{eq:A_susy_11d}\\
\delta_{S}\chi & =\frac{1}{4}\Gamma^{AB}\hat{F}_{AB}S\label{eq:chi_susy_11d}
\end{align}

\section[Kaluza-Klein Reduction Over a Calabi-Yau Threefold]{Kaluza-Klein Reduction Over a Calabi-Yau\protect \linebreak{}
Threefold\label{sec:Kaluza-Klein}}

To perform a dimensional reduction we consider the spacetime of a
full theory in some higher dimension as the product of the lower dimensional
spacetime of the reduced theory with the extra dimensions which we
are hiding. The essence of a Kaluza-Klein reduction is that we take
the extra dimensions to be compact and further to be small. Since
the extra dimensions are compact the momenta of the fields of the
full theory in those directions are quantized and the fields of the
full theory can be expressed as a Fourier series: a sum over the possible
momenta in the extra dimensions of fields depending only on the co-ordinates
of the reduced theory. From the point of view of the reduced theory
these Fourier modes are independent fields with various masses, which
depend on their momenta in the extra directions. As the extra dimensions
are small, the mass associated with a quantum of extra dimensional
momentum is large and so at low energies compared to that mass scale
we can truncate the reduced theory and consider only the less massive
fields. Indeed we will take the size of the Calabi-Yau three-fold
which describes the extra dimensions to be small enough that we need
consider only the leading terms, i.e.\ those which have zero momentum
in the Calabi-Yau directions.

The momentum operator is just the spacetime derivative, so zero momentum
modes are those which are covariantly constant. The fields of the
reduced theory then are those given by the fields of the full theory
whose extra dimensional part is a constant tensor (for bosonic fields)
or spinor (for fermionic fields). To identify the reduced fields we
need to know all the constant tensors and spinors of the extra dimensions
and then to see which of those can contribute to each of the fields
of the full theory. 

Calabi-Yau manifolds are those which have a single covariantly constant
spinor, which we call $u_{A}$. This is the reason for choosing them
for the reduction, since the fermionic supersymmetry parameter of
the eleven dimensional theory then has a single zero mode given by
that single constant spinor and so we have a single supersymmetry
parameter in the five dimensional theory, which therefore possesses
$\m N=1$ supersymmetry. One can also construct constant spinors by
acting on $u_{A}$ with gamma matrices, $\gamma_{a}u_{A}$ for example.
The tensor structure precludes such objects from contributing to the
supersymmetry parameter (there are no indices there to contract with
those of the gamma matrices) but they can, and do, give contributions
to the other fermionic fields.

A Calabi-Yau manifold is a complex manifold, so we categorize differential
forms by the number of holomorphic and anti-holomorphic components:
a $(p,q)$-form has $p$ holomorphic and $q$ anti-holomorphic components.
There is no covariantly constant vector on a Calabi-Yau space by definition.
There is a single holomorphic three-form, $\Omega_{abc}$: there can
only be one since a holomorphic index can only take the three values
$1,2,3$ and so a totally anti-symmetric object with three indices
only has one independent component; thus all constant holomorphic
three-forms can only differ by an overall factor. There are no constant
(2,0)-forms because if there were we could contract them the three-form
$\Omega$ to get a constant vector, which there cannot be. The number
of independent, constant (2,1)- and (1,1)-forms are properties of
particular Calabi-Yau manifolds, known as the Hodge numbers $h^{2,1}$
and $h^{1,1}$ respectively. We consider only cases with $h^{2,1}=0$
and so we have no (2,1)-forms, nor do we have their complex conjugates,
(1,2)-forms. We do however allow $h^{1,1}$ to be arbitrary and so
we have a basis of (1,1) forms, $\omega_{ia\bar{b}}$ with $h^{1,1}$
elements labelled by $i$. (3,1)-forms are Hodge duals of (0,2)-forms
so there are none. (2,2)-forms are dual to (1,1)-forms so the basis
$\omega_{ia\bar{b}}$ gives us a dual basis $\nu_{a\bar{b}c\bar{d}}^{i}$
with $h^{1,1}$ elements. (3,2)-forms are dual to vectors and thus
absent. (3,3)-forms are proportional to the volume element of the
Calabi-Yau and dual to scalars. In summary the constant tensors we
have available to construct the Calabi-Yau parts of our eleven dimensional
fields are scalars, $\Omega_{abc}$ and $\omega_{ia\bar{b}}$.

In performing the reduction we need to consider two types of object.
The action (or parts thereof), which is an integral expression, and
the boundary conditions and supersymmetry transformations which are
tensor or spinor equations. Once we have truncated to just the zero
energy modes, the eleven dimensional action is just a sum of scalars
which are constant over the Calabi-Yau, so once we have contracted
all the Calabi-Yau tensor indices and spinors the only part remaining
is the volume element, which gives $\int_{CY}dv=v$ with $v$ the
volume of the reference Calabi-Yau whose spacetime dependence has
been scaled out into the volume modulus $V$ and the $h^{1,1}$ moduli
$b^{i}$. On the other hand for the boundary condition and supersymmetry
transformations, each eleven dimensional equation will generally give
us several five dimensional equations since it must be satisfied separately
for each independent Calabi-Yau tensor or spinor which is present.

\section{Reduction Ansatz\label{sec:Reduction-Ansatz}}

We now move on to find the appropriate ansatz for our reduction of
the improved Heterotic M-Theory from an eleven dimensional bulk with
ten dimensional boundaries to a five dimensional bulk, $\m M$, with
four dimensional boundaries, $\partial M_{1}$ and $\partial\m M_{2}$.
The reduction is performed over a Calabi-Yau threefold, $X$, which
is taken to have $h^{1,1}$ $(1,1)$ moduli and no $(2,1)$ moduli.
As in 11 dimensions the normal vector is inward pointing on $\partial\m M_{1}$
and outward pointing on $\partial\m M_{2}$. The spin connection is
embedded in the $E_{8}$ gauge group on $\partial\m M_{1}$, breaking
it to $E_{6}\otimes SU(3)$. To find the appropriate ansatz, essentially
we need to find all the five dimensional fields that we can build
from the eleven dimensional ones in this framework.

\subsection{The Metric}

Since we split the eleven dimensional spacetime into a direct product
of the five dimensional spacetime and the Calabi-Yau space, the metric
must be a sum of spacetime and Calabi-Yau parts, with no cross terms.
The Calabi-Yau moduli appear in the ansatz in the Calabi-Yau components
of the metric. In particular, we wish to scale out the Volume modulus,
$V$, so that we integrate over a Calabi-Yau with a fixed reference
volume $v$ and so that the Calabi-Yau components of the metric are
given by the metric of the reference Calabi-Yau multiplied by an appropriate
power of the volume modulus. In order for the five dimensional action
to be in the Einstein frame and for the volume modulus to have a correctly
normalized kinetic term, the spacetime components of the metric must
also contain a power of the volume modulus. Einstein frame is the
choice of scaling of the metric which gives us the five dimensional
Einstein-Hilbert action with no conformal factor, $S_{EH,5}=\int_{\m M_{5}}dvR(g_{\alpha\beta})$.
We then have a line element of the form
\begin{align}
ds^{2} & =V^{n}g_{\alpha\beta}dx^{\alpha}dx^{\beta}+V^{m}\left(g_{a\bar{b}}dx^{a}dx^{\bar{b}}+g_{\bar{a}b}dx^{\bar{a}}dx^{b}\right)
\end{align}
where we must find the values of $n$ and $m$ which satisfy these
requirements. \linebreak{}
$\alpha,\beta,\gamma,\ldots=0,\ldots,4$ are five dimensional spacetime
indices. $a,b,c,\ldots=1,2,3$ and \textbf{$\bar{a},\bar{b},\bar{c},\ldots=\bar{1},\bar{2},\bar{3}$}
are respectively holomorphic and anti-holomorphic indices on $X$.
$g_{a\bar{b}}$ is given in terms of the (1,1) moduli, $b^{i}$, as
$ig_{a\bar{b}}=b^{i}\omega_{ia\bar{b}}$ where $\omega_{ia\bar{b}}$
is the basis of (1,1)-forms on $X$. With this form the metric determinant
is 
\begin{equation}
\sqrt{-g_{(11)}}=V^{\frac{5}{2}n+\frac{6}{2}m}\sqrt{-g_{(5)}}\sqrt{g_{CY}}
\end{equation}
and the eleven dimensional Ricci scalar gives
\begin{align}
R(g_{(11)}) & =V^{-n}R(g_{(5)})+\left(4n-3n^{2}-\frac{3}{2}m^{2}\right)V^{-2-n}\partial_{\alpha}V\partial^{\alpha}V\nonumber \\
 & \quad+\left(3m^{2}-\frac{15}{2}nm\right)V^{-1-m}\partial_{\alpha}V\partial^{\alpha}V\nonumber \\
 & \quad-\left(4n+3m\right)V^{-1-n}\nabla^{2}V-3mV^{-m}\nabla^{2}V+\ldots
\end{align}
The extra terms involve $\partial b^{i}$ and we do not need them
here. We must have $m=n+1$ for the $V$ factors to be consistent
and, after integrating by parts, the kinetic term for $V$ is
\begin{equation}
\left(36n^{2}+\frac{99}{2}n+\frac{33}{2}\right)V^{\frac{9}{2}n+1}\partial_{\alpha}V\partial^{\alpha}V
\end{equation}
In order for the coefficient to be $-\frac{1}{2}$, either $n=-\frac{2}{3}$
or $n=-\frac{17}{24}$. Finally, to be in Einstein frame, $V^{-n+\left(\frac{5}{2}n+\frac{6}{2}m\right)}=1$
and so we must take $n=-\frac{2}{3}$ and $m=\frac{1}{3}$. So the
metric ansatz is
\begin{align}
ds^{2} & =V^{-\frac{2}{3}}g_{\alpha\beta}dx^{\alpha}dx^{\beta}+V^{\frac{1}{3}}\left(g_{a\bar{b}}dx^{a}dx^{\bar{b}}+g_{\bar{a}b}dx^{\bar{a}}dx^{b}\right)
\end{align}

Once we allow $h^{1,1}>1$ the metric, $g_{a\bar{b}}$, of the Calabi-Yau
is no longer constant since the moduli $b^{i}$ are spacetime dependent
scalars: $\partial_{\alpha}g_{a\bar{b}}=\omega_{ia\bar{b}}\partial_{\alpha}b^{i}$.
This results in many extra terms in the action, etc., but we find
that almost all of them contribute just to connection terms in the
derivatives of fields with $i,j,k,\ldots$ indices as $D_{\alpha}\phi^{i}=\partial_{\alpha}\phi^{i}+\partial_{\alpha}b^{j}?\Gamma^{i}{}_{jk}?\phi^{k}$.
The exceptions are a single term from the kinetic term of the eleven
dimensional gravitino, $\frac{i}{2}G_{ij}\bar{\lambda}_{A}^{i}\gamma^{\alpha}\gamma^{\beta}\psi_{\alpha}^{A}\partial_{\beta}b^{j}$,
and the kinetic term for the moduli, $-G_{ij}\partial_{\alpha}b^{i}\partial^{\alpha}b^{j}$.

\subsection{The Three-Form $C$}

Each part of the ansatz for $C$ is the product of a spacetime tensor
with a tensor on the Calabi-Yau space. The available tensors are those
given at the end of Section \ref{sec:Kaluza-Klein}.

The constant scalar field over the Calabi-Yau, gives us a three-form
in five dimensions from the pure spacetime component of $C$. We will
see in Section \ref{sec:Dualization} how to dualize this three-form
to a scalar $\sigma$ which will form part of the hypermultiplet of
the five dimensional supergravity, Section \ref{sub:Hypermultiplet-Quaternionic-Structure}.
After this dualization we will have
\begin{align}
G_{\alpha\beta\gamma\delta} & =V^{-2}?\epsilon_{\alpha\beta\gamma\delta}^{\epsilon}?\mathcal{D}_{\alpha}\sigma
\end{align}

We get a spacetime scalar from the Calabi-Yau holomorphic three-form
$\Omega_{abc}$, and another from its complex conjugate, the anti-holomorphic
three-form $\bar{\Omega}_{\bar{a}\bar{b}\bar{c}}=-i\epsilon_{\bar{a}\bar{b}\bar{c}}$,
which will both be part of the hypermultiplet
\begin{align}
C_{abc} & =\frac{i}{2}\xi\epsilon_{abc}\\
C_{\bar{a}\bar{b}\bar{c}} & =-\frac{i}{2}\bar{\xi}\epsilon_{\bar{a}\bar{b}\bar{c}}
\end{align}

The basis of (1,1)-forms is $\omega_{ia\bar{b}}$ with $i=1,\ldots,h^{1,1}$
and the Kähler form is picked out by the Kähler moduli $b^{i}$ as
$ig_{a\bar{b}}=\omega_{a\bar{b}}=b^{i}\omega_{ia\bar{b}}$. These
(1,1)-forms give us $h^{1,1}$ spacetime vectors
\begin{align}
C_{\alpha a\bar{b}} & =\frac{1}{\sqrt{2}}\m A_{\alpha}^{i}\omega_{ia\bar{b}}
\end{align}
These vectors gauge the shift isometry of $\sigma$, as we will see
in Section \ref{sub:Hypermultiplet-Quaternionic-Structure}. The graviphoton
$b_{i}\m A_{\alpha}^{i}$, which belongs to the gravity multiplet,
is the part corresponding to the Kähler form and the remaining $h^{1,1}-1$
vectors belong to vector multiplets.

These are all the constant tensors on the Calabi-Yau. However, the
boundary conditions on $C$ contain (2,2)-form sources for its field
strength $G$ and so, as explained in Section \ref{sub:Spin_embedding},
we also have a `non-zero mode'
\begin{align}
G_{a\bar{b}c\bar{d}} & =-\sqrt{2}\alpha_{i}\nu_{a\bar{b}c\bar{d}}^{i}=-\frac{\sqrt{2}}{2}\alpha_{i}*(\omega^{i})_{a\bar{b}c\bar{d}}\label{eq:G_abcd}
\end{align}
where $\nu^{i}$ are the basis of (2,2)-forms dual to $\omega_{i}$
\eqref{eq:v}.

\subsection{The Gravitino}

As the metric is just the sum of the five dimensional metric and the
Calabi-Yau metric, the gamma matrices split into direct products of
a square root of unity with the five dimensional and Calabi-Yau gamma
matrices. There are volume factors given by the square root of the
volume factors in the metric so that the eleven dimensional Dirac
algebra, $\left\{ \Gamma_{I},\Gamma_{J}\right\} =2g_{(11)IJ}$, reduces
to give $\left\{ \gamma_{\alpha},\gamma_{\beta}\right\} =2g_{\alpha\beta}$
and $\left\{ \gamma_{a},\gamma_{\bar{b}}\right\} =2g_{a\bar{b}}$.
Thus we have 
\begin{align}
\Gamma_{\alpha} & =V^{-\frac{1}{3}}\gamma_{\alpha}\otimes\gamma_{7}\\
\Gamma_{a} & =V^{\frac{1}{6}}\id\otimes\gamma_{a}
\end{align}
The choice of $\gamma_{7}$ in $\Gamma_{\alpha}$ and $\id$ in $\Gamma_{a}$
is made so that $\gamma_{\mu}^{*}=\gamma_{\mu}$, $\gamma_{5}^{*}=-\gamma_{5}$
and $\gamma_{7}^{*}=-\gamma_{7}$ where $\gamma_{5}$ is the chirality
matrix for the 4d spacetime manifold and $\gamma_{7}$ is the chirality
matrix for $X$.

An eleven dimensional spinor gives us a direct product of a five dimensional
spinor with a Calabi-Yau spinor, so to find the ansatz for $\Psi_{I}$
we use the constant spinor, $u_{A}$, and the vector-spinors, $\gamma_{a}u_{A}$
and $\gamma_{\bar{a}}u_{A}$, on the Calabi-Yau. So at first glance
$\Psi_{I}$ just gives us a five dimensional vector-spinor from $u_{A}$
\begin{align}
\Psi_{\alpha} & =\theta_{\alpha}^{A}\otimes u_{A}
\end{align}
and a spinor from $\gamma_{a}u_{A}$
\begin{align}
\Psi_{a} & =\zeta^{A}\otimes\gamma_{a}u_{A}
\end{align}
However, this only gives us enough fermions for the gravity multiplet
and hypermultiplet. There are also $h^{1,1}-1$ vector multiplets
which need fermion components. In fact we have one more tool in the
basis of (1,1)-forms, $\omega_{i}$, which we can use to construct
$h^{1,1}$ vector-spinors on the Calabi-Yau instead of just one:
\begin{align}
\Psi_{a} & =\frac{1}{2}V^{\frac{1}{3}}\lambda^{Ai}\otimes\omega_{ia\bar{b}}\gamma^{\bar{b}}u_{A}\label{eq:Psi_a_ansatz}
\end{align}
There are no more possibilities since the other distinct tensor we
have available, $\Omega_{abc}$ does not give us anything new since
$\Omega_{abc}\gamma^{bc}u_{A}=-2\gamma_{a}u_{A}$. As we explain in
Section \ref{sec:The-5d-Gravitino} the fermion, $\zeta^{A}$, which
comes from the Kähler form part of the basis, $\omega_{a\bar{b}}\gamma^{\bar{b}}$,
and is part of the hypermultiplet, is $\sqrt{2}b_{i}\lambda^{Ai}$,
while $\theta_{\alpha}^{A}$ is in fact not the five dimensional gravitino,
$\psi_{\alpha}^{A}$, but contains $\zeta^{A}$ as well. We will show
there that the ansatz in terms of the five dimensional gravitino,
$\psi_{\alpha}^{A}$, is
\begin{align}
\Psi_{\alpha} & =V^{-\frac{1}{6}}\left(\psi_{\alpha}^{A}-\frac{\sqrt{2}i}{3}\gamma_{\alpha}\zeta^{A}\right)\otimes u_{A}\label{eq:Psi_alpha_ansatz}
\end{align}

It might seem more natural to maintain everywhere the distinction
between $\zeta^{A}$ and the other $h^{1,1}-1$ components, $\lambda^{\perp Ai}$,
rather than grouping them together into $\lambda^{Ai}$, as they do
belong to different multiplets. However, we ultimately decided that
the benefits of more concise presentation outweighed this desire.

\subsection{The $E_{8}$ Boundary }

On $\partial\m M_{2}$, where the $E_{8}$ symmetry is not broken,
the story is simple. There is no constant vector on the Calabi-Yau
and the only constant spinor is $u_{A}$, so the gauge fields and
gauginos just give their four dimensional counterparts
\begin{align}
A_{\mu}^{(10)} & =A_{\mu}\\
A_{a}^{(10)} & =0\label{eq:A_a-ansatz-2}
\end{align}
and
\begin{align}
\chi & =\chi^{A}\otimes u_{A}
\end{align}
with $A_{\mu}$ and $\chi^{A}$ belonging to the adjoint representation
of $E_{8}$ just as the eleven dimensional fields do.

\subsection{The $E_{6}$ Boundary}

On $\partial\m M_{1}$ the gauge group $E_{8}$ is broken to $E_{6}$
by the embedding of the spin connection of the Calabi-Yau into the
$SU(3)$ part of the subgroup $E_{6}\times SU(3)$. Here situation
is somewhat more complicated than on $\partial\m M_{2}$. We have
a gauge multiplet just as on $\partial\m M_{2}$, though in the adjoint
of $E_{6}$ instead of $E_{8}$. The part of the gauge field proportional
to the generators of the $SU(3)$ subgroup is identified with the
spin connection of the Calabi-Yau, which is a one-form (though not
globally defined) giving $\left.A_{a}^{(10)}\right|_{SU(3)}=\omega_{a}$.
However, there are also generators of the off-block-diagonal components
of $E_{8}$. The decomposition of the $E_{8}$ adjoint representation
$\mathbf{248}$ is $\mathbf{248}=\left(\mathbf{8},\mathbf{1}\right)\oplus\left(\mathbf{1},\mathbf{78}\right)\oplus\left(\mathbf{3},\mathbf{27}\right)\oplus\left(\mathbf{\bar{3}},\overline{\mathbf{27}}\right)$
where $\left(\mathbf{8},\mathbf{1}\right)$ is the adjoint representation
of $SU(3)$ and $\left(\mathbf{1},\mathbf{78}\right)$ that of $E_{6}$
while $\left(\mathbf{3},\mathbf{27}\right)$ and $\left(\mathbf{\bar{3}},\overline{\mathbf{27}}\right)$
are the off-block-diagonal parts which are in the fundamental and
anti-fundamental respectively of both $SU(3)$ and $E_{6}$. These
off-diagonal generators, $T_{ap}$ and $T^{ap}$, have an $E_{6}$
gauge index, $p$, and also an $SU(3)$ index which is equivalent
to a holomorphic or anti-holomorphic index on the Calabi-Yau space.
We can have an additional contribution to $A_{a}^{(10)}$ by contracting
the $SU(3)$ index with one index of a (1,1)-form, $A_{a}^{(10)}\sim?\omega_{ia}^{b}?T_{bp}$.
This gives us a set of $h^{1,1}$ spacetime scalars which are charged
under the $E_{6}$, $C^{ip}$, and their complex conjugates, $\bC ip$.
The complete ansatz for the gauge field is then
\begin{align}
A_{\mu}^{(10)} & =A_{\mu}\\
A_{a}^{(10)} & =\omega_{a}+?\omega_{ia}^{b}?T_{bp}C^{ip}\label{eq:A_a-ansatz-1}
\end{align}
where $A_{\mu}$ belongs to the adjoint representation of $E_{6}$
and $\omega_{a}$ is the spin connection of the Calabi-Yau space considered
as an $SU(3)$ adjoint-valued (1,0)-form.

As for the gaugino, we can introduce an extra fermion in the fundamental
representation of $E_{6}$ by contracting the $SU(3)$ index of $T_{ap}$
with $\gamma^{a}u_{A}$. However, in a similar way to the gravitino,
we have $h^{1,1}$ multiplets (chiral multiplets this time) which
need fermions, not just one. We find these in much the same way, by
using the basis of (1,1)-forms to give us more vector-spinors on the
Calabi-Yau, $\omega_{ia\bar{b}}\gamma^{\bar{b}}u_{A}$. In this way
we have $h^{1,1}$ Majorana fermions $\eta^{Aip}$, in the fundamental
of $E_{6}$ to go along with the five dimensional gaugino in the adjoint:
\begin{align}
\chi & =\chi^{A}\otimes u_{A}+\frac{1}{2}\omega_{ia\bar{b}}T^{ap}?\eta_{R}^{i}{}_{p}?\otimes\gamma^{\bar{b}}u_{1}+\frac{1}{2}?\omega_{i}^{a}{}_{b}?T_{ap}?\eta_{L}^{ip}?\otimes\gamma^{b}u_{2}
\end{align}
$\eta_{L}^{i}$ and $\eta_{R}^{i}$ are the left and right handed
components of $\eta^{Ai}$: $\eta_{L}^{i}=\eta^{1i}$ and $\eta_{R}^{i}=\eta^{2i}$.
We label them thus because the chirality in ten dimensions of $\chi$,
$P_{-}^{(11)}\chi=\frac{1}{2}\left(\id-\Gamma_{N}\right)\chi=0$,
tells us that if we define $P_{L}=\frac{1}{2}\left(\id+\gamma_{5}\right)$
and $P_{R}=\frac{1}{2}\left(\id-\gamma_{5}\right)$ then $P_{R}\eta_{L}^{i}=0$
and $P_{L}\eta_{R}^{i}=0$ (since $\gamma_{7}u_{1}=u_{1}$ and $\gamma_{7}u_{2}=-u_{2}$)
and so, for example, $?{\t{\bar{\eta}}}_{A}^{i}?P_{L}\eta^{Aj}=\bar{\eta}_{L}^{i}\eta_{L}^{j}$.
Defining $?P_{+}^{A}{}_{B}?=?{\left(\begin{array}{cc}
P_{L} & 0\\
0 & P_{R}
\end{array}\right)}^{A}{}_{B}?$ and $?P_{-}^{A}{}_{B}?=?{\left(\begin{array}{cc}
P_{R} & 0\\
0 & P_{L}
\end{array}\right)}^{A}{}_{B}?$ the chirality of the 4d Majorana fermions is $?P_{-}^{A}{}_{B}?\chi^{B}=0$
and $?P_{+}^{A}{}_{B}?\eta^{iB}=0$.

\section{Five Dimensional Multiplets\label{sec:Five-Dimensional-Multiplets}}

Now that we have the field content of the five dimensional theory,
we need to know the supersymmetry structure. In any supersymmetric
theory, the component fields are grouped into multiplets whose fields
transform among themselves under supersymmetry. The supersymmetry
transformations of the five dimensional fields are fixed by the eleven
dimensional supersymmetry transformations (\ref{eq:E_susy}-\ref{eq:chi_susy_11d})
and the ansatz described above. (We do need to do some extra work
to obtain the supersymmetry transformation of $\sigma$ which is found
as part of the dualization process by which we turn the three-form
$C_{\alpha\beta\gamma}$ into the scalar $\sigma$.)

Let us start by considering the gravity multiplet. By definition it
contains the five dimensional vielbein $?e^{\hat{\alpha}}{}_{\alpha}?$
whose supersymmetry transformation is
\begin{align}
\delta?e^{\hat{\alpha}}{}_{\alpha}? & =-\frac{1}{2}\bar{s}_{A}\gamma^{\hat{\alpha}}\psi_{\alpha}^{A}
\end{align}
$\psi_{\alpha}^{A}$ appears and must therefore be the second component
of the gravity multiplet, confirming its identification as the gravitino
which comes from examining the kinetic terms in the action (Section
\ref{sec:The-5d-Gravitino}). The transformation of the gravitino
is
\begin{align}
\delta\psi_{\alpha}^{A} & =\m D_{\alpha}s^{A}-\frac{\sqrt{2}i}{12}\left(?\gamma_{\alpha}^{\beta\gamma}?-4\delta_{\alpha}^{\beta}\gamma^{\gamma}\right)b_{i}\m F_{\beta\gamma}^{i}s^{A}-\frac{\sqrt{2}}{6}V^{-1}\alpha?\tau^{A}{}_{B}?\gamma_{\alpha}s^{B}\label{eq:psi_susy_1}
\end{align}
in which there is one new field $b_{i}\m A_{\alpha}^{i}\equiv\m A_{\alpha}$,
appearing as its field strength $b_{i}\m F_{\alpha\beta}^{i}=2b_{i}\partial_{[\alpha}\m A_{\beta]}^{i}\equiv2\partial_{[\alpha}\m A_{\beta]}$.
$\m A_{\alpha}$ is the graviphoton, which is the final component
of the gravity multiplet as we confirm by its supersymmetry transformation
\begin{align}
\delta\m A_{\alpha} & =-\frac{3\sqrt{2}i}{4}?\tau^{A}{}_{B}?\bar{s}_{A}\psi_{\alpha}^{B}
\end{align}
which contains no new fields. Thus the set $\left\{ ?e^{\hat{\alpha}}{}_{\alpha}?,\psi_{\alpha}^{A},\m A_{\alpha}\right\} $
is closed under supersymmetry; it is the gravity multiplet.

With the $b_{i}\m A_{\alpha}^{i}$ component taken care of by the
gravity multiplet, the $h^{1,1}-1$ remaining vectors, $\m A_{\alpha}^{\perp i}$
form multiplets which we can see from the variations
\begin{align}
\delta\m A_{\alpha}^{\perp i} & =\frac{\sqrt{2}}{2}?\tau^{A}{}_{B}?\bar{s}_{A}\gamma_{\alpha}\lambda^{\perp Bi}\\
\delta\lambda^{\perp iA} & =-\frac{i}{2}\partial_{\alpha}b^{i}\gamma^{\alpha}s^{A}+\frac{\sqrt{2}}{4}\m F_{\alpha\beta}^{i\perp}\gamma^{\alpha\beta}s^{A}-\frac{\sqrt{2}i}{4}V^{-1}G^{\perp ij}\alpha_{j}?\tau^{A}{}_{B}?s^{B}\\
\delta b^{i} & =-\frac{i}{2}?\tau^{A}{}_{B}?\bar{s}_{A}\lambda^{\perp iB}\label{eq:delta-b^i}
\end{align}
Varying $\m A_{\alpha}^{\perp i}$ brings in $\lambda^{\perp Ai}$
which brings $b^{i}$ and then the set closes, so we have $h^{1,1}-1$
vector multiplets $\left\{ \m A_{\alpha}^{\perp i},\lambda^{\perp iA},b^{\perp i}\right\} $.

There is one more fermion (in the bulk) and so we have one more multiplet,
which includes $\zeta^{A}$.
\begin{align}
\delta\zeta^{A} & =?{\left(\begin{array}{cc}
{\scriptstyle \frac{i}{48\sqrt{2}}VG_{\alpha\beta\gamma\delta}\gamma^{\alpha\beta\gamma\delta}-\frac{i}{2\sqrt{2}}V^{-1}\partial_{\alpha}V\gamma^{\alpha}} & {\scriptstyle \frac{i}{8\sqrt{2}}V^{-\frac{1}{2}}\partial_{\alpha}\xi\gamma^{\alpha}}\\
{\scriptstyle -\frac{i}{8\sqrt{2}}V^{-\frac{1}{2}}\partial_{\alpha}\bar{\xi}\gamma^{\alpha}} & {\scriptstyle -\frac{i}{48\sqrt{2}}VG_{\alpha\beta\gamma\delta}\gamma^{\alpha\beta\gamma\delta}-\frac{i}{2\sqrt{2}}V^{-1}\partial_{\alpha}V\gamma^{\alpha}}
\end{array}\right)}^{A}{}_{B}?s^{B}\label{eq:zeta-susy}\\
\delta V & =-\frac{i}{\sqrt{2}}?\tau^{A}{}_{B}?\bar{s}_{A}\zeta^{B}\label{eq:V-susy}\\
\delta\xi & =-\frac{\sqrt{2}i}{2}V^{\frac{1}{2}}\bar{s}_{2}\zeta^{1}\label{eq:xi-susy}\\
\delta\bar{\xi} & =\frac{\sqrt{2}i}{2}V^{\frac{1}{2}}\bar{s}_{1}\zeta^{2}\label{eq:barxi-susy}\\
\delta C_{\alpha\beta\gamma} & =-\frac{3}{2}\bar{s}_{A}\gamma_{[\alpha\beta}\left(\psi_{\gamma]}^{A}-\frac{\sqrt{2}}{3}\gamma_{\gamma]}\zeta^{A}\right)
\end{align}
The variation of $\zeta^{A}$ connects it to $V$, $\xi$, $\bar{\xi}$
and $G_{\alpha\beta\gamma\delta}=4\partial_{[\alpha}C_{\beta\gamma\delta]}$.
The variations of $V$, $\xi$ and $\bar{\xi}$ connect back to $\zeta^{A}$,
which is as expected. However, the variation of $C_{\alpha\beta\gamma}$
also involves $\psi_{\alpha}^{A}$ and in any case a three-form does
not fit into the possible five dimensional supersymmetry multiplets.
Both issues are resolved by dualizing the three-form $C$ to a scalar
$\sigma$ which, as we will see in Section \ref{sec:Dualization},
has a variation
\begin{align}
\delta_{s}\sigma & =-\frac{1}{\sqrt{2}}?\tau^{A}{}_{B}?\bar{\eta}_{A}\zeta^{B}+\frac{1}{\sqrt{2}}V^{\frac{1}{2}}?{\left(\begin{array}{cc}
0 & \xi\\
\bar{\xi} & 0
\end{array}\right)}^{A}{}_{B}?\bar{\eta}_{A}\zeta^{B}
\end{align}
which does not involve $\psi_{\alpha}^{A}$, and so $\left\{ \zeta^{A},V,\sigma,\xi,\bar{\xi}\right\} $
form a hypermultiplet.

The gauge fields $A_{\mu}$ on the $E_{8}$ boundary $\partial\m M_{2}$
simply transform into the gauginos $\chi^{A}$ and vice versa, 
\begin{align}
\delta A_{\mu}^{I} & =\frac{1}{2}V^{-\frac{1}{2}}\bar{s}_{A}\gamma_{\mu}\chi^{A}\\
\delta\chi^{IA} & =\frac{1}{4}V^{\frac{1}{2}}\gamma^{\mu\nu}F_{\mu\nu}^{I}s^{A}
\end{align}
so we have a vector multiplet $\left\{ A_{\mu},\chi^{A}\right\} $.

On the $E_{6}$ boundary $\partial\m M_{1}$ there are a set of $h^{1,1}$
complex scalars $C^{i}$ and fermions $\eta^{Aip}$ which transform
as
\begin{align}
\delta?\eta_{L}^{ip}? & =-\m D_{\mu}C^{ip}\gamma^{\mu}s^{2}-\frac{3\sqrt{10}}{2}V^{-\frac{1}{2}}\m K^{-1}?{\m K}^{i}{}_{jk}?d^{pqr}\bC jq\bC krs^{1}\\
\delta?\eta_{R}^{i}{}_{p}? & =\m D_{\mu}\bC ip\gamma^{\mu}s^{1}-\frac{3\sqrt{10}}{2}V^{-\frac{1}{2}}\m K^{-1}?{\m K}^{i}{}_{jk}?d_{pqr}C^{jq}C^{kr}\\
\delta C^{ip} & =-\frac{1}{2}\bar{s}_{2}?\eta_{L}^{ip}?-\frac{i}{4}?\Gamma^{i}{}_{jk}?C^{jp}\bar{s}_{A}\lambda^{\perp Ak}\\
\delta\bC ip & =-\frac{1}{2}\bar{s}_{1}?\eta_{R}^{i}{}_{p}?-\frac{i}{4}?\Gamma^{i}{}_{kj}?\bC jp\bar{s}_{A}\lambda^{\perp Ak}
\end{align}
so we have $h^{1,1}$ chiral multiplets $\left\{ \eta^{Ai},C^{i},\bar{C}^{i}\right\} $.
The appearance of the fermions $\lambda^{\perp Ai}$ is due to the
fact that the basis over which the $i,j,k,\ldots$ indices run depends
on the Calabi-Yau moduli $b^{i}$. The variation of the $b^{i}$,
\eqref{eq:delta-b^i}, then gives rise to these terms which would
not be present were the basis fixed. There is also a vector multiplet,
$\left\{ A_{\mu},\chi^{A}\right\} $, similar to that on the $E_{8}$
boundary but with gauge group $E_{6}$ whose transformations are

\begin{align}
\delta A_{\mu}^{I} & =\frac{1}{2}V^{-\frac{1}{2}}\bar{s}_{A}\gamma_{\mu}\chi^{A}\\
\delta\chi^{IA} & =\frac{1}{4}V^{\frac{1}{2}}\gamma^{\mu\nu}F_{\mu\nu}^{I}s^{A}+V^{-\frac{1}{2}}G_{ij}\bar{C}^{i}\Lambda^{I}C^{j}?\tau^{A}{}_{B}?s^{B}
\end{align}
The scalar fields $C^{i}$ enter here because in the four dimensional
Yang-Mills theory coupled to chiral multiplets the equation of motion
for the auxiliary field of the vector multiplet depends on the chiral
multiplets. When the auxiliary field which appears `legitimately'
in the supersymmetry transformations of the vector multiplet is replaced
by its (algebraic) equation of motion, it introduces this dependence
on the scalars of the chiral multiplet.

\section{The Bulk\label{sec:The-Bulk}}

We find the action in the bulk of the reduced theory by substituting
the ansatzes from Section \ref{sec:Reduction-Ansatz} into the eleven
dimensional action \eqref{eq:11d-bulk-action} and also dualizing
the three-form $C_{\alpha\beta\gamma}$ to a scalar $\sigma$ (Section
\ref{sec:Dualization}) to show the five dimensional supergravity
structure.

\subsection{Einstein-Hilbert Term}

After integrating out the Calabi-Yau modes, the eleven dimensional
Einstein-Hilbert term gives rise to the five dimensional Einstein-Hilbert
term as well as the kinetic terms for the Calabi-Yau moduli. The eleven
dimensional metric determinant is $\sqrt{-g_{\left(11\right)}}=V^{-\frac{2}{3}}\sqrt{-g_{\left(5\right)}}\sqrt{g_{CY}}$
which tells us the relation between the eleven and five dimensional
volume elements: $dv_{\left(11\right)}=\sqrt{-g_{\left(11\right)}}d^{11}x=V^{-\frac{2}{3}}\sqrt{-g_{\left(5\right)}}d^{5}x\sqrt{g_{CY}}d^{6}x=V^{-\frac{2}{3}}dv_{\left(5\right)}dv_{CY}$.
We also need the connection coefficients, the non-zero components
of which are
\begin{align}
?{\mathsf{\Gamma}}_{\left(11\right)}^{\alpha}{}_{\beta\gamma}? & =?{\mathsf{\Gamma}}_{\left(5\right)}^{\alpha}{}_{\beta\gamma}?-\frac{1}{3}V^{-1}\left(\delta_{\beta}^{\alpha}\partial_{\gamma}V+\delta_{\gamma}^{\alpha}\partial_{\beta}V-g_{\beta\gamma}\partial^{\alpha}V\right)\\
?{\mathsf{\Gamma}}_{\left(11\right)}^{\alpha}{}_{b\bar{c}}? & =-\frac{1}{6}g_{b\bar{c}}\partial^{\alpha}V+\frac{i}{2}V\omega_{ib\bar{c}}\partial^{\alpha}b^{i}\\
?{\mathsf{\Gamma}}_{\left(11\right)}^{a}{}_{\beta c}? & =\frac{1}{6}V^{-1}\delta_{c}^{a}\partial_{\beta}V-\frac{i}{2}?\omega_{ib}^{a}?\partial_{\beta}b^{i}\\
?{\mathsf{\Gamma}}_{\left(11\right)}^{\bar{a}}{}_{\beta\bar{c}}? & =\frac{1}{6}V^{-1}\delta_{\bar{c}}^{\bar{a}}\partial_{\beta}V-\frac{i}{2}?\omega_{i}^{\bar{a}}{}_{\bar{b}}?\partial_{\beta}b^{i}
\end{align}
and we find
\begin{align}
\frac{1}{2\kappa_{11}^{2}}\int_{\m M_{11}}dvR_{\left(11\right)}=\; & \frac{1}{2\kappa_{5}^{2}}\int_{\m M}dv\left(R-\frac{1}{2}V^{-2}\partial_{\alpha}V\partial^{\alpha}V-G_{ij}\partial_{\alpha}b^{i}\partial^{\alpha}b^{j}\right)\nonumber \\
 & -\frac{1}{2\kappa_{5}^{2}}\int_{\partial\m M_{1}}dv\frac{2}{3}V^{-1}\partial_{z}V+\frac{1}{2\kappa_{5}^{2}}\int_{\partial\m M_{2}}dv\frac{2}{3}V^{-1}\partial_{z}V\label{eq:R_11}
\end{align}
where the five dimensional and eleven dimensional gravitational couplings
are related by the volume of the reference Calabi-Yau, $v$, as $\kappa_{5}^{2}=\frac{\kappa_{11}^{2}}{v}$.
The boundary terms which depend on $\partial_{z}V$ will cancel against
terms coming from the Gibbons-Hawking term, \eqref{eq:K_10,1} and
\eqref{eq:K_10,2}, as we would expect since the consistency of the
variational principle in eleven dimensions implies its consistency
in five and so we cannot have such a term, whose variation would include
$\partial_{z}\delta V$.

\subsection{The Five Dimensional Gravitino\label{sec:The-5d-Gravitino}}

The eleven dimensional gravitino gives rise to a number of five dimensional
fermion fields from its spacetime and Calabi-Yau components. We can
identify the proper ansatz by demanding that the five dimensional
fields have canonical kinetic terms. Suppose we start with the straightforward
ansatz
\begin{align}
\Psi_{\alpha} & =V^{-\frac{1}{6}}\theta_{\alpha}^{A}\otimes u_{A}\\
\Psi_{a} & =\beta V^{\frac{1}{3}}\lambda^{Ai}\otimes\omega_{ia\bar{b}}\gamma^{\bar{b}}u_{A}
\end{align}
then reducing the eleven dimensional kinetic term we get
\begin{align}
V^{-\frac{2}{3}}\bar{\Psi}_{I}\Gamma^{IJK}D_{J}\Psi_{K} & =\left(\bar{\theta}_{\alpha A}+\frac{4i}{3}\beta b_{i}\bar{\lambda}_{A}^{i}\gamma_{\alpha}\right)\gamma^{\alpha\beta\gamma}D_{\beta}\left(\theta_{\gamma}^{A}+\frac{4i}{3}\beta b_{i}\gamma_{\gamma}\lambda^{iA}\right)\nonumber \\
 & \quad+4\beta^{2}G_{ij}^{\perp}\bar{\lambda}_{A}^{i}\gamma^{\beta}D_{\beta}\lambda^{jA}+8\beta^{2}b_{i}\bar{\lambda}_{A}^{i}\gamma^{\beta}D_{\beta}\left(b_{j}\lambda^{jA}\right)\nonumber \\
 & \quad-4i\beta b_{i}\bar{\lambda}_{A}^{i}\gamma^{\alpha}\gamma^{\beta}\left(\theta_{\alpha}^{A}+\frac{4i}{3}\beta b_{i}\gamma_{\alpha}\lambda^{iA}\right)\partial_{\beta}V+\frac{i}{2}G_{ij}\bar{\lambda}_{A}^{i}\gamma^{\alpha}\gamma^{\beta}\psi_{\alpha}^{A}\partial_{\beta}b^{j}\label{eq:psi_kinetic}
\end{align}
so we see that $\beta=\frac{1}{2}$ to correctly normalize the $b^{i}$
superpartners, $\lambda^{\perp iA}={\delta^{\perp}}_{j}^{i}\lambda^{jA}$;
$b_{i}\lambda^{Ai}=\frac{1}{\sqrt{2}}\zeta^{A}$ ($\zeta^{A}$ is
the fermion in the hypermultiplet); and the 5d gravitino $\psi_{\alpha}^{A}=\theta_{\alpha}^{A}+\frac{\sqrt{2}i}{3}\zeta^{A}$.
Thus we see the ansatz \eqref{eq:Psi_a_ansatz}, \eqref{eq:Psi_alpha_ansatz}
was indeed the correct choice.

\subsection{Three-Form Terms}

The kinetic term for $C$, $-\frac{1}{48}G_{IJKL}G^{IJKL}$, gives
\begin{align}
S_{C1} & =\frac{1}{2\kappa_{5}^{2}}\int_{\m M}dv\left(-V^{2}G_{\alpha\beta\gamma\delta}G^{\alpha\beta\gamma\delta}-2V^{-1}\partial_{\alpha}\bar{\xi}\partial^{\alpha}\xi-\frac{1}{4}\m F_{i\alpha\beta}\m F^{i\alpha\beta}-\frac{1}{2}V^{-2}\alpha_{i}\alpha^{i}\right)\label{eq:S_C1}
\end{align}
where $\m F^{i}$ is the curvature of $\m A^{i}$, $\m F_{\alpha\beta}^{i}=2\partial_{[\alpha}\m A_{\beta]}^{i}$.

The Chern-Simons term, $\frac{1}{12^{4}}\epsilon^{I_{1}\ldots I_{11}}G_{I_{1}I_{2}I_{3}I_{4}}G_{I_{5}I_{6}I_{7}I_{8}}C_{I_{9}I_{10}I_{11}}$,
gives
\begin{align}
S_{C2}=\; & \frac{1}{2\kappa_{5}^{2}}\int_{\m M}dv\bigg(\frac{1}{4!}\epsilon^{\alpha\beta\gamma\delta\epsilon}G_{\alpha\beta\gamma\delta}\left(i\left(\xi\partial_{\epsilon}\bar{\xi}-\bar{\xi}\partial_{\epsilon}\xi\right)+\alpha_{i}\m A_{\epsilon}^{i}\right)\nonumber \\
 & \qquad\qquad\quad+\frac{1}{8\sqrt{2}}\m K^{-1}\m K_{ijk}\epsilon^{\alpha\beta\gamma\delta\epsilon}\m A_{\alpha}^{i}\m F_{\beta\gamma}^{j}\m F_{\delta\epsilon}^{k}\bigg)\nonumber \\
 & -\frac{1}{2\kappa_{5}^{2}}\int_{\partial\m M_{1}}dv\frac{1}{6}\epsilon^{\mu\nu\rho\sigma}C_{\mu\nu\rho}\left(i\left(\xi\partial_{\sigma}\bar{\xi}-\bar{\xi}\partial_{\sigma}\xi\right)+\alpha_{i}\m A_{\sigma}^{i}\right)\nonumber \\
 & +\frac{1}{2\kappa_{5}^{2}}\int_{\partial\m M_{2}}dv\frac{1}{6}\epsilon^{\mu\nu\rho\sigma}C_{\mu\nu\rho}\left(i\left(\xi\partial_{\sigma}\bar{\xi}-\bar{\xi}\partial_{\sigma}\xi\right)+\alpha_{i}\m A_{\sigma}^{i}\right)
\end{align}
and the gravitino bilinear term $-\frac{1}{96}\left(\bar{\Psi}_{M}\Gamma^{IJKLMO}\Psi_{O}+12\bar{\Psi}^{I}\Gamma^{JK}\Psi^{L}\right)G_{IJKL}$
gives
\begin{align}
S_{C3}=\frac{1}{2\kappa_{5}^{2}}\int_{\m M}dv & \left(-\frac{1}{96}VG^{\alpha\beta\gamma\delta}?\epsilon_{\alpha\beta\gamma\delta}^{\epsilon}?\Big(2\sqrt{2}?\tau^{A}{}_{B}?\bar{\zeta}_{A}\psi_{\epsilon}^{B}+3i?\tau^{A}{}_{B}?\bar{\zeta}_{A}\gamma_{\epsilon}\zeta^{B}\right.\nonumber \\
 & \qquad\qquad\qquad\qquad\qquad+iG_{ij}^{\perp}?\tau^{A}{}_{B}?\bar{\lambda}_{A}^{i}\gamma_{\epsilon}\lambda^{jB}\Big)\nonumber \\
 & \quad-\frac{1}{96}VG^{\alpha\beta\gamma\delta}\left(-8\sqrt{2}i?\tau^{A}{}_{B}?\bar{\zeta}_{A}\gamma_{\alpha\beta\gamma}\psi_{\delta}^{B}+12?\tau^{A}{}_{B}?\bar{\psi}_{\alpha A}\gamma_{\beta\gamma}\psi_{\delta}^{B}\right)\nonumber \\
 & \quad+\bar{\psi}_{A\alpha}\gamma^{\alpha\beta\gamma}\psi_{\gamma}^{B}?{\left(\begin{array}{cc}
0 & -\partial_{\beta}\xi\\
\partial_{\beta}\bar{\xi} & 0
\end{array}\right)}^{A}{}_{B}?\nonumber \\
 & \quad-\sqrt{2}i\bar{\zeta}_{A}\gamma^{\alpha}\gamma^{\beta}\psi_{\alpha}^{B}?{\left(\begin{array}{cc}
0 & \partial_{\beta}\xi\\
-\partial_{\beta}\bar{\xi} & 0
\end{array}\right)}^{A}{}_{B}?\nonumber \\
 & \quad+G_{ij}^{\perp}\bar{\lambda}_{A}^{i}\gamma^{\beta}\lambda^{Bj}?{\left(\begin{array}{cc}
0 & -\partial_{\beta}\xi\\
\partial_{\beta}\bar{\xi} & 0
\end{array}\right)}^{A}{}_{B}?\nonumber \\
 & \quad-\frac{\sqrt{2}i}{8}\left(\bar{\psi}_{A\gamma}\gamma^{\alpha}\gamma^{\gamma\delta}\gamma^{\beta}\psi_{\delta}^{A}-\bar{\zeta}_{A}\gamma^{\alpha\beta}\zeta^{A}-\frac{1}{3}G_{jk}^{\perp}\bar{\lambda}_{A}^{j}\gamma^{\alpha\beta}\lambda^{Ak}\right)b_{i}\m F_{\alpha\beta}^{i}\nonumber \\
 & \quad-\frac{\sqrt{2}}{4}G_{ij}^{\perp}\bar{\lambda}_{A}^{i}\gamma^{\gamma}\gamma^{\alpha\beta}\psi_{\gamma}^{A}\m F_{\alpha\beta}^{j}+\frac{3\sqrt{2}i}{8}\m K^{-1}\m K_{ijk}^{\perp}\bar{\lambda}_{A}^{i}\gamma^{\alpha\beta}\lambda^{Aj}\m F_{\alpha\beta}^{k}\nonumber \\
 & \quad-\frac{\sqrt{2}}{4}V^{-1}\alpha?\tau^{A}{}_{B}?\bar{\psi}_{A\alpha}\gamma^{\alpha\beta}\psi_{\beta}^{B}-\frac{\sqrt{2}i}{2}V^{-1}\alpha_{i}^{\perp}?\tau^{A}{}_{B}?\bar{\lambda}_{A}^{i}\gamma^{\alpha}\psi_{\alpha}^{B}\nonumber \\
 & \quad+iV^{-1}\alpha?\tau^{A}{}_{B}?\bar{\zeta}_{A}\gamma^{\alpha}\psi_{\alpha}^{B}\nonumber \\
 & \quad+\frac{3\sqrt{2}}{4}V^{-1}\left(\m K^{-1}?{\m K}^{\perp i}{}_{jk}?+\frac{1}{9}b^{i}G_{jk}^{\perp}\right)\alpha_{i}?\tau^{A}{}_{B}?\bar{\lambda}_{A}^{j}\lambda^{kB}\nonumber \\
 & \quad-2V^{-1}\alpha_{i}^{\perp}?\tau^{A}{}_{B}?\bar{\zeta}_{A}\lambda^{Bi}+\frac{3\sqrt{2}}{4}V^{-1}\alpha?\tau^{A}{}_{B}?\bar{\zeta}_{A}\zeta^{B}\bigg)\label{eq:psi^2G}
\end{align}
where we define $\alpha=\alpha_{i}b^{i}$. Much use is made in the
calculation of $S_{C3}$ of the identities in Appendix \ref{cha:Calabi-Yau_Geometry}.

We now have the ingredients in the action needed to dualize the four-form
field strength $G$ to a scalar field $\sigma$ we turn our attention
to this process first, followed by the elucidation of the quaternionic
structure of the hypermultiplet fields which simplifies these terms
in the action somewhat.

\subsection{Aside on Auxiliary Fields}

To manifest the hypermultiplet structure of the five dimensional theory
we wish to dualize the three-form $C$ to a scalar $\sigma$. I was
puzzled by the fact that the equation relating $\sigma$ and $G_{\alpha\beta\gamma\delta}$
is not supersymmetry invariant, yet one uses it to go from an action
in terms of $C$ to one in terms of $\sigma$ and both actions are
supersymmetric. Since this confusion has apparently manifested elsewhere
(e.g.\ in Weinberg's book \cite{Weinberg:2000cr} following equation
(26.4.7) ``With the auxiliary fields eliminated in this way, the action
is no longer invariant under the supersymmetry transformations\ldots{}'')
it seems worth setting down the elegant resolution to this `paradox',
explained to me by Ian Moss, before going into the details of the
present case.

Suppose we have a supersymmetric Lagrangian $\m L_{0}\left(\phi^{i},F\right)$
which depends on some fields $\phi^{i}$ and an auxiliary field $F$.
We wish to replace $F$ by its (algebraic) equation of motion. The
supersymmetric variation of $\m L_{0}$ vanishes
\begin{align}
\delta_{s}\m L_{0}\left(\phi^{i},F\right) & =\frac{\delta\m L_{0}}{\delta\phi^{i}}\delta_{s}\phi^{i}+\frac{\delta\m L_{0}}{\delta F}\delta_{s}F=0\label{eq:auxL0_susy}
\end{align}
and since $F$ is an auxiliary field, its equation of motion is does
not contain derivatives of $F$ and so it is just
\begin{align}
\frac{\delta\m L_{0}}{\delta F} & =0\label{eq:auxF_eom}\\
\Rightarrow F & =F\left(\phi^{i}\right)\label{eq:F_phi}
\end{align}
which we substitute into $\m L_{0}$ to get a Lagrangian depending
only on $\phi^{i}$
\begin{align}
\m L_{1}\left(\phi^{i}\right) & =\left.\m L_{0}\left(\phi^{i},F\right)\right|_{F=F\left(\phi^{i}\right)}
\end{align}
Now the supersymmetric variation of $\m L_{1}$ is
\begin{align}
\delta_{s}\m L_{1}\left(\phi^{i}\right) & =\left.\frac{\delta\m L_{0}}{\delta\phi^{i}}\delta_{s}\phi^{i}\right|_{F=F\left(\phi^{i}\right)}+\left.\frac{\delta\m L_{0}}{\delta F}\frac{\delta F\left(\phi^{i}\right)}{\delta\phi^{i}}\delta_{s}\phi^{i}\right|_{F=F\left(\phi^{i}\right)}
\end{align}
but evaluating \eqref{eq:auxL0_susy} on $F=F\left(\phi^{i}\right)$
tells us that
\begin{align}
\left.\frac{\delta\m L_{0}}{\delta\phi^{i}}\delta_{s}\phi^{i}\right|_{F=F\left(\phi^{i}\right)} & =-\left.\frac{\delta\m L_{0}}{\delta F}\delta_{s}F\right|_{F=F\left(\phi^{i}\right)}
\end{align}
and so we see that $\delta_{s}\m L_{1}\left(\phi^{i}\right)=0$ even
though generally \eqref{eq:F_phi} is not supersymmetric, i.e.\ $\delta_{s}F\neq\frac{\delta F\left(\phi^{i}\right)}{\delta\phi^{i}}\delta_{s}\phi^{i}$.

In our case there are boundaries as well, so we cannot drop total
derivatives when integrating out the auxiliary field. Rather the total
derivatives contribute to the process of imposing the boundary condition
of $G_{\alpha\beta\gamma\delta}$.

\subsection{Dualization\label{sec:Dualization}}

As we saw in Section \ref{sec:Five-Dimensional-Multiplets}, in order
to manifest the multiplet structure of the five dimensional theory
we must dualize the three-form $C_{\alpha\beta\gamma}$ to a scalar
$\sigma$. We introduce $\sigma$ as a Lagrange multiplier enforcing
the Bianchi identity on the field strength $dG=0$ and the boundary
conditions on $G_{\alpha\beta\gamma\delta}$.

$\sigma$ is given in relation to the eleven dimensional fields by
the equation of motion for $g_{\alpha\beta\gamma\delta}$, as we will
describe below. However, as just mentioned this equation of motion
is not supersymmetric and so we cannot use it to find the supersymmetry
transformation of $\sigma$ from the reduction. Instead, it arises
when we impose supersymmetry on the action after adding $\sigma$
as a Lagrange multiplier, which fixes the variation to be \eqref{eq:sigma-susy}.

The terms with $G_{\alpha\beta\gamma\delta}$ in the five dimensional
Lagrangian are
\begin{align}
\m L_{G}=\; & -\frac{1}{48}V^{2}G_{\alpha\beta\gamma\delta}G^{\alpha\beta\gamma\delta}+\frac{1}{4!}\epsilon^{\alpha\beta\gamma\delta\epsilon}G_{\alpha\beta\gamma\delta}\left(i\left(\xi\partial_{\epsilon}\bar{\xi}-\bar{\xi}\partial_{\epsilon}\xi\right)+\alpha_{i}\m A^{i}\right)\nonumber \\
 & -\frac{1}{96}VG^{\alpha\beta\gamma\delta}?\epsilon_{\alpha\beta\gamma\delta}^{\epsilon}?\left(2\sqrt{2}?\tau^{A}{}_{B}?\bar{\zeta}_{A}\psi_{\epsilon}^{B}+3i?\tau^{A}{}_{B}?\bar{\zeta}_{A}\gamma_{\epsilon}\zeta^{B}+iG_{ij}^{\perp}?\tau^{A}{}_{B}?\bar{\lambda}_{A}^{i}\gamma_{\epsilon}\lambda^{jB}\right)\nonumber \\
 & -\frac{1}{96}VG^{\alpha\beta\gamma\delta}\left(-8\sqrt{2}i?\tau^{A}{}_{B}?\bar{\zeta}_{A}\gamma_{\alpha\beta\gamma}\psi_{\delta}^{B}+12?\tau^{A}{}_{B}?\bar{\psi}_{\alpha A}\gamma_{\beta\gamma}\psi_{\delta}^{B}\right)
\end{align}
the integration by parts to eliminate $C_{\alpha\beta\gamma}$ from
the bulk action also gives boundary terms
\begin{align}
S_{C2}(\partial\m M)=\; & -\frac{1}{2\kappa_{5}^{2}}\int_{\partial\m M_{1}}dv\frac{1}{6}\epsilon^{\mu\nu\rho\sigma}C_{\mu\nu\rho}\left(i\left(\xi\partial_{\sigma}\bar{\xi}-\bar{\xi}\partial_{\sigma}\xi\right)+\alpha_{i}\m A_{\sigma}^{i}\right)\nonumber \\
 & +\frac{1}{2\kappa_{5}^{2}}\int_{\partial\m M_{2}}dv\frac{1}{6}\epsilon^{\mu\nu\rho\sigma}C_{\mu\nu\rho}\left(i\left(\xi\partial_{\sigma}\bar{\xi}-\bar{\xi}\partial_{\sigma}\xi\right)+\alpha_{i}\m A_{\sigma}^{i}\right)\label{eq:S_Cboundary}
\end{align}
and the boundary condition on $G$ is
\begin{align}
G_{\mu\nu\rho\sigma} & =\mp f_{\mu\nu\rho\sigma}\label{eq:G_bc}
\end{align}
where we find from the reduction of \eqref{eq:bulkC_bc} that $f_{\mu\nu\rho\sigma}$
is given by
\begin{align}
f_{\mu\nu\rho\sigma} & =\frac{\kappa_{5}^{2}}{g^{2}}\left(6F_{[\mu\nu}^{I}F_{\rho\sigma]}^{I}-\partial_{[\mu}\left(V^{-1}\bar{\chi}_{A}^{I}\gamma_{\nu\rho\sigma]}\chi^{IA}\right)-\partial_{[\mu}\tr\left(V^{-1}\bar{\eta}_{iA}\gamma_{\nu\rho\sigma]}\eta^{iA}\right)\right)\nonumber \\
 & =4\frac{\kappa_{5}^{2}}{g^{2}}\partial_{[\mu}\left(\omega_{\nu\rho\sigma]}^{Y}-\frac{1}{4}V^{-1}\bar{\chi}_{A}^{I}\gamma_{\nu\rho\sigma]}\chi^{IA}-\frac{1}{4}V^{-1}\tr\left(\bar{\eta}_{iA}\gamma_{\nu\rho\sigma]}\eta^{iA}\right)\right)\label{eq:f_mu-nu-rho-sigma}
\end{align}
where the $I$ index runs over the adjoint of $E_{6}$ on $\partial\m M_{1}$
and $E_{8}$ on $\partial\m M_{2}$ and there is no chiral multiplet
on $\partial\m M_{2}$ (i.e.\ $\eta^{iA}=0$ there). To dualize we
replace the field strength $G_{\alpha\beta\gamma\delta}$ with a generic
4-form $g_{\alpha\beta\gamma\delta}$ which has the same supersymmetry
transformation and we impose the Bianchi identity $\partial_{[\alpha}g_{\beta\gamma\delta\epsilon]}=0$
and the boundary condition \eqref{eq:G_bc} by introducing $\sigma$
as a Lagrange multiplier 
\begin{align}
S_{\text{Lagrange}}=\; & \frac{1}{2\kappa_{5}^{2}}\int_{\mathcal{M}}dv\left(-\frac{1}{24}\epsilon^{\alpha\beta\gamma\delta\epsilon}\sigma\partial_{\alpha}g_{\beta\gamma\delta\epsilon}\right)\nonumber \\
 & +\frac{1}{2\kappa_{5}^{2}}\int_{\partial\m M_{1}}\left(\frac{1}{24}\epsilon^{\mu\nu\rho\sigma}\left(g_{\mu\nu\rho\sigma}+f_{\mu\nu\rho\sigma}\right)\sigma\right)\nonumber \\
 & +\frac{1}{2\kappa_{5}^{2}}\int_{\partial\m M_{2}}\left(-\frac{1}{24}\epsilon^{\mu\nu\rho\sigma}\left(g_{\mu\nu\rho\sigma}-f_{\mu\nu\rho\sigma}\right)\sigma\right)
\end{align}
The equation of motion for $g_{\alpha\beta\gamma\delta}$ is then
\begin{align}
g_{\alpha\beta\gamma\delta}=\; & V^{-2}?\epsilon_{\alpha\beta\gamma\delta}^{\epsilon}?\left(\partial_{\epsilon}\sigma-i\left(\xi\partial_{\epsilon}\bar{\xi}-\bar{\xi}\partial_{\epsilon}\xi\right)-\alpha_{i}\m A_{\epsilon}^{i}\right)\nonumber \\
 & -V^{-1}?\epsilon_{\alpha\beta\gamma\delta}^{\epsilon}?\left(\frac{\sqrt{2}}{2}?\tau^{A}{}_{B}?\bar{\zeta}_{A}\psi_{\epsilon}^{B}+\frac{3i}{4}?\tau^{A}{}_{B}?\bar{\zeta}_{A}\gamma_{\epsilon}\zeta^{B}+\frac{i}{4}G_{ij}^{\perp}?\tau^{A}{}_{B}?\bar{\lambda}_{A}^{i}\gamma_{\epsilon}\lambda^{jB}\right)\nonumber \\
 & -V^{-1}?\epsilon_{\alpha\beta\gamma\delta}^{\epsilon}?\left(\frac{i}{4}?\tau^{A}{}_{B}?\bar{\psi}_{A\zeta}?\gamma_{\epsilon}^{\zeta\eta}?\psi_{\eta}^{B}-\frac{\sqrt{2}}{2}?\tau^{A}{}_{B}?\bar{\zeta}_{A}?\gamma_{\epsilon}^{\zeta}?\psi_{\zeta}^{B}\right)\\
\equiv\; & V^{-2}?\epsilon_{\alpha\beta\gamma\delta}^{\epsilon}?\left(\mathcal{D}_{\epsilon}\sigma-\frac{i}{4}V?\tau^{A}{}_{B}?\bar{\psi}_{A\zeta}?\gamma_{\epsilon}^{\zeta\eta}?\psi_{\eta}^{B}-\frac{3i}{4}V?\tau^{A}{}_{B}?\bar{\zeta}_{A}\gamma_{\epsilon}\zeta^{B}\right.\nonumber \\
 & \qquad\qquad\quad\left.-\frac{i}{4}VG_{ij}^{\perp}?\tau^{A}{}_{B}?\bar{\lambda}_{A}^{i}\gamma_{\epsilon}\lambda^{jB}-\frac{\sqrt{2}}{2}V?\tau^{A}{}_{B}?\bar{\zeta}_{A}\gamma^{\zeta}\gamma_{\epsilon}\psi_{\zeta}^{B}\right)
\end{align}
where we define
\begin{align}
\m D_{\alpha}\sigma & =\partial_{\alpha}\sigma-i\left(\xi\partial_{\alpha}\bar{\xi}-\bar{\xi}\partial_{\alpha}\xi\right)-\alpha_{i}\m A_{\alpha}^{i}\label{eq:curlyD_sigma}
\end{align}
consistent with the quaternionic structure in Section \eqref{sub:Hypermultiplet-Quaternionic-Structure}.

The action for $G$ before we introduced the Lagrange multiplier term
was supersymmetric by virtue of being the reduction over a Calabi-Yau
space of a supersymmetric eleven dimensional theory. However, the
Lagrange multiplier clearly varies under supersymmetry itself, as
in the bulk we have
\begin{equation}
\delta S_{Lagrange}=\frac{1}{2\kappa_{5}^{2}}\int_{\m M}dv\left(\frac{1}{24}\epsilon^{\alpha\beta\gamma\delta\epsilon}\partial_{\alpha}\sigma\delta g_{\beta\gamma\delta\epsilon}-\frac{1}{24}\epsilon^{\alpha\beta\gamma\delta\epsilon}\delta\sigma\partial_{\alpha}g_{\beta\gamma\delta\epsilon}\right)
\end{equation}
and we also have the terms which arise because $\partial_{[\alpha}g_{\beta\gamma\delta\epsilon]}\neq0$
when we do not integrate out $\sigma$. Demanding that the action
with both $g_{\alpha\beta\gamma\delta}$ and $\sigma$ be supersymmetric
fixes the variation of $\sigma$. Then eliminating $g_{\alpha\beta\gamma\delta}$
from this variation by its equation of motion we find that
\begin{align}
\delta_{s}\sigma & =-\frac{1}{\sqrt{2}}V?\tau^{A}{}_{B}?\bar{s}_{A}\zeta^{B}+\frac{1}{\sqrt{2}}V^{\frac{1}{2}}?{\left(\begin{array}{cc}
0 & \xi\\
\bar{\xi} & 0
\end{array}\right)}^{A}{}_{B}?\bar{s}_{A}\zeta^{B}\label{eq:sigma-susy}
\end{align}

Substituting the equation of motion back into the action gives
\begin{align}
S_{\sigma} & =\frac{1}{2\kappa_{5}^{2}}\int_{\m M}dv\left(-\frac{1}{2}\m D_{\alpha}\sigma\m D^{\alpha}\sigma\right)\nonumber \\
 & +\frac{1}{2\kappa_{5}^{2}}\int_{\m M}dv\left(\frac{i}{4}V?\tau^{A}{}_{B}?\bar{\psi}_{A\beta}?\gamma_{\alpha}^{\beta\gamma}?\psi_{\gamma}^{B}+\frac{3i}{4}V?\tau^{A}{}_{B}?\bar{\zeta}_{A}\gamma_{\alpha}\zeta^{B}+\frac{i}{4}VG_{ij}^{\perp}?\tau^{A}{}_{B}?\bar{\lambda}_{A}^{i}\gamma_{\alpha}\lambda^{jB}\right)\m D^{\alpha}\sigma\nonumber \\
 & +\frac{1}{2\kappa_{5}^{2}}\int_{\m M}dv\left(\frac{\sqrt{2}}{2}V?\tau^{A}{}_{B}?\bar{\zeta}_{A}\gamma^{\beta}\gamma_{\alpha}\psi_{\beta}^{B}\right)\m D^{\alpha}\sigma\nonumber \\
 & -\frac{1}{2\kappa_{5}^{2}}\int_{\partial\m M_{1}}dv\frac{1}{6}\epsilon^{\mu\nu\rho\sigma}\frac{\kappa_{5}^{2}}{g^{2}}\left(\omega_{\mu\nu\rho}^{Y}-\frac{1}{4}V^{-1}\bar{\chi}_{A}^{I}\gamma_{\mu\nu\rho}\chi^{IA}-\frac{1}{4}V^{-1}\tr\left(\bar{\eta}_{iA}\gamma_{\mu\nu\rho}\eta^{iA}\right)\right)\partial_{\sigma}\sigma\nonumber \\
 & -\frac{1}{2\kappa_{5}^{2}}\int_{\partial\m M_{2}}dv\frac{1}{6}\epsilon^{\mu\nu\rho\sigma}\frac{\kappa_{5}^{2}}{g^{2}}\left(\omega_{\mu\nu\rho}^{Y}-\frac{1}{4}V^{-1}\bar{\chi}_{A}^{I}\gamma_{\mu\nu\rho}\chi^{IA}\right)\partial_{\sigma}\sigma\nonumber \\
 & -\frac{1}{2\kappa_{5}^{2}}\int_{\partial\m M_{1}}dv\frac{1}{6}\epsilon^{\mu\nu\rho\sigma}C_{\mu\nu\rho}\left(i\left(\xi\partial_{\sigma}\bar{\xi}-\bar{\xi}\partial_{\sigma}\xi\right)+\alpha_{i}\m A_{\sigma}^{i}\right)\nonumber \\
 & +\frac{1}{2\kappa_{5}^{2}}\int_{\partial\m M_{2}}dv\frac{1}{6}\epsilon^{\mu\nu\rho\sigma}C_{\mu\nu\rho}\left(i\left(\xi\partial_{\sigma}\bar{\xi}-\bar{\xi}\partial_{\sigma}\xi\right)+\alpha_{i}\m A_{\sigma}^{i}\right)
\end{align}
and finally by replacing $C_{\mu\nu\rho}$ by its boundary condition
using \eqref{eq:f_mu-nu-rho-sigma} we find that the terms in the
action involving $\sigma$ are 
\begin{align}
S_{\sigma} & =\frac{1}{2\kappa_{5}^{2}}\int_{\m M}dv\left(-\frac{1}{2}\m D_{\alpha}\sigma\m D^{\alpha}\sigma\right)\nonumber \\
 & +\frac{1}{2\kappa_{5}^{2}}\int_{\m M}dv\left(\frac{i}{4}V?\tau^{A}{}_{B}?\bar{\psi}_{A\beta}?\gamma_{\alpha}^{\beta\gamma}?\psi_{\gamma}^{B}+\frac{3i}{4}V?\tau^{A}{}_{B}?\bar{\zeta}_{A}\gamma_{\alpha}\zeta^{B}+\frac{i}{4}VG_{ij}^{\perp}?\tau^{A}{}_{B}?\bar{\lambda}_{A}^{i}\gamma_{\alpha}\lambda^{jB}\right)\m D^{\alpha}\sigma\nonumber \\
 & +\frac{1}{2\kappa_{5}^{2}}\int_{\m M}dv\left(\frac{\sqrt{2}}{2}V?\tau^{A}{}_{B}?\bar{\zeta}_{A}\gamma^{\beta}\gamma_{\alpha}\psi_{\beta}^{B}\right)\m D^{\alpha}\sigma\nonumber \\
 & -\frac{1}{2\kappa_{5}^{2}}\int_{\partial\m M_{1}}dv\frac{1}{6}\epsilon^{\mu\nu\rho\sigma}\frac{\kappa_{5}^{2}}{g^{2}}\left(\omega_{\mu\nu\rho}^{Y}-\frac{1}{4}V^{-1}\bar{\chi}_{A}^{I}\gamma_{\mu\nu\rho}\chi^{IA}-\frac{1}{4}V^{-1}\tr\left(\bar{\eta}_{iA}\gamma_{\mu\nu\rho}\eta^{iA}\right)\right)\m D_{\sigma}\sigma\nonumber \\
 & -\frac{1}{2\kappa_{5}^{2}}\int_{\partial\m M_{2}}dv\frac{1}{6}\epsilon^{\mu\nu\rho\sigma}\frac{\kappa_{5}^{2}}{g^{2}}\left(\omega_{\mu\nu\rho}^{Y}-\frac{1}{4}V^{-1}\bar{\chi}_{A}^{I}\gamma_{\mu\nu\rho}\chi^{IA}\right)\m D_{\sigma}\sigma\label{eq:S_sigma}
\end{align}

\subsection{Hypermultiplet Quaternionic Structure\label{sub:Hypermultiplet-Quaternionic-Structure}}

The hypermultiplet scalars can be regarded as the coordinates of a
quaternionic manifold, as a particular case of the set-up described
in Appendix \ref{cha:Quaternionic_Appendix}. The kinetic terms for
the scalars in the hypermultiplet are
\begin{align}
S_{sk} & =\frac{1}{2\kappa_{5}^{2}}\int_{\m M}\left(-\frac{1}{2}V^{-2}\dd V\wedge*\dd V-2V^{-1}\dd\bar{\xi}\wedge\dd\xi-\frac{1}{2}V^{-2}\m D\sigma\wedge*\m D\sigma\right)\\
 & =\frac{1}{2\kappa_{5}^{2}}\int_{\m M}\left(-?f_{x}^{A}{}_{B}??f_{y}^{\dagger B}{}_{A}?Dq^{x}\wedge Dq^{y}\right)
\end{align}
with $?f_{x}^{A}{}_{B}?=?{\left(\begin{array}{cc}
\bar{v}_{x} & u_{x}\\
-\bar{u}_{x} & v_{x}
\end{array}\right)}^{A}{}_{B}?$ and
\begin{align}
u & =u_{x}\dd q^{x}=V^{-\frac{1}{2}}\dd\xi\\
v & =v_{x}\dd q^{x}=\frac{V^{-1}}{2}\left(\dd V+i\dd\sigma-\bar{\xi}\dd\xi+\xi\dd\bar{\xi}\right)\\
Dq^{x} & =\left(\dd V,\m D\sigma,\dd\xi,\dd\bar{\xi}\right)
\end{align}
so explicitly
\begin{align}
?f^{A}{}_{B}? & =?{\left(\begin{array}{cc}
\frac{1}{2}V^{-1}\left(\dd V-i\m D\sigma\right) & V^{-\frac{1}{2}}\dd\xi\\
-V^{-\frac{1}{2}}\dd\bar{\xi} & \frac{1}{2}V^{-1}\left(\dd V+i\m D\sigma\right)
\end{array}\right)}^{A}{}_{B}?
\end{align}
$\m D\sigma$ being given by \eqref{eq:curlyD_sigma}. The isometries
here are shifts in $\sigma$, so the Killing vectors are $gk_{i}=-\alpha_{i}\frac{\partial}{\partial\sigma}$
which give the prepotential $g\m P_{i}=-\frac{i\alpha_{i}}{4V}\left(\begin{array}{cc}
1 & 0\\
0 & -1
\end{array}\right)$ and hence, from \eqref{eq:omega_L} and \eqref{eq:omega_R}, we have
gauged connections
\begin{align}
?\omega_{\left(L\right)}^{A}{}_{B}? & =?{\left(\begin{array}{cc}
\frac{i}{4}V^{-1}\m D\sigma & -V^{-\frac{1}{2}}\dd\xi\\
V^{-\frac{1}{2}}\dd\bar{\xi} & -\frac{i}{4}V^{-1}\m D\sigma
\end{array}\right)}^{A}{}_{B}?\\
?\omega_{\left(R\right)}^{A}{}_{B}? & =?{\left(\begin{array}{cc}
\frac{3i}{4}V^{-1}\m D\sigma & 0\\
0 & -\frac{3i}{4}V^{-1}\m D\sigma
\end{array}\right)}^{A}{}_{B}?
\end{align}
We may use these connections to write covariant fermion derivatives
\begin{align}
\m D_{\alpha}\psi_{\beta}^{A} & =D_{\alpha}\psi_{\beta}^{A}+?\omega_{\left(L\right)}^{A}{}_{B}?\psi_{\beta}^{B}\\
\m D_{\alpha}\lambda^{iA} & =D_{\alpha}\lambda^{iA}+?\omega_{\left(L\right)}^{A}{}_{B}?\lambda^{iB}\\
\m D_{\alpha}\zeta^{A} & =D_{\alpha}\zeta^{A}-?\omega_{\left(R\right)}^{A}{}_{B}?\zeta^{B}
\end{align}
so that we can package up the fermion terms in the bulk action from
\eqref{eq:psi_kinetic}, \eqref{eq:psi^2G} and \eqref{eq:S_sigma}
into
\begin{align}
S_{C3}=\frac{1}{2\kappa_{5}^{2}}\int_{\m M}dv & \left(\bar{\psi}_{A\alpha}\gamma^{\alpha\beta\gamma}\m D_{\beta}\psi_{\gamma}^{A}+\bar{\zeta}_{A}\gamma^{\beta}\m D_{\beta}\zeta^{A}+G_{ij}^{\perp}\bar{\lambda}_{A}^{i}\gamma^{\beta}\m D_{\beta}\lambda^{jA}\vphantom{\Bigg)}\right.\nonumber \\
 & \quad-\sqrt{2}i\bar{\zeta}_{A}\gamma^{\alpha}\gamma^{\beta}\psi_{\alpha}^{B}?f_{\beta}^{A}{}_{B}?+\frac{i}{2}G_{ij}\bar{\lambda}_{A}^{i}\gamma^{\alpha}\gamma^{\beta}\psi_{\alpha}^{A}\partial_{\beta}b^{j}\nonumber \\
 & \quad-\frac{\sqrt{2}i}{8}\left(\bar{\psi}_{A\gamma}\gamma^{\alpha}\gamma^{\gamma\delta}\gamma^{\beta}\psi_{\delta}^{A}-\bar{\zeta}_{A}\gamma^{\alpha\beta}\zeta^{A}-\frac{1}{3}G_{jk}^{\perp}\bar{\lambda}_{A}^{j}\gamma^{\alpha\beta}\lambda^{Ak}\right)b_{i}\m F_{\alpha\beta}^{i}\nonumber \\
 & \quad-\frac{\sqrt{2}}{4}G_{ij}^{\perp}\bar{\lambda}_{A}^{i}\gamma^{\gamma}\gamma^{\alpha\beta}\psi_{\gamma}^{A}\m F_{\alpha\beta}^{j}+\frac{3\sqrt{2}i}{8}\m K^{-1}\m K_{ijk}^{\perp}\bar{\lambda}_{A}^{i}\gamma^{\alpha\beta}\lambda^{Aj}\m F_{\alpha\beta}^{k}\nonumber \\
 & \quad-\frac{\sqrt{2}}{4}V^{-1}\alpha?\tau^{A}{}_{B}?\bar{\psi}_{A\alpha}\gamma^{\alpha\beta}\psi_{\beta}^{B}-\frac{\sqrt{2}i}{2}V^{-1}\alpha_{i}^{\perp}?\tau^{A}{}_{B}?\bar{\lambda}_{A}^{i}\gamma^{\alpha}\psi_{\alpha}^{B}\nonumber \\
 & \quad+iV^{-1}\alpha?\tau^{A}{}_{B}?\bar{\zeta}_{A}\gamma^{\alpha}\psi_{\alpha}^{B}\nonumber \\
 & +\frac{3\sqrt{2}}{4}V^{-1}\left(\m K^{-1}?{\m K}^{\perp i}{}_{jk}?+\frac{1}{9}b^{i}G_{jk}^{\perp}\right)\alpha_{i}?\tau^{A}{}_{B}?\bar{\lambda}_{A}^{j}\lambda^{kB}\nonumber \\
 & \quad-2V^{-1}\alpha_{i}^{\perp}?\tau^{A}{}_{B}?\bar{\zeta}_{A}\lambda^{Bi}+\frac{3\sqrt{2}}{4}V^{-1}\alpha?\tau^{A}{}_{B}?\bar{\zeta}_{A}\zeta^{B}\Bigg)
\end{align}
The quaternionic structure also appears in the supersymmetry transformations
that we find by reducing \eqref{eq:Psi_susy} and \eqref{eq:C_susy_11d}.
The individual components, (\ref{eq:zeta-susy}-\ref{eq:barxi-susy})
and \eqref{eq:sigma-susy}, can be grouped together to find that 
\begin{align}
\delta q^{x} & =-\frac{\sqrt{2}i}{2}?f^{xA}{}_{B}?\bar{s}_{A}\zeta^{B}\\
\delta\zeta^{A} & =-\frac{\sqrt{2}i}{2}\gamma^{\alpha}s^{B}?f_{xB}^{A}?\m D_{\alpha}q^{x}
\end{align}
while \eqref{eq:psi_susy_1} becomes
\begin{align}
\delta\psi_{\alpha}^{A} & =\m D_{\alpha}s^{A}-\frac{\sqrt{2}i}{12}\left(?\gamma_{\alpha}^{\beta\gamma}?-4\delta_{\alpha}^{\beta}\gamma^{\gamma}\right)b_{i}\m F_{\beta\gamma}^{i}s^{A}-\frac{2\sqrt{2}i}{3}gb^{i}?{\m P}_{i}^{A}{}_{B}?\gamma_{\alpha}s^{B}
\end{align}
with 
\begin{align}
\m D_{\alpha}s^{A} & =D_{\alpha}s^{A}+?\omega_{\left(L\right)}^{A}{}_{B}?s^{B}
\end{align}

One might ask why we must gauge the isometry of the quaternionic manifold.
From the point of view of the reduction of the eleven dimensional
theory, it just turns out to be that way. From the perspective of
five-dimensional supergravity however we see that the gauging is intimately
bound up with the presence of the non-zero mode (which is forced upon
us by the sources of $G$-flux on the boundaries): if $\alpha_{i}$
were zero we would not have a gauged theory. The reason for the gauging
is that the non-zero $\alpha_{i}$ gives a potential for $V$, which
is $-\frac{1}{2}V^{-2}\alpha_{i}\alpha^{i}$ as we see in \eqref{eq:S_C1}.
A potential for hypermultiplet terms appears in gauged supergravity
theories, such as the $N=2$ four dimensional theories described in
\cite{Andrianopoli:1996vr} which are rather similar to our five dimensional
theory since they have the same amount of supersymmetry. We can see
why such a potential term necessitates a gauged theory if we consider
its supersymmetry transformation. Clearly the theory with $\alpha_{i}=0$
is supersymmetric. So the only terms which might cancel the variation
of the potential, $V\alpha^{2}\delta V=-\frac{i}{\sqrt{2}}V\alpha^{2}?\tau^{A}{}_{B}?\bar{s}_{A}\zeta^{B}$,
are those which appear when we switch on $\alpha_{i}$ and these are
just $\alpha_{i}\m A_{\alpha}^{i}\m D^{\alpha}\sigma$. The only terms
in the supersymmetry variation of this which contain $?\tau^{A}{}_{B}?\bar{s}_{A}\zeta^{B}$
are $\alpha_{i}\m A_{\alpha}^{i}\m D^{\alpha}\left(\delta\sigma\right)$,
since $\m A_{\alpha}^{i}$ is not part of the hypermultiplet. This
does not have the right form to cancel the variation of the potential
directly. Therefore we see that the potential can only be present
as part of a gauged theory in which the extra variations when $\alpha_{i}\neq0$
can be identified as a gauge transformation of the hypermultiplet.

This completes the reduction of the theory in the bulk. We now turn
to the description of the gauge theories on the boundaries and their
couplings to the bulk fields.

\section{The Boundaries\label{sec:The-Boundaries}}

We find the boundary action and the boundary conditions coming from
that of the eleven dimensional three-form $C_{ABC}$ by substituting
the ansatzes of Section \ref{sec:Reduction-Ansatz} into the eleven-dimensional
boundary action \eqref{eq:11d-boundary-action} and the boundary condition
\eqref{eq:bulkC_bc}. It is easiest to find the remaining boundary
conditions by varying the five dimensional action: these are given
in Appendix \ref{sec:Summary-Boundary-Conditions}.

\subsection{Boundary Action for the Gravity Multiplet}

The ten dimensional metric determinant is $\sqrt{-g_{\left(10\right)}}=V^{-\frac{1}{3}}\sqrt{-g_{\left(4\right)}}\sqrt{g_{CY}}$
and so the volume element is $dv_{\left(10\right)}=V^{-\frac{1}{3}}dv_{\left(4\right)}dv_{CY}$.
The unit vector normal to the boundaries (normalized by the five dimensional
metric) is given in terms of the eleven dimensional normal vector
by $n^{\alpha}=V^{-\frac{1}{3}}n_{\left(11\right)}^{\alpha}$. Thus
we find that the Gibbons-Hawking term gives
\begin{align}
-\frac{1}{2\kappa_{11}^{2}}\int_{\partial\m M_{10,1}}dv\left(2K_{(10)}\right) & =-\frac{1}{2\kappa_{5}^{2}}\int_{\partial\m M_{1}}dv\left(2K-\frac{2}{3}V^{-1}\partial_{z}V\right)\label{eq:K_10,1}
\end{align}
and
\begin{align}
\frac{1}{2\kappa_{11}^{2}}\int_{\partial\m M_{10,2}}dv\left(2K_{(10)}\right) & =\frac{1}{2\kappa_{5}^{2}}\int_{\partial\m M_{2}}dv\left(2K-\frac{2}{3}V^{-1}\partial_{z}V\right)\label{eq:K_10,2}
\end{align}
As promised above the $V^{-1}\partial_{z}V$ contributions are appropriate
to cancel those which come from the Einstein-Hilbert term \eqref{eq:R_11}.

The gravitino bilinear term in the boundary action gives
\begin{align}
 & \pm\frac{1}{2\kappa_{11}^{2}}\int_{\partial\m M_{10}}dv\left(\frac{1}{2}\bar{\Psi}_{A}\Gamma^{AB}\Psi_{B}\right)\nonumber \\
 & =\pm\frac{1}{2\kappa_{5}^{2}}\int_{\partial\m M}dv\left(\frac{1}{2}?\tau^{A}{}_{B}?\bar{\psi}_{A\mu}\gamma^{\mu\nu}\psi_{\nu}^{B}+\frac{1}{2}?\tau^{A}{}_{B}?\bar{\zeta}_{A}\zeta^{B}+\frac{1}{2}G_{ij}^{\perp}?\tau^{A}{}_{B}?\bar{\lambda}_{A}^{i}\lambda^{Bj}\right)
\end{align}

The term with the component of the gravitino normal to the boundary,
$\mp\bar{\Psi}_{A}\Gamma^{A}\Psi_{N}$ is a fermionic torsion term
which just cancels a total derivative of the torsion from the bulk
Ricci scalar and never appears in the field equations. It is given
explicitly by $\m T=\mp\bar{\psi}_{A\mu}\gamma^{\mu}\left(\psi_{z}^{A}-\frac{\sqrt{2}i}{3}\gamma_{z}\zeta^{A}\right)\pm\frac{7\sqrt{2}i}{3}\bar{\zeta}_{A}\psi_{z}^{A}$
.

\subsection{Boundary Sources for $G$ and the Embedding of the Spin Connection\label{sub:Spin_embedding}}

As the only non-vanishing components of the curvature tensor on a
Calabi-Yau manifold are $R_{a\bar{b}c\bar{d}}$, the part of $\tr R^{2}$
in the tangent space of the Calabi-Yau manifold is a (2,2)-form and
so we can expand it on the basis $\nu^{i}$ of (2,2) forms as
\begin{align}
\tr R^{2} & =-\frac{2\sqrt{2}g^{2}}{\kappa_{5}^{2}}\alpha_{i}\nu^{i}\label{eq:tr-R^2}
\end{align}
with the coefficient being chosen to give \eqref{eq:G_abcd} (which
is in turn chosen to make $\alpha_{i}$ the charge associated with
$\m A^{i}$ in the covariant derivative $\m D_{\alpha}\sigma$). The
integrals of $\tr R^{2}$ over (2,2)-cycles of the Calabi-Yau, $\beta_{i}=-\frac{1}{8\pi}\int_{C_{i}}\tr(R\wedge R)$,
are integers characterizing the first Pontrjagin class of the particular
Calabi-Yau. These fix $\alpha_{i}$ to be $\alpha_{i}=\frac{2\sqrt{2}\pi\kappa_{5}^{2}}{g^{2}}\beta_{i}$.

Due to the spin embedding, restricting to the Calabi-Yau components
of the curvatures, $\tr F^{2}=\tr R^{2}$ on $\partial\m M_{1}$,
as we see from \eqref{eq:A_a-ansatz-1}, while $\tr F^{2}=0$ on $\partial\m M_{2}$,
from \eqref{eq:A_a-ansatz-2}. Therefore the boundary condition \eqref{eq:bulkC_bc}
gives $G_{a\bar{b}c\bar{d}}=-\sqrt{2}\alpha_{i}\nu_{a\bar{b}c\bar{d}}^{i}$
on both boundaries and we see that in the compactification the lowest
energy mode is $G_{a\bar{b}c\bar{d}}=-\sqrt{2}\alpha_{i}\nu_{a\bar{b}c\bar{d}}^{i}$
everywhere; we cannot have $G_{a\bar{b}c\bar{d}}=0$.

\subsection{Reducing the Lorentz Multiplet\label{sub:Reducing-the-Lorentz-Multiplet}}

The eleven dimensional theory includes $R^{2}$ terms on the boundaries
in a supersymmetric fashion by constructing a Lorentz multiplet from
the spin connection and the gravitino. This transforms similarly under
supersymmetry to a Yang-Mills multiplet and its action can thereby
be constructed by analogy, though with some extra complications involving
components of the gravitino normal to the boundary and $G$-flux fields.
We have not, however, included this Lorentz multiplet in the reduction,
except for the brane charges $\alpha_{i}$ described in the previous
section. The reason for this is essentially that the analogy with
the Yang-Mills multiplet, which in the eleven dimensional theory dictates
the structure of the Lorentz multiplet, fails completely in the context
of the reduction. One can find the contribution of the Lorentz multiplet
to the ten dimensional boundary action and boundary conditions in
a way which is in a sense independent of the bulk fields. In contrast
once we have given the reduction ansatz for the metric, gravitino
and three form in the bulk we have no freedom to choose an ansatz
for the Lorentz multiplet by analogy with the Yang-Mills ansatz. Of
course this does not prohibit simply using the ansatz for the bulk
fields to calculate the contribution of the Lorentz multiplet terms
to the five dimensional theory and indeed we attempted this, for contributions
at leading order in the derivative expansion of \cite{Moss:2008ng}.
The problem with this approach is twofold: firstly the $SO(1,9)$
traces in the ten dimensional terms give many different combinations
of five dimensional fields arising from the permutations of spacetime
and Calabi-Yau components; secondly we have lost any organizing principle
to assemble the terms to exhibit some sort of multiplet structure
in the reduced theory. The best candidate for such a structure would
be a reprise in four dimensions of the ten dimensional Lorentz multiplet.
However, such a four dimensional Lorentz multiplet could only include
the five dimensional gravitino, $\psi_{\alpha}^{A}$, and not the
other fermions, $\zeta^{A}$ and $\lambda^{\perp Ai}$, which come
from the eleven dimensional gravitino but are not part of the gravity
multiplet in five dimensions. Unfortunately upon examining the results
no alternative structure suggested itself to us either. The utility
of such an unordered set of terms composed of bulk fermion fields
in the boundary action and boundary conditions seems doubtful and
so it seems better to us to adopt an approximation (of small spacetime
curvature on the boundaries) which excludes them altogether.

\subsection{Yang-Mills Action on the $E_{8}$ Boundary}

On $\partial\m M_{2}$ where the $E_{8}$ gauge group is unbroken
the action is much simpler, so we will first examine that. We just
have an $E_{8}$ Yang-Mills vector multiplet. The approximation made
here is that $R_{a\bar{b}c\bar{d}}R^{a\bar{b}c\bar{d}}$ is the only
non-negligible part of $R^{2}$ (i.e.\ the four dimensional spacetime
curvature is small). So using \eqref{eq:tr-R^2}, the bosonic part
of the Yang-Mills action is just
\begin{align}
 & \frac{1}{\lambda^{2}}\int_{\partial\m M_{10,2}}dv\Bigg(-\frac{1}{4}\left(\tr\left(F_{AB}F^{AB}\right)-\frac{1}{2}\tr\left(R_{AB}R^{AB}\right)\right)\Bigg)\nonumber \\
 & =\int_{\partial\m M_{2}}dv\left(-\frac{1}{4g^{2}}VF_{\mu\nu}^{I}F^{I\mu\nu}+\frac{\sqrt{2}}{2\kappa_{5}^{2}}V^{-1}\alpha\right)
\end{align}
where the four dimensional gauge coupling is $g^{2}=\lambda^{2}/v$.
The fermionic part is
\begin{align}
 & \frac{1}{\lambda^{2}}\int_{\partial\m M_{10,2}}dv\left(\frac{1}{2}\tr\left(\bar{\chi}\Gamma^{A}D_{A}\chi\right)-\frac{1}{4}\bar{\Psi}_{C}\Gamma^{AB}\Gamma^{C}\tr\left(F_{AB}\chi\right)\right)\nonumber \\
 & =\frac{1}{g^{2}}\int_{\partial\m M_{2}}dv\left(\frac{1}{2}\bar{\chi}_{A}^{I}\gamma^{\mu}D_{\mu}\chi^{IA}+\bar{\psi}_{A\mu}j^{A\mu}+\bar{\zeta}_{A}j^{A}\right)
\end{align}
where the supercurrent is
\begin{align}
j^{A\mu} & =-\frac{1}{4}V^{\frac{1}{2}}F_{\nu\rho}^{I}\gamma^{\nu\rho}\gamma^{\mu}\chi^{IA}
\end{align}
and the current which couples to $\zeta^{A}$ is
\begin{align}
j^{A} & =-\frac{\sqrt{2}i}{2}V^{\frac{1}{2}}F_{\mu\nu}^{I}\gamma^{\mu\nu}\chi^{IA}
\end{align}

\subsection{Chiral Multiplets\label{sub:Chiral-Multiplets}}

Given the field content in the reduction ansatz and the fact that
reduction on a Calabi-Yau gives us a supersymmetric four dimensional
theory on the boundary, we know that we must find an instance of a
four dimensional super-Yang-Mills theory coupled to chiral multiplets,
as described in the general case in, for example, \cite{Weinberg:2000cr}.
The challenge in the reduction then is to arrange into this known
structure the fields arising from substituting the ansatz into the
eleven dimensional action, and in the process to identify the superpotential
and Kähler potential which specify the particular four dimensional
theory. This is indeed a challenge since the four dimensional supersymmetry
structure is not manifest upon performing the reduction directly.
For instance, the potential for the scalars arising from the reduction
is
\begin{align}
 & V^{-1}\left(?\omega_{ia}^{b}??\omega_{jb}^{c}??\omega_{kc}^{d}??\omega_{ld}^{a}?\right)\left(\frac{1}{4}C^{ip}\bC jpC^{kr}\bC lr-5d_{pst}d^{qrt}C^{ip}\bC jqC^{ks}\bC lr\right)\nonumber \\
 & +V^{-1}\left(?\omega_{ia}^{b}??\omega_{lb}^{a}?\right)\left(?\omega_{jc}^{d}??\omega_{kd}^{c}?\right)\left(\frac{1}{4}C^{ip}\bC jpC^{kr}\bC lr+\frac{5}{2}d_{pst}d^{qrt}C^{ip}\bC jqC^{ks}\bC lr\right)\label{eq:scalar-potential-reduction}
\end{align}
which should be
\begin{equation}
-\frac{1}{2}V^{-1}G^{ij}\frac{\partial W}{\partial C^{ip}}\frac{\partial\bar{W}}{\partial\bC jp}-2V^{-1}D^{I}D^{I}\label{eq:scalar-potential-susy}
\end{equation}
where $W$ is the superpotential, which we must determine, and $D^{I}=G_{ij}\bC ip?\Lambda^{Ip}{}_{q}?C^{jq}$
with $\Lambda^{I}$ the generators of the fundamental of $E_{6}$.
The $E_{6}$ adjoint projector $\Lambda^{I}\Lambda^{I}$ is given
in equation \eqref{eq:E6-adjoint-projector}.

To reconcile these two forms we need some identities for contractions
of strings of the (1,1)-form basis elements. Those identities we could
find are summarized at the end of Appendix \ref{cha:Calabi-Yau_Geometry}.
Unfortunately, for the contraction of four basis elements we could
find only an identity for a symmetrized version and so in this one
place we cannot complete the entire reduction directly. We do however
have enough information to calculate the superpotential and Kähler
potential and to cross-check.

The Kähler potential, $K(C,\bar{C})$ gives the coefficients of the
kinetic terms as $-\frac{\partial^{2}K}{\partial C^{ip}\partial\bC jq}\m D_{\mu}C^{ip}\m D^{\mu}\bC jq$
and since from the reduction this is $-2G_{ij}\m D_{\mu}C^{ip}\m D^{\mu}\bC jp$
the Kähler potential is just $2G_{ij}C^{ip}\bC jp$.

We can find the superpotential most easily from the mass terms for
the $\eta$'s, which are given in terms of the superpotential by
\begin{align}
 & \frac{1}{2}\frac{\partial^{2}W}{\partial C^{ip}\partial C^{jq}}?{\t{\bar{\eta}}}_{R}^{jq}??\eta_{L}^{kr}?+\frac{1}{2}\frac{\partial^{2}\bar{W}}{\partial\bC ip\partial\bC jq}?{\t{\bar{\eta}}}_{L}^{j}{}_{q}??\eta_{R}^{k}{}_{r}?
\end{align}
and from the reduction by
\begin{align}
 & \frac{3\sqrt{10}}{2}V^{-\frac{1}{2}}\m K^{-1}\m K_{ijk}\left(d_{pqr}C^{ip}?{\t{\bar{\eta}}}_{R}^{jq}??\eta_{L}^{kr}?+d^{pqr}\bC ip?{\t{\bar{\eta}}}_{L}^{j}{}_{q}??\eta_{R}^{k}{}_{r}?\right)\label{eq:eta-mass-reduction}
\end{align}
and so we find that the superpotential is 
\begin{align}
W & =3\sqrt{10}\m K^{-1}\m K_{ijk}d_{pqr}C^{ip}C^{jq}C^{kr}\label{eq:superpotential}
\end{align}

The algebra of the gauge group generators $T_{ap}$ and $T^{ap}$,
the $E_{6}$ generators $X^{I}$ and the $SU(3)$ generators $S^{i}$
is vital to this chiral multiplet part of the reduction. To find \eqref{eq:scalar-potential-reduction}
we need $\tr\left([T_{ap},T^{bq}][T_{cr},T^{ds}]\right)$ and to find
\eqref{eq:eta-mass-reduction} we need $\tr\left([T_{ap},T_{bq}]T_{cr}\right)$
and so to find the correct coefficient of the superpotential requires
that we know the algebra of the generators which is consistent with
the normalizations we take for the traces. The algebra and normalizations
we used are described in Appendix \ref{cha:E8-Gauge-Theory}.

The consistency check is that $D^{I}D^{I}$ does not contain any terms
with coefficient $\m K_{ijm}?{\m K}^{m}{}_{kl}?d_{pst}d^{qrt}$ and
so the only such term in \eqref{eq:scalar-potential-susy} is $\left|\frac{\partial W}{\partial C}\right|^{2}$.
On the other hand $d_{pst}d^{qrt}C^{ip}\bC jqC^{ks}\bC lr$ imposes
enough symmetry on $\omega^{4}$ to give the part we can calculate
and so we know the corresponding term in \eqref{eq:scalar-potential-reduction}.
Indeed comparing these using the superpotential calculated from the
fermion mass terms, \eqref{eq:superpotential}, we find that the coefficients
match correctly.

\subsection{Yang-Mills Action on the $E_{6}$ Boundary}

On $\partial\m M_{1}$ we have an $E_{6}$ Yang-Mills multiplet with
the same structure as the $E_{8}$ multiplet on $\partial\m M_{2}$
but we also have a set of $h^{1,1}$ chiral multiplets from the $\left(\mathbf{3},\mathbf{27}\right)$
and $\left(\overline{\mathbf{3}},\overline{\mathbf{27}}\right)$ parts
of the eleven dimensional $E_{8}$ fields. We find, as described in
the previous section, that the bosonic part of the action is
\begin{align}
 & \frac{1}{\lambda^{2}}\int_{\partial\m M_{10,1}}dv\Bigg(-\frac{1}{4}\left(\tr\left(F_{AB}F^{AB}\right)-\frac{1}{2}\tr\left(R_{AB}R^{AB}\right)\right)\Bigg)\nonumber \\
 & =\frac{1}{g^{2}}\int_{\partial\m M_{1}}dv\left(-\frac{1}{4}VF_{\mu\nu}^{I}F^{I\mu\nu}-2G_{ij}\m D_{\mu}C^{ip}\m D^{\mu}\bC jp-\frac{\sqrt{2}g^{2}}{2\kappa_{5}^{2}}V^{-1}\alpha\vphantom{\frac{\partial\bar{W}}{\partial\bC jp}}\right.\nonumber \\
 & \qquad\qquad\qquad\quad\left.-\frac{1}{2}V^{-1}G^{ij}\frac{\partial W}{\partial C^{ip}}\frac{\partial\bar{W}}{\partial\bC jp}-2V^{-1}\bar{C}^{i}\Lambda^{I}C_{i}\bar{C}^{j}\Lambda^{I}C_{j}\right)
\end{align}
with the superpotential $W=3\sqrt{10}\m K^{-1}\m K_{ijk}d_{pqr}C^{ip}C^{jq}C^{kr}$,
and the fermionic part is
\begin{align}
 & \frac{1}{\lambda^{2}}\int_{\partial\m M_{10,1}}dv\left(\frac{1}{2}\tr\left(\bar{\chi}\Gamma^{A}D_{A}\chi\right)-\frac{1}{4}\bar{\Psi}_{C}\Gamma^{AB}\Gamma^{C}\tr\left(F_{AB}\chi\right)\right)\nonumber \\
 & =\frac{1}{g^{2}}\int_{\partial\m M_{1}}dv\bigg(\frac{1}{2}\bar{\chi}_{A}^{I}\gamma^{\mu}\m D_{\mu}\chi^{IA}+\frac{1}{2}G_{ij}\tr\left(?{\t{\bar{\eta}}}_{A}^{i}?\gamma^{\mu}\m D_{\mu}\eta^{Aj}\right)\nonumber \\
 & \qquad\qquad\qquad\quad+\frac{3\sqrt{10}}{2}V^{-\frac{1}{2}}\m K^{-1}\m K_{ijk}\left(d_{pqr}C^{ip}?{\t{\bar{\eta}}}_{R}^{jq}??\eta_{L}^{kr}?+d^{pqr}\bC ip?{\t{\bar{\eta}}}_{L}^{j}{}_{q}??\eta_{R}^{k}{}_{r}?\right)\nonumber \\
 & \qquad\qquad\qquad\quad+2G_{ij}V^{-\frac{1}{2}}\left(\bar{\chi}_{L}^{I}?\eta_{R}^{i}{}_{q}?C^{jp}?\Lambda^{Iq}{}_{p}?-\bar{\chi}_{R}^{I}?\eta_{L}^{ip}?\bC jq?\Lambda^{Iq}{}_{p}?\right)\nonumber \\
 & \qquad\qquad\qquad\quad+\bar{\psi}_{A\mu}j^{A\mu}+\bar{\zeta}_{A}j^{A}+G_{ij}^{\perp}\bar{\lambda}_{A}^{i}j^{Aj}\bigg)
\end{align}
The current coupling to the gravitino $\psi_{\mu}^{A}$ is
\begin{align}
j^{A\mu}=\;- & \frac{1}{4}V^{\frac{1}{2}}\gamma^{\nu\rho}\gamma^{\mu}F_{\nu\rho}^{I}\chi^{I}\nonumber \\
 & -\gamma^{\nu}\gamma^{\mu}P_{R}?\eta^{A}{}_{ip}?\m D_{\nu}C^{ip}+\gamma^{\nu}\gamma^{\mu}P_{L}?\eta^{A}{}_{i}^{p}?\m D_{\nu}\bC ip\nonumber \\
 & +\frac{3\sqrt{10}}{2}V^{-\frac{1}{2}}\gamma^{\mu}\m K^{-1}\m K_{ijk}\left(d_{pqr}P_{L}\eta^{Aip}C^{jq}C^{kr}+d^{pqr}P_{R}?\eta^{Ai}{}_{p}?\bC jq\bC kr\right)
\end{align}
which is, as one would expect, the supercurrent. The supercurrent
is the part of the variation of the action due to the spacetime dependence
of the supersymmetry parameter in a locally supersymmetric theory.
By calculating directly the variation of the boundary Yang-Mills action
we compute the supercurrent as the $D_{\mu}\bar{s}_{A}$ part of the
variation and this indeed matches the current given above.

The current coupling to $\zeta^{A}$ is
\begin{align}
j^{A}=\; & -\frac{\sqrt{2}i}{4}V^{\frac{1}{2}}\gamma^{\mu\nu}F_{\mu\nu}^{I}\chi^{IA}\nonumber \\
 & +\frac{2\sqrt{2}i}{9}b_{i}b_{j}\gamma^{\mu}\left(P_{R}?\eta^{Aj}{}_{p}?\m D_{\mu}C^{ip}-P_{L}\eta^{Aip}\m D_{\mu}\bC jp\right)\nonumber \\
 & -\frac{\sqrt{2}i}{3}V^{-\frac{1}{2}}\left(G_{ij}+\frac{4}{3}b_{i}b_{j}\right)?\tau^{A}{}_{B}?\chi^{IB}\bar{C}^{j}\Lambda^{I}C^{i}\nonumber \\
 & -3\sqrt{5}iV^{-\frac{1}{2}}\m K^{-1}\m K_{ijk}\left(d_{pqr}C^{ip}C^{jq}P_{L}\eta^{Akr}+d^{pqr}\bC ip\bC jqP_{R}?\eta^{Ak}{}_{r}?\right)
\end{align}
and that coupling to $\lambda^{Ai}$ is
\begin{align}
j^{Ai}=\; & i\left(?\Gamma^{\perp i}{}_{jk}?-\frac{4}{3}{\delta^{\perp}}_{(j}^{i}b_{k)}^{\vphantom{i}}\right)\gamma^{\mu}\left(P_{R}?\eta^{Ak}{}_{p}?\m D_{\mu}C^{jp}-P_{L}\eta^{Ajp}\m D_{\mu}\bC kp\right)\nonumber \\
 & -2iV^{-\frac{1}{2}}\left(?\Gamma^{\perp i}{}_{jk}?-\frac{4}{3}{\delta^{\perp}}_{(j}^{i}b_{k)}^{\vphantom{i}}\right)?\tau^{A}{}_{B}?\chi^{IB}\bar{C}^{k}\Lambda^{I}C^{j}
\end{align}

\subsection{Boundary Conditions}

As in the eleven dimensional theory it is also necessary to specify
some boundary conditions in addition to the action to complete the
description of the theory, namely those arising from the eleven dimensional
boundary condition on the three-form $C$. The $C_{\alpha\beta\gamma}$
parts have already been built in to the action in the process of dualizing
$C_{\alpha\beta\gamma}$ to the scalar $\sigma$, the rest give the
boundary conditions on $\xi$, $\bar{\xi}$ and $\m A_{\mu}^{i}$,
which must be imposed separately. On $\partial\m M_{2}$:
\begin{align}
\xi & =\frac{\kappa_{5}^{2}}{2g^{2}}V^{\frac{1}{2}}\bar{\chi}_{2}^{I}\chi^{I1}\\
\bar{\xi} & =-\frac{\kappa_{5}^{2}}{2g^{2}}V^{\frac{1}{2}}\bar{\chi}_{1}^{I}\chi^{I2}\\
\m A_{\mu}^{i} & =-\frac{\sqrt{2}i\kappa_{5}^{2}}{4g^{2}}b^{i}?\tau^{A}{}_{B}?\bar{\chi}_{A}^{I}\gamma_{\mu}\chi^{IB}
\end{align}
and on $\partial\m M_{1}$:
\begin{align}
\xi & =\frac{\kappa_{5}^{2}}{g^{2}}\left(\sqrt{10}\m K^{-1}\m K_{ijk}d_{pqr}C^{ip}C^{jq}C^{kr}+\frac{1}{2}V^{\frac{1}{2}}\bar{\chi}_{1}^{I}\chi^{I2}\right)\label{eq:xi-bc}\\
\bar{\xi} & =\frac{\kappa_{5}^{2}}{g^{2}}\left(\sqrt{10}\m K^{-1}\m K_{ijk}d^{pqr}\bC ip\bC jq\bC kr-\frac{1}{2}V^{\frac{1}{2}}\bar{\chi}_{2}^{I}\chi^{I1}\right)\label{eq:barxi-bc}\\
\m A_{\mu}^{i} & =-\frac{\sqrt{2}\kappa_{5}^{2}}{g^{2}}\bigg(i?\Gamma^{i}{}_{jk}?\left(\bC kp\m D_{\mu}C^{jp}-C^{jp}\m D_{\mu}\bC kp\right)-\frac{\sqrt{2}i}{4}b^{i}?\tau^{A}{}_{B}?\bar{\chi}_{A}^{I}\gamma_{\mu}\chi^{IB}\nonumber \\
 & \qquad\quad-\frac{i}{4}\left(?\Gamma^{i}{}_{jk}?-b^{i}G_{jk}\right)\left(?{\t{\bar{\eta}}}_{L}^{jp}?\gamma_{\mu}?\eta_{R}^{k}{}_{p}?-?{\t{\bar{\eta}}}_{L}^{k}{}_{p}?\gamma_{\mu}?\eta_{L}^{jp}?\right)\bigg)
\end{align}

The full set of boundary conditions, including those derived from
the action, is given in the Appendix, in Section \ref{sec:Summary-Boundary-Conditions}.

\section{Gaugino Condensation\label{sec:Gaugino-Condensation}}

Gaugino condensation (the outcome of the gauge theory becoming strongly
coupled) provides a good example of the phenomenological impact of
the improved version of Heterotic M-Theory, since the improved theory
includes a contribution of the condensate to the boundary conditions
which was missed by the original version. Gaugino condensation in
the dimensional reduction of the improved Heterotic M-Theory and its
contribution to the vacuum energy were considered, before the full
reduction discussed here was completed, by Ahmed and Moss \cite{Ahmed:2008jz,Ahmed:2009ty}.
We will discuss some of their results here, the main aim being to
illustrate the effect of the improved theory.

If a gaugino condensate forms its contribution is
\begin{align}
\bar{\chi}\Gamma_{abc}\chi & =\Lambda\Omega_{abc}\label{eq:condensate}
\end{align}
or in terms of the four dimensional fields
\begin{align}
\bar{\chi}_{2}\chi^{1} & =\Lambda
\end{align}

This contributes to the boundary conditions \eqref{eq:bulkC_bc} or
equivalently \eqref{eq:xi-bc} and \eqref{eq:barxi-bc} giving a background
flux of $\xi$ and $\bar{\xi}$ in terms of the fields of the reduced
theory described above. For the background described in \cite{Ahmed:2008jz},
this gives a contribution to the superpotential of the four dimensional
theory of $W_{g}\sim\Lambda_{1}+\Lambda_{2}$, with $\Lambda_{1}$
and $\Lambda_{2}$ being the condensates on $\partial\m M_{1}$ and
$\partial\m M_{2}$ respectively.

If we continue the calculation of the boundary condition on the gravitino
which gives \eqref{eq:Psi_bc_2} to include all the fermion terms
then we have
\begin{align}
\Gamma^{AB}\left(P_{-}+\frac{\kappa_{11}^{2}}{\lambda^{2}}\Gamma P_{+}\right)\Psi_{A} & =\mp\frac{\kappa_{11}^{2}}{4\lambda^{2}}\Gamma^{BC}\Gamma^{A}\tr\left(F_{BC}^{*}\chi\right)\label{eq:Psi-bc-with-twist-term}
\end{align}
where $F_{AB}^{*}=\frac{1}{2}\left(F_{AB}+\hat{F}_{AB}\right)$ with
$\hat{F}_{AB}$ the supercovariant version of $F_{AB}$. The operator
$\Gamma$ is given by
\begin{align}
\Gamma & =\frac{1}{96}\tr\left(\bar{\chi}\Gamma_{ABC}\chi\right)\Gamma^{ABC}\label{eq:twist}
\end{align}
and represents a twist in the chirality condition which $\Psi_{A}$
must obey on the boundaries. This contribution from $\Gamma$ was
missed by the original formulation of Heterotic M-Theory, so its effects
are a novelty of the improved version. Clearly a non-zero gaugino
condensate \eqref{eq:condensate} gives us a source of such a twist.

Suppose we have background solutions to the five dimensional theory,
such as those described in \cite{Ahmed:2008jz}, which give us a four
dimensional model with a potential which stabilizes the moduli fields
in some minimum. The four dimensional moduli in that case were the
values of the Calabi-Yau volume modulus on either boundary, $V_{1}$
and $V_{2}$. The potential which fixes the moduli is negative at
its minima, i.e.\ its vacua are anti-de Sitter. Therefore some contribution
is needed to give the small positive vacuum energy that we observe
in the real universe.

The twist in the chirality of the gravitino breaks the supersymmetry
of the background. It also changes the quantum vacuum energy of the
gravitino. Without the condensate the vacuum energy contributions
of the bosonic and fermionic fields cancel each other exactly due
to the supersymmetry. The condensate does not affect the vacuum energy
of the bosons, but changes the vacuum energy of the gravitino, thus
giving a change in the overall vacuum energy. The vacuum energy due
to the twist given in \eqref{eq:Psi-bc-with-twist-term} by the gaugino
condensate \eqref{eq:condensate} was calculated in \cite{Ahmed:2009ty}
using the results of \cite{Flachi:2001ke,Flachi:2001bj}. It was found
to be always positive and could, for some choices of Calabi-Yau in
the reduction, be large enough to cancel the negative vacuum energy
resulting from the moduli stabilization.

The conclusion of \cite{Ahmed:2009ty} is then that the improved version
of Heterotic M-Theory unveils a new mechanism for the uplift of the
vacuum from anti-de Sitter to de Sitter.

The fuller reduction described in this chapter, as compared to that
used in \cite{Ahmed:2008jz,Ahmed:2009ty}, sheds a little additional
light on this mechanism, as was discussed in  \cite{Moss:2011pi}.
The only fermion considered in \cite{Ahmed:2008jz,Ahmed:2009ty} was
the five dimensional gravitino $\psi_{\alpha}^{A}$. Now that we have
the ansatz for all the fermion fields coming from the eleven dimensional
gravitino $\Psi_{I}$ we can see that $\psi_{\alpha}$ is the only
one which receives a contribution of the gaugino condensate to its
boundary condition from \eqref{eq:twist}. Recall from \eqref{eq:Psi_a_ansatz}
that the fermions $\zeta^{A}$ and $\lambda^{\perp Ai}$, those which
were omitted before, are given by the parts of $\Psi_{I}$ whose Calabi-Yau
spinor components are $\gamma^{a}u_{A}$ or $\gamma^{\bar{a}}u_{A}$.
Now for the Calabi-Yau gamma matrices $\gamma^{1}\gamma^{1}=0$, etc.
and there are only three values $a,b,c,d$ can take so $\gamma^{abc}\gamma^{d}u_{A}=0$
while we also have that $\gamma^{abc}\gamma^{\bar{d}}u_{A}=\gamma^{abc}\gamma^{\bar{d}}u_{1}=2g^{c\bar{d}}\gamma^{ab}u_{1}=0$
as $\gamma^{a}u_{1}=0$ and $\gamma^{\bar{a}}u_{2}=0$. Thus the contribution
of the condensate \eqref{eq:condensate} to the twist \eqref{eq:twist}
annihilates the parts of $\Psi_{I}$ which give $\zeta^{A}$ and $\lambda^{\perp Ai}$
and thus the condensate cannot affect their boundary conditions directly.
We see then that the analysis of the Casimir energy of \cite{Ahmed:2009ty}
should carry over to our more complete reduction with its conclusions
intact.

\section{Conclusions\label{sec:Reduction_Conclusions}}

We have described here the reduction to five dimensions of an improved
version of Heterotic M-Theory over a Calabi-Yau manifold with (1,1)
moduli only. Comparing to the reduction of Ho\v{r}ava and Witten's
original formulation of Heterotic M-Theory, the bosonic sector is
largely unaffected, and along with it most of the existing work on
the phenomenology of Heterotic M-Theory.

Although phenomenology based upon the bosonic sector should not be
much affected by the improved version of Heterotic M-Theory, that
previous phenomenological work is now on a firmer footing; we can
now see that it can be built upon a theory which is consistent to
all orders in the gravitational coupling and the fact that the analogous
four dimensional expansion parameter $\epsilon=\kappa^{2/3}\rho V^{-2/3}$
seems to be of order one, when chosen to give reasonable values for
Newton's constant and the GUT scale \cite{Banks:1996ss}, is no longer
a problem. On the other hand, work that has been done using explicitly
terms higher order in the gravitational coupling, such as on the back
reaction of anti-branes \cite{Gray:2007zza} and gaugino condensation
\cite{Gray:2007qy} probably needs to be reconsidered in the light
of the improved theory. 

The inclusion of five-branes, and the further inclusion of anti-five-branes,
has not yet been considered in the improved theory. Though such objects
are vital to Heterotic M-Theory phenomenology, it is not possible
with our current understanding of them to incorporate them in a fully
consistent way which includes the backreaction of their matter and
curvature, in the manner in which we have treated the boundary branes
here. In the absence of such understanding, one might adopt a hybrid
approach with boundary conditions at either end of the bulk, as we
describe here, and junction conditions across the five-branes, as
have been used in the Ho\v{r}ava-Witten theory.

Gaugino condensation has been considered already in the improved theory
in the context of a much simpler reduction \cite{Ahmed:2008jz,Ahmed:2009ty}.
The condensate appears in the boundary condition for the scalars $\xi$
and $\bar{\xi}$ and so acts as a source of flux. It also appears
in the three-fermion terms in the fermion boundary conditions, its
effect being to introduce a `twist' in the chirality of the fermions
which contributes to the vacuum energy. By including just those higher
fermion terms necessary for gaugino condensation into the full reduction
given here, one can see that the gravitino contribution considered
in \cite{Ahmed:2009ty} is the dominant one \cite{Moss:2011pi}. Therefore
the calculation there of the vacuum energy and the conclusion that
it is a candidate to uplift a negative cosmological constant, as left
by moduli stabilization, to a positive one, in accordance with observation,
still hold here.

A full treatment of moduli stabilization in the improved theory would
require the inclusion of the (2,1) moduli for the Calabi-Yau space.
There are no obvious obstacles to this other than complexity so we
would expect it to be feasible if there is sufficient interest to
warrant it.

Leaving aside the M-Theory context, the topic of supergravities with
boundary matter is worthy of study of itself, particularly for its
connection to braneworld models. Examples are hard to come by in dimensions
greater than three, where there is an off-shell formulation to assist.
Thus in providing a five dimensional example with four dimensional
boundaries the theory described here may be of some interest.

\chapter{Big Crunch from Colliding Branes\label{cha:Colliding-Branes}}

Interest in braneworld scenarios began with attempts to confine matter
to field theory domain walls \cite{Akama:1982jy,Rubakov:1983bb}.
Suppose we have a scalar field, $\phi$, coupled to a fermion, $\psi$,
by a Yukawa coupling $\sim\phi\bar{\psi}\psi$ and we have a domain
wall solution with $\phi=0$ on some surface and $\phi=m$ far from
the surface. Then the fermion is massless on the wall but massive
away from it, so that it is energetically favourable for the fermion
to be close to the wall. Indeed one may show \cite{Jackiw:1975fn}
that in such a case there is a fermion zero mode which is both normalizable
and confined to the wall.

Such scenarios attracted further interest more recently once it was
realized that gravity could also be confined to the brane, as was
shown by the construction of Randall and Sundrum \cite{Randall:1999vf}.
They considered a brane separating two anti-de Sitter spacetimes so
that the metric is warped in the $z$ direction, perpendicular to
the brane, as $ds^{2}=e^{-\kappa\left|z\right|}\eta_{\mu\nu}dx^{\mu}dx^{\nu}+dz^{2}$.
The warping suppresses the Kaluza-Klein modes of the gravitational
field and means that the zero mode, i.e.\ the four dimensional graviton,
dominates the gravitational interactions. Although the $z$ direction
is infinite in extent it actually has a finite volume so that the
zero mode is normalizable, and in fact gives general relativity on
the brane \cite{Garriga:1999yh}. The possibility of confining gravity
opened up the possibility of large extra dimensions and hence the
possibility of building cosmological models using the dynamics of
branes moving in them \cite{Langlois:2002bb}.

In a braneworld cosmological model it is natural to ask whether the
big bang could have been caused by a collision of branes \cite{Khoury:2001wf,Bucher:2001it,Gen:2001bx,Langlois:2001uq}
and in this context the scalar field domain wall returns as a convenient
model of a braneworld whose detailed behaviour can be studied using
numerical simulations. Particle production in a five dimensional model
was studied in \cite{Takamizu:2004rq} and, with the inclusion of
fermions on the branes, in \cite{Gibbons:2006ge,Saffin:2007ja,Saffin:2007qa,Takamizu:2004rq}
but in both cases without gravitational backreaction. When the backreaction
was included in the scalar field theory model it was found \cite{Takamizu:2006gm,Takamizu:2007ks}
that, as one might have expected, a singularity forms if the collision
is energetic enough. The domain walls are assumed to be spatially
homogeneous and isotropic, which reduces the 4+1 dimensional theory
to a 1+1 dimensional problem, which is much more amenable to numerical
solution. Given this symmetry the singularity formed was naturally
assumed \cite{Takamizu:2007ks} to be the AdS black brane.

However, the asymptotic spacetimes of the domain wall solutions and
the AdS black brane are not compatible, as we will show in Section
\ref{sec:Asymptotic-Structure}. This prompted us to further investigate
the structure of the singularity, as reported in \cite{Omotani:2011un}.
The results of our simulations support the conclusion that the singularity
formed cannot be a black brane; they suggest instead a big crunch
singularity which ends the spacetime.

\section{Static Kinks}

The model we use is a single real scalar canonically coupled to gravity
in five dimensions
\begin{align}
\m L & =\frac{m_{p}^{3}}{2}R-\frac{1}{2}\partial_{a}\phi\partial^{a}\phi-\m V(\phi)
\end{align}
where $m_{p}$ is the Planck mass.

A domain wall is a topological defect: we have one vacuum solution
at $r\rightarrow-\infty$ and a different one at $r\rightarrow+\infty$.
The transition region between the two vacua has a high cost in terms
of the potential energy of the fields which are forced to leave the
potential minima which they occupy in the vacuum. This cost is minimized
by making the region small, so we have a localized structure: a domain
wall. The domain wall is stable because of its topological character:
neither vacuum could ever fill the whole spacetime on its own and
so the transition between the two, the wall, must persist. In order
to have a domain wall like this we must choose an appropriate potential
which has two distinct minima, which give us the two vacua (with a
potential barrier between them in field space) which we need to support
a domain wall.

We examine domain walls which are spatially flat so for a static domain
wall the metric and the scalar field only depend on the co-ordinate
normal to the brane, $r$. We can thus describe the metric with an
ansatz
\begin{align}
ds^{2} & =e^{2U(r)}\eta_{\mu\nu}dx^{\mu}dx^{\nu}+dr^{2}
\end{align}
while the scalar field is a function of $r$ only
\begin{align}
\phi & =\phi(r)
\end{align}

With this ansatz the non-zero components of the Ricci tensor are
\begin{align}
R_{tt} & =e^{2U}\left(\partial_{r}^{2}U+4\left(\partial_{r}U\right)^{2}\right)\\
R_{xx} & =-e^{2U}\left(\partial_{r}^{2}U+4\left(\partial_{r}U\right)^{2}\right)\\
R_{rr} & =-4\partial_{r}^{2}U-4\left(\partial_{r}U\right)^{2}
\end{align}
and the Ricci scalar is
\begin{align}
R & =-8\partial_{r}^{2}U-20\left(\partial_{r}U\right)^{2}
\end{align}
and so the equations of motion, $R_{ab}=\kappa^{2}\left(\partial_{a}\phi\partial_{b}\phi+\frac{2}{3}g_{ab}\m V\right)$
and $\nabla_{a}\nabla^{a}\phi-\frac{\partial V}{\partial\phi}=0$,
are
\begin{align}
\left(\partial_{r}U\right)^{2} & =\frac{1}{6m_{p}^{3}}\left(\frac{1}{2}\left(\partial_{r}\phi\right)^{2}-\m V\right)\\
\partial_{r}^{2}U & =-\frac{1}{3m_{p}^{3}}\left(\partial_{r}\phi\right)^{2}\\
\partial_{r}^{2}\phi+4\partial_{r}U\partial_{r}\phi & =\frac{\partial V}{\partial\phi}
\end{align}

In order to be able to find an exact solution, it is convenient to
take a supergravity-inspired form for the potential \cite{Chamblin:1999cj}
\begin{align}
\m V & =\frac{1}{2}\left(\left(\frac{\partial W}{\partial\phi}\right)^{2}-\frac{4}{3m_{p}^{3}}W^{2}\right)
\end{align}
with a superpotential $W(\phi)$ because if this is the case, then
the equations of motion are solved if the BPS equations,
\begin{align}
\frac{\partial\phi}{\partial r} & =\mp\frac{\partial W}{\partial\phi}\nonumber \\
\frac{\partial U}{\partial r} & =\pm\frac{1}{3m_{p}^{3}}W\label{eq:BPS_equations}
\end{align}
are solved, where the upper and lower signs will give kinks and anti-kinks
respectively. This is a great simplification as we need only solve
the first order BPS equations rather than solving the second order
equations of motion directly. Moreover, we do not need to know $U$
to solve the BPS equation for $\phi$ and so we can solve for $\phi$
and $U$ consecutively, rather than simultaneously as we would have
to for the equations of motion.

Of course we must also pick an appropriate superpotential: we choose
the sine-Gordon model
\begin{align}
W & =\mu^{4}-\frac{4m}{\beta^{2}}\cos\left(\frac{\beta\phi}{2}\right)
\end{align}
which gives a potential
\begin{align}
\m V & =\left(\frac{2m^{2}}{\beta^{2}}-\frac{2\mu^{8}}{3m_{p}^{3}}\right)+\frac{16m\mu^{4}}{3m_{p}^{3}\beta^{2}}\cos\left(\frac{\beta\phi}{2}\right)-\left(\frac{2m^{2}}{\beta^{2}}+\frac{32m^{2}}{3m_{p}^{3}\beta^{4}}\right)\cos^{2}\left(\frac{\beta\phi}{2}\right)
\end{align}
whose parameters are: $m$ which is the mass of the scalar field in
the non-gravitating limit $m_{p}\rightarrow\infty$, and controls
the curvature of the potential at the minima; $\beta$ which sets
the separation of the vacua in field space; and $\mu$ which controls
the constant part of the potential and will allow us to make some
of the minima be Minkowski vacua.

This potential has two sets of minima: one set at $\frac{\beta\phi}{2}=2n\pi$
and another at $\frac{\beta\phi}{2}=\left(2n+1\right)\pi$. At the
former the value of the potential is
\begin{align}
\m V\left(\frac{\beta\phi}{2}=2n\pi\right) & =-\frac{32}{3m_{p}^{3}\beta^{4}}\left(1-\frac{\mu^{4}\beta^{2}}{4m}\right)^{2}\label{eq:V-minimum}
\end{align}
so we can make these Minkowski vacua by setting $\mu^{4}=\frac{4m}{\beta^{2}}$.
With this choice its value at the latter is
\begin{align}
\m V\left(\frac{\beta\phi}{2}=\left(2n+1\right)\pi\right) & =-6m_{p}^{3}\left(\frac{8m}{3\beta^{2}m_{p}^{3}}\right)^{2}=m_{p}^{3}\Lambda
\end{align}
giving us the value of the (negative) cosmological constant $\Lambda$
in the AdS vacua.

\begin{figure}
\begin{centering}
\includegraphics[height=9cm]{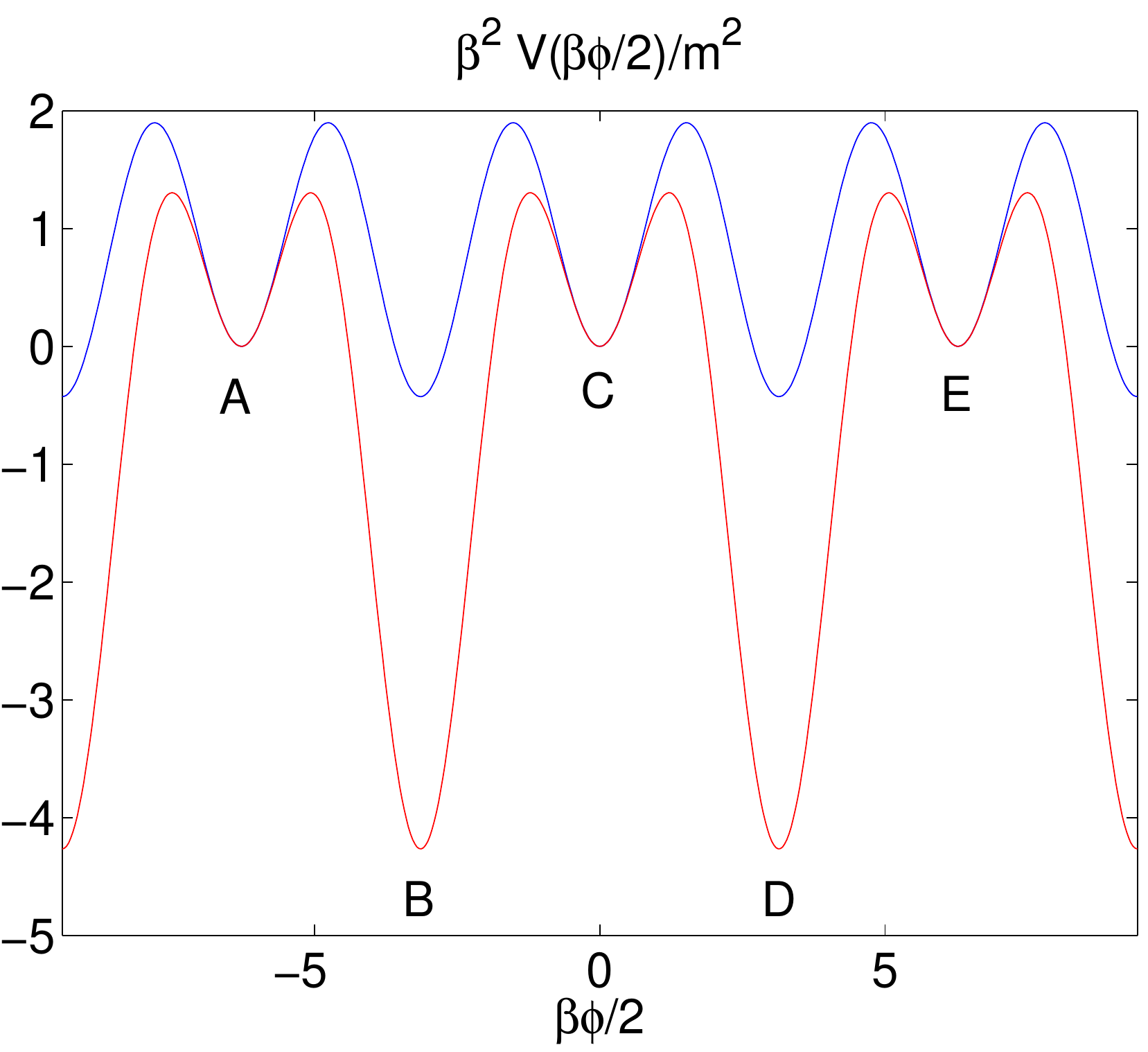}
\par\end{centering}

\caption{The potential with $\beta^{2}m_{p}^{3}=10$ (red, lower curve) and
$\beta^{2}m_{p}^{3}=100$ (blue, upper curve).\label{fig:Potential}}
\end{figure}

The potential is plotted in Figure \ref{fig:Potential} where the
Minkowski vacua are labelled A, C and E and the AdS vacua are B and
D. Our simulations are performed with $\beta^{2}m_{p}^{3}=100$, which
is shown as the blue, upper curve.

The reason for the choice of this particular form of the sine-Gordon
potential is this: we choose to have AdS vacua outside the domain
walls so that gravity is confined, à la Randall-Sundrum, to the region
around and between the domain walls. A priori we could choose the
intermediate vacuum to be either AdS or Minkowski, noting from \eqref{eq:V-minimum}
that $\m V\leq0$ at a minimum so we cannot have dS vacua. However,
in order to be able to construct a two-wall solution by superposition
of the solutions for single walls we are forced to take the intermediate
vacuum to be Minkowski. To superpose the solutions we must match their
asymptotic spaces: the $r\rightarrow+\infty$ limit of the wall at
smaller $r$ and the $r\rightarrow-\infty$ limit of the wall at larger
$r$. Since the solution corresponding to one of the walls enters
the intermediate vacuum as $r$ increases while that corresponding
to the other leaves it, one wall must be a kink while the other must
be an anti-kink. Since they share the intermediate vacuum, the value
of $\phi$ there is the same, and constant asymptotically, and similarly
for the superpotential $W(\phi)$. However, this being the case we
can see from the BPS equations, \eqref{eq:BPS_equations}, that if
the intermediate vacuum is AdS, with $W<0$, then for a kink $U$
is monotonically decreasing with $r$ while for the anti-kink $U$
is monotonically increasing (or vice versa if $W>0$). Thus there
is no way that the solutions for $U(r)$ for the kink and anti-kink
can match asymptotically in the intermediate region unless $U$ is
constant: we must have a Minkowski vacuum (with $W=0$) in the intermediate
region in order to superpose analytic single wall solutions to find
a two wall solution.

Our initial set-up then will be a spacetime with two parallel domain
walls which is asymptotically AdS$^{5}$ and approaches Minkowski
space between the domain walls. The scalar field evolves through space,
in the direction normal to the domain walls, from $-\frac{4\pi}{\beta}$
(as $r\rightarrow-\infty$) through 0 (between the domain walls) to
$\frac{4\pi}{\beta}$ (as $r\rightarrow+\infty$), that is from B
through C to D in Figure \ref{fig:Potential}. Since we have a Minkowski
intermediate vacuum we can, as we have just seen, construct a two-wall
solution by superposition of single-wall solutions. We will construct
our initial conditions by adding the analytic solutions for a kink
(with $\phi$ going from B to C) and an anti-kink (with $\phi$ going
from C to D).

A BC kink is a solution to the BPS equations \eqref{eq:BPS_equations}
with the upper sign, which is given by
\begin{align}
\frac{\beta\phi}{2} & =2\tan^{-1}\Bigg(\tanh\left(\frac{m(r-r_{0})}{2}\right)\Bigg)-\frac{\pi}{2}\nonumber \\
U & =-\frac{4}{3\beta^{2}m_{p}^{3}}\Bigg(\ln\Big(\cosh\big(m\left(r-r_{0}\right)\big)\Big)-\frac{\beta^{2}\mu^{4}}{4}\left(r-r_{0}\right)\Bigg)\label{eq:static_kink}
\end{align}
and a CD anti-kink is a solution to the equations with the lower sign,
given by
\begin{align}
\frac{\beta\phi}{2} & =2\tan^{-1}\Bigg(\tanh\left(\frac{m(r-r_{0})}{2}\right)\Bigg)+\frac{\pi}{2}\nonumber \\
U & =-\frac{4}{3\beta^{2}m_{p}^{3}}\Bigg(\ln\Big(\cosh\big(m\left(r-r_{0}\right)\big)\Big)+\frac{\beta^{2}\mu^{4}}{4}\left(r-r_{0}\right)\Bigg)\label{eq:static_anti-kink}
\end{align}

To construct the initial conditions which will give us colliding domain
walls we will need to boost these static solutions, but first a comment
on the asymptotic structure of systems such as these.

\section{Asymptotic Structure\label{sec:Asymptotic-Structure}}

In the limit $r\rightarrow+\infty$ we are in the AdS$^{5}$ region
of the anti-kink solution so the metric is
\begin{align}
\lim_{r\rightarrow+\infty}ds^{2} & =e^{-2\alpha r}\eta_{\mu\nu}dx^{\mu}dx^{\nu}+dr^{2}
\end{align}
where $\alpha=\frac{8m}{3\beta^{2}m_{p}^{3}}$ and making the co-ordinate
transformation defined by $e^{\alpha r}=\alpha Z$ we see that the
metric of the asymptotic region is
\begin{align}
\lim_{Z\rightarrow+\infty}ds^{2} & =\frac{1}{\alpha^{2}Z^{2}}\left(\eta_{\mu\nu}dx^{\mu}dx^{\nu}+dZ^{2}\right)\label{eq:anti-kink_asymptotic_metric}
\end{align}
This is a portion of AdS$^{5}$ that, in particular, does not include
the AdS$^{5}$ boundary, which is at $Z=0$.

On the other hand, the metric of the AdS$^{5}$ black brane is
\begin{align}
ds_{bb}^{2} & =-f(R)dT^{2}+\frac{1}{f(R)}dR^{2}+R^{2}\delta_{ij}dx^{i}dx^{j}
\end{align}
with
\begin{align}
f(R) & =-\frac{M}{R^{2}}-\frac{\Lambda R^{2}}{6}
\end{align}
and there is a version of Birkhoff's theorem \cite{Charmousis:2002rc,Zegers:2005vx}
which assures us that this is the only black brane solution which
shares the symmetry of our domain walls, i.e.\ plane symmetry orthogonal
to the $r$ direction. For comparison with the domain wall solution
\eqref{eq:anti-kink_asymptotic_metric} let us make the co-ordinate
change
\begin{align}
 & t=\sqrt{-\frac{\Lambda}{6}}T,\quad R=\frac{1}{\sqrt{-\frac{\Lambda}{6}}Z}
\end{align}
Then, remembering that the black brane is at $R=0$ so the asymptotic
region, far from the brane, is $Z\rightarrow0$, the asymptotic metric
of the black brane is
\begin{align}
\lim_{Z\rightarrow0}ds_{bb}^{2} & =-\frac{6}{\Lambda Z^{2}}\left(\eta_{\mu\nu}dx^{\mu}dx^{\nu}+dZ^{2}\right)
\end{align}
which is the same form as the asymptotic metric of the domain wall,
but in a different region of AdS$^{5}$: The asymptotic region of
the domain wall is far from the AdS$^{5}$ boundary but the asymptotic
region of the black brane is the region near the boundary. Thus it
would be distinctly odd if a black brane were formed by the collision
of domain walls.

\section{Dynamical Solutions\label{sec:Dynamical-Solutions}}

If we define dimensionless variables by
\begin{align}
 & x=\frac{\tilde{x}}{m},\quad\phi=\frac{\tilde{\phi}}{\beta}
\end{align}
then we are left with just a single parameter, $\beta^{2}m_{p}^{3}$.
Henceforth we will use the dimensionless parameters, but drop the
tildes. In other words we will choose to measure distances and the
Planck mass in units of $m$, and $\phi$ in units of $\beta$.

We obtain solutions representing moving kinks and anti-kinks just
by boosting the static ones given in \eqref{eq:static_kink} and \eqref{eq:static_anti-kink}.
To do so it is convenient to change to co-ordinates which make manifest
the $SO(1,1)$ symmetry between the temporal direction and the direction
perpendicular to the domain walls, so that the metric has the form
\begin{align}
ds^{2} & =e^{2A(t,z)}\left(-dt^{2}+dz^{2}\right)+e^{2B(t,z)}\delta_{ij}dx^{i}dx^{j}\label{eq:Dynamical_metric}
\end{align}
This is accomplished for a static kink or anti-kink by 
\begin{equation}
e^{U}dz=dr\label{eq:z-r}
\end{equation}
 since in the static case $A(t,z)=B(t,z)=U\big(r(z)\big)$. In the
new frame the equations of motion are
\begin{align}
\partial_{\tilde{\mu}}\partial^{\tilde{\mu}}\phi+3\partial_{\tilde{\mu}}B\partial^{\tilde{\mu}}\phi & =e^{2A}\frac{\partial\m V}{\partial\phi}\label{eq:phi-eom-tz}\\
\partial_{\tilde{\mu}}\partial^{\tilde{\mu}}A-3\partial_{\tilde{\mu}}B\partial^{\tilde{\mu}}B & =\frac{1}{m_{p}^{3}}\left(-\frac{1}{2}\partial_{\tilde{\mu}}\phi\partial^{\tilde{\mu}}\phi+\frac{1}{3}e^{2A}\m V\right)\label{eq:A-eom-tz}\\
\partial_{\tilde{\mu}}\partial^{\tilde{\mu}}B+3\partial_{\tilde{\mu}}B\partial^{\tilde{\mu}}B & =-\frac{2}{3m_{p}^{3}}e^{2A}\label{eq:B-eom-tz}
\end{align}
where $\tilde{\mu},\tilde{\nu},\ldots=\left\{ t,z\right\} $ and the
constraint equations are
\begin{align}
 & \partial_{\tilde{\mu}}\partial_{\tilde{\nu}}B+\partial_{\tilde{\mu}}B\partial_{\tilde{\nu}}B+\eta_{\tilde{\mu}\tilde{\nu}}\partial_{\tilde{\rho}}B\partial^{\tilde{\rho}}B-\partial_{\tilde{\mu}}A\partial_{\tilde{\nu}}B-\partial_{\tilde{\mu}}B\partial_{\tilde{\nu}}A+\eta_{\tilde{\mu}\tilde{\nu}}\partial_{\tilde{\rho}}A\partial^{\tilde{\rho}}B\nonumber \\
 & =-\frac{1}{3m_{p}^{3}}\left(\partial_{\tilde{\mu}}\phi\partial_{\tilde{\nu}}\phi-\frac{1}{2}\eta_{\tilde{\mu}\tilde{\nu}}\partial_{\tilde{\rho}}\phi\partial^{\tilde{\rho}}\phi+\eta_{\tilde{\mu}\tilde{\nu}}e^{2A}\m V\right)\label{eq:constraint-tz}
\end{align}
Clearly $A$, $B$ and $\phi$ are scalars under $SO(1,1)$ boosts
and so it is trivial to transform between frames: if $\m O^{\prime}$
is moving at velocity $v$ in the $z$ direction with respect to $\m O$,
then
\begin{align}
t^{\prime} & =\gamma\left(t-vz\right)\\
z^{\prime} & =\gamma\left(z-vt\right)
\end{align}
with the standard $\gamma$-factor $\gamma=\frac{1}{\sqrt{1-v^{2}}}$
and for a generic scalar $\Psi$
\begin{align}
\Psi^{\prime}(t^{\prime},z^{\prime}) & =\Psi(t,z)
\end{align}
These co-ordinates also have the advantage of making it easy to picture
the causal structure, since null rays are just $45^{\circ}$ lines
in the $t-z$ plane.

\begin{wrapfigure}{O}{0.3\columnwidth}%
\noindent \begin{centering}
\includegraphics[bb=205bp 720bp 295bp 775bp,clip]{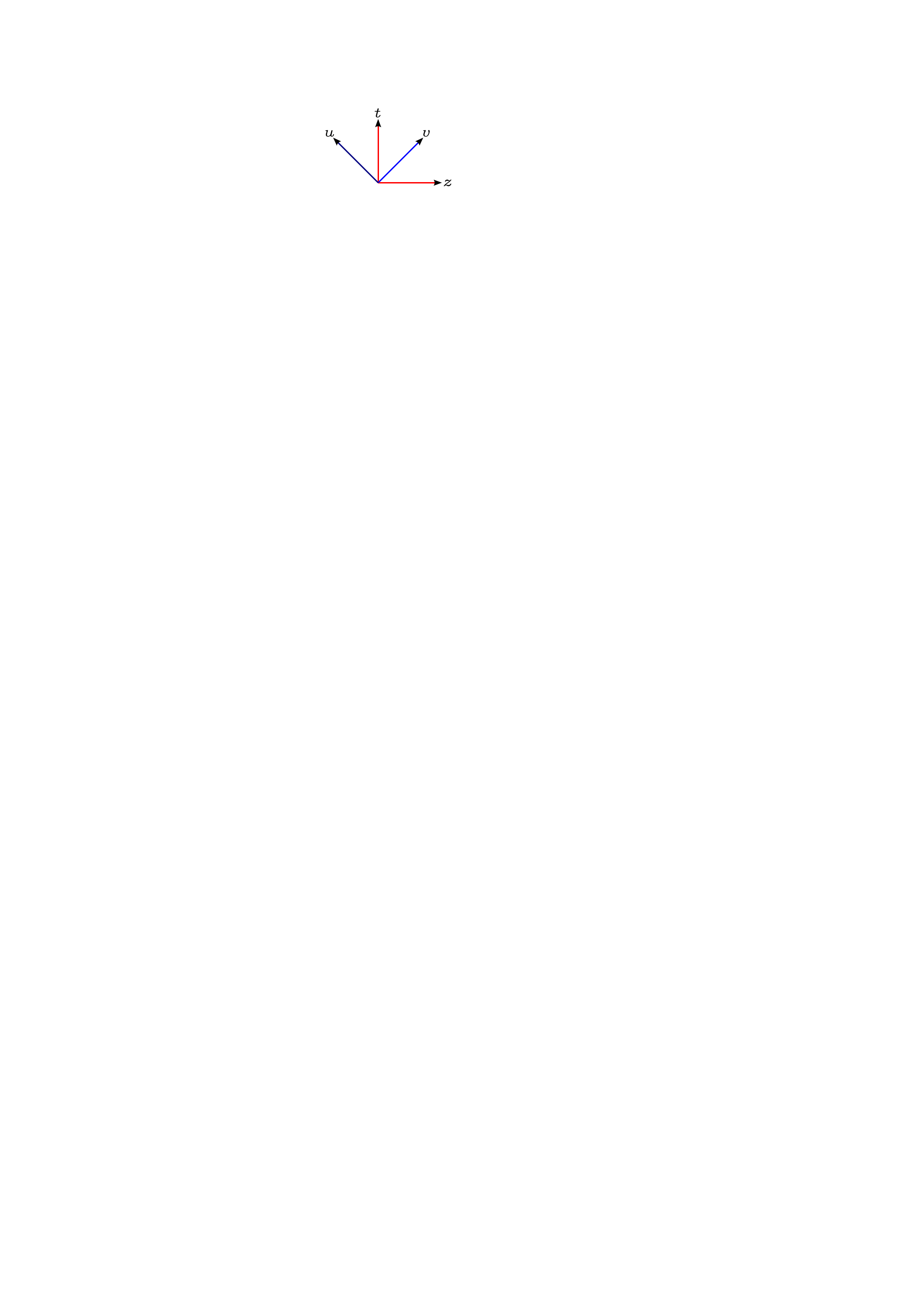}
\par\end{centering}

\caption{Double-null co-ordinates}

\end{wrapfigure}%
In fact it is more convenient numerically to use double-null co-ordinates
\cite{Takamizu:2007ks,Burko:1997tb}
\begin{align}
u & =\frac{1}{\sqrt{2}}\left(t-z\right),\quad v=\frac{1}{\sqrt{2}}\left(t+z\right)
\end{align}
because we want to investigate the behaviour when a singularity forms;
such co-ordinates allow us to cut off the simulation when the singularity
is reached, but continue the simulation in the regions outside of
causal contact with the singularity. Thus we can try to investigate
the asymptotic properties of the spacetime numerically despite the
presence of the singularity. In double-null co-ordinates the equations
of motion (\ref{eq:phi-eom-tz}-\ref{eq:B-eom-tz}) are
\begin{align}
\partial_{u}\partial_{v}\phi+\frac{3}{2}\left(\partial_{u}B\partial_{v}\phi+\partial_{v}B\partial_{u}\phi\right) & =-\frac{1}{2}e^{2A}\frac{\partial\m V}{\partial\phi}\label{eq:phi-eom-uv}\\
\partial_{u}\partial_{v}A-3\partial_{u}B\partial_{v}B & =-\frac{1}{2m_{p}^{3}}\left(\partial_{u}\phi\partial_{v}\phi+\frac{1}{3}e^{2A}\m V\right)\label{eq:A-eom-uv}\\
\partial_{u}\partial_{v}B+3\partial_{u}B\partial_{v}B & =\frac{1}{3m_{p}^{3}}e^{2A}\label{eq:B-eom-uv}
\end{align}
and the constraints \eqref{eq:constraint-tz} are
\begin{align}
\partial_{u}\partial_{u}B+\partial_{u}B\partial_{u}B-\partial_{u}A\partial_{u}B-\partial_{u}B\partial_{u}A & =-\frac{1}{3m_{p}^{3}}\partial_{u}\phi\partial_{u}\phi\\
\partial_{v}\partial_{v}B+\partial_{v}B\partial_{v}B-\partial_{v}A\partial_{v}B-\partial_{v}B\partial_{v}A & =-\frac{1}{3m_{p}^{3}}\partial_{v}\phi\partial_{v}\phi\\
\partial_{u}\partial_{v}B-\partial_{u}B\partial_{v}B-2\left(\partial_{u}A\partial_{v}B+\partial_{u}B\partial_{v}A\right) & =\frac{1}{3m_{p}^{3}}e^{2A}\m V
\end{align}

It is straightforward to solve the relation \eqref{eq:z-r} between
$r$ and $z$ numerically. This allows us to translate the analytic
solutions \eqref{eq:static_kink} and \eqref{eq:static_anti-kink}
into the static kink and anti-kink solutions in the new frame. We
then construct the initial conditions for colliding domain walls by
adding boosted kink and anti-kink solutions equidistant from $z=0$
and with equal and opposite velocities to allow us to exploit the
reflection symmetry of the problem in the centre of mass frame.

\section{Simulation Methods}

Our first set of simulations were performed in $t-z$ co-ordinates
using a fourth order Runge-Kutta method. These showed us the kink
and anti-kink approaching and bouncing and the subsequent formation
of a singularity at the origin. However, this set up has some limitations
which prompted the later move to $u-v$ co-ordinates.

As the grid on which we run the simulation is finite in extent in
the $z$ direction, we are forced to apply some boundary condition
at the edge. Since this boundary condition is not a physical one,
it introduces errors into the numerical solution. In order to prevent
these errors from affecting the region of physical interest, the grid
has to be much larger than that region so that they do not have time
during the length of the simulation to propagate to there from the
boundary. This is not ideal from the point of view of computational
efficiency, but is not a fundamental problem.

\begin{wrapfigure}{O}{0.4\columnwidth}%
\noindent \begin{centering}
\includegraphics[bb=190bp 660bp 350bp 725bp,clip]{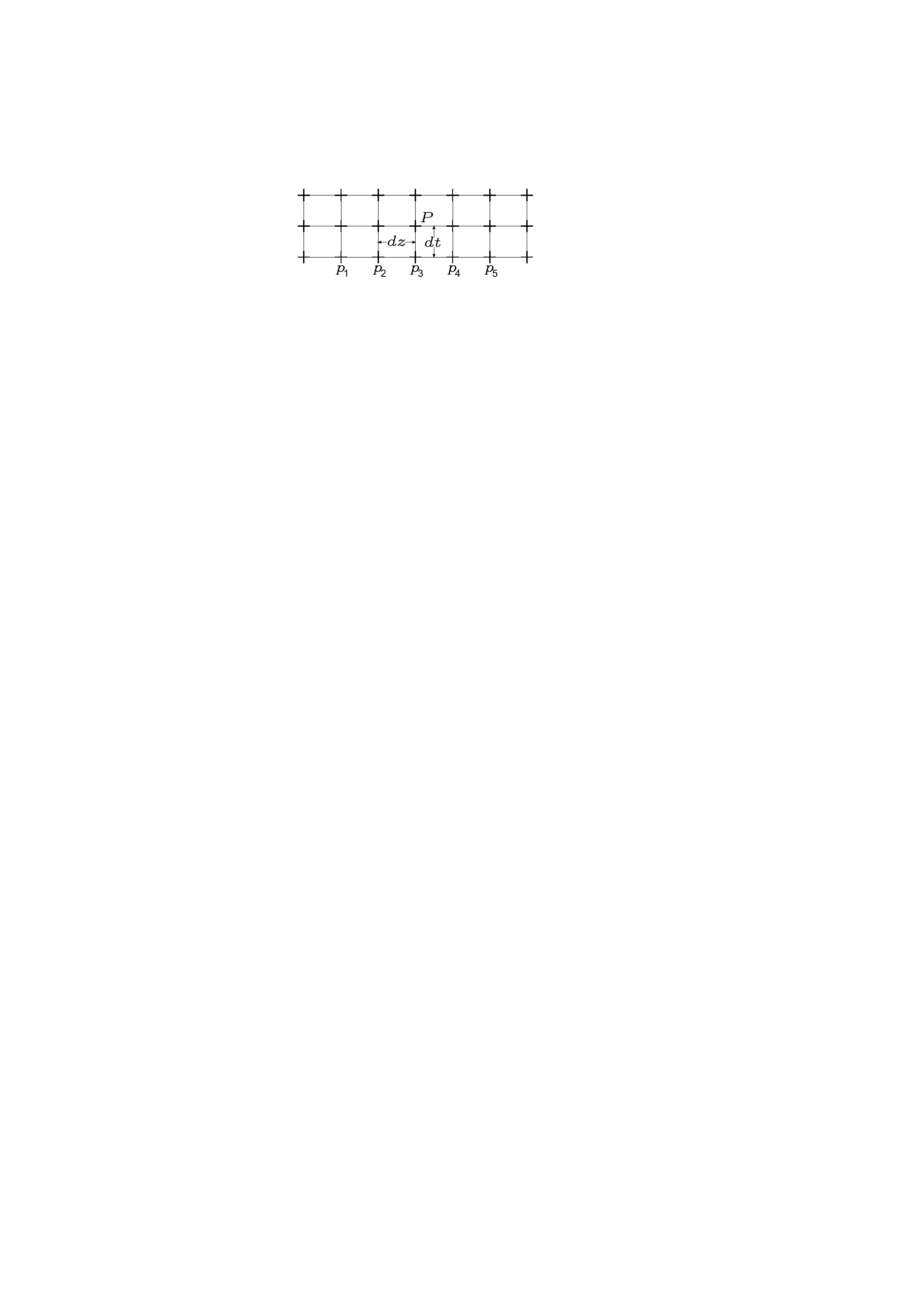}
\par\end{centering}

\caption{$t-z$ grid\label{fig:tz-grid}}

\end{wrapfigure}%
The more important issue is that since the method is fourth order,
in order to calculate the fields at $P$ (see Figure \ref{fig:tz-grid})
we use the values at the points $p_{1}$ to $p_{4}$ which are up
to two spatial steps away from $P$. This means that errors propagate
through the simulation rather quickly. This is merely inconvenient
for the errors introduced at the boundary but it is a fundamental
problem for the errors introduced by the formation of a singularity,
near which the accuracy of the simulation obviously breaks down. Indeed
the time interval of the grid, $dt$, must be small (with the speed
of light taken to be unity, $c=1$) compared to the spatial interval,
$dz$. The timestep being small means that the effects of the singularity
propagate superluminally and so prevents us from investigating with
our simulations regions which are \emph{not} in causal contact with
the singularity and which should therefore be physically well behaved.
This problem is resolved by our later simulation in double-null co-ordinates
which, as we shall see, explicitly respects the causality structure
of the spacetime. The reason the timestep must be small is in order
to ensure that we satisfy the Courant condition (which states that
we must have $dt<Cdz$ for a scheme dependent constant $C$ of order
unity): otherwise the Runge-Kutta method would be unstable. This condition
arises from the fact that in one timestep this sort of numerical scheme
can only transmit information a certain number of spatial points away
(two in our case); if the timestep were too large then information
would propagate faster physically than it could in the numerical scheme,
which could lead to large errors. 

Despite these issues, there is a window in which we have a numerical
solution for the region between the kink and anti-kink and away from
the singularity. Under the assumption that the singularity is a black
brane we would expect that the formation of the singularity just replaces
the initial Minkowski space between the walls with a black brane solution.
In particular we expected to be able to measure the tension of the
black brane. In fact the apparent `tension' of the `black brane' never
settled down and so we were unable to measure it. This was our first
hint that the singularity formed is not in fact a black brane, as
we confirmed by examination of the asymptotic structure, described
above, and further simulations, described below.

\begin{wrapfigure}{O}{0.5\columnwidth}%
\noindent \begin{centering}
\includegraphics[bb=195bp 620bp 390bp 815bp,clip]{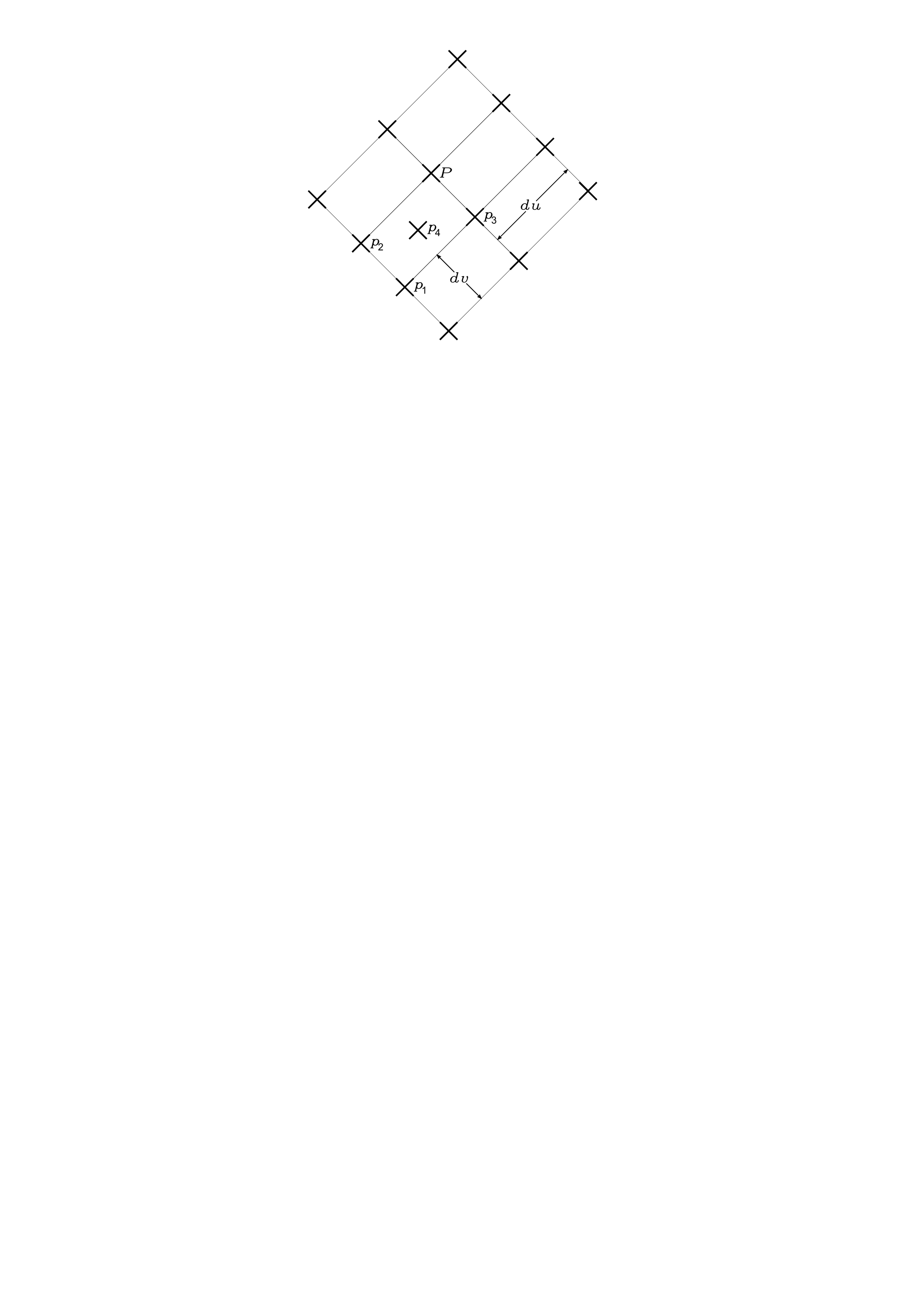}
\par\end{centering}

\caption{$u-v$ grid\label{fig:uv-grid}}
\end{wrapfigure}

In order to examine the structure of the singularity in more detail
in our simulations, we switch to a scheme using double null co-ordinates
\cite{Burko:1997tb} which alleviates the problems just described
and also allows some useful optimizations. In order to describe this
scheme let us denote the various fields by $h^{i}=\left\{ \phi,A,B\right\} $,
$i=1,2,3$. To calculate the values of $h^{i}$ at $P$ (Figure \ref{fig:uv-grid})
we now depend only on the values at $p_{1}$, $p_{2}$ and $p_{3}$,
which are all in the past light cone of $P$. To cover the space we
calculate $h^{i}$ along each ingoing ray at constant $v$, starting
at $u=0$ and continuing to $z=0$ or until we get too close to the
singularity, before moving on to the next ray at $v+dv$. In this
way the only boundary condition we need is the initial data along
$u=0$, which we know exactly so we have no boundary errors as in
the $t-z$ case above. More importantly, this scheme respects the
causal structure: Therefore the singularity does not disrupt the simulation
in regions that are outside of causal contact with it and so we can
continue to simulate the whole external spacetime even after the singularity
is formed.

The method we use to calculate $h^{i}$ at $P$ from the values at
$p_{1}$, $p_{2}$ and $p_{3}$ (as shown in Figure \ref{fig:uv-grid})
is second order \cite{Burko:1997tb}. We take 
\begin{equation}
h^{i}(P)=-h^{i}(p_{1})+h^{i}(p_{2})+h^{i}(p_{3})+F^{i}(p_{4})dudv\label{eq:h^i-next}
\end{equation}
with $du$ and $dv$ the lattice spacings in $u$ and $v$ and defining
$F^{i}(p)=\partial_{u}\partial_{v}h^{i}$ as given by the equations
of motion (\ref{eq:phi-eom-uv}-\ref{eq:B-eom-uv}); $F^{i}$ expresses
the second derivative of $h^{i}$ in terms of the fields and their
first derivatives: $h^{j}$, $\partial_{u}h^{k}$ and $\partial_{v}h^{l}$.
We need to know the values of these at $p_{4}$ to evaluate \eqref{eq:h^i-next}.
We calculate these using the standard ``predictor-corrector'' method
to iterate until the corrections are below the chosen threshold.%

This numerical scheme only depends upon a rectangular set of grid
points with one step in the $u$ direction and one in the $v$ direction.
This makes it relatively straightforward to adapt the size of the
grid, as compared to, for example, the Runge-Kutta method we used
previously. To do so there we would have to take into account the
changing step size between say $p_{1}\rightarrow p_{2}$ and $p_{2}\rightarrow p_{3}$
of Figure \ref{fig:tz-grid} within the single calculation of the
point at $P$, whereas here we only have one step size in the calculation
of a point, since we only have to consider one step. We make use of
this property to implement an adaptive grid spacing, via `point splitting'
and `point reduction'. Near the domain walls and especially as we
approach the singularity the fields are changing rapidly and so we
need a fine lattice spacing to maintain the accuracy of our simulation.
To maintain this fine spacing everywhere would be prohibitively expensive
computationally, which creates the need for adaptive grid spacing.

When the difference between any $h^{i}$ at adjacent points, $(u_{0},v_{0})$
and $(u_{0}+du,v_{0})$, exceeds some threshold, we know that we need
a finer grid. We obtain this by adding an extra point to our lattice
at $(u_{0}+\frac{1}{2}du,v_{0})$ and calculating the values of $h^{i}$
there by interpolation so that when we reach the same region on the
next pass at $v_{0}+dv$ we can use $h^{i}(u_{0}+\frac{1}{2}du,v_{0})$
to calculate $h^{i}(u_{0}+\frac{1}{2}du,v_{0}+dv)$ by our usual scheme
described above. In this way regions where the fields are changing
rapidly acquire more grid points so that we have a grid spacing fine
enough to maintain the desired accuracy. This is what is called point
splitting.

Points at a given value of $u$ move away from the origin (and so
from the singularity) at the speed of light as we increase $v$. Therefore
if the influence of the singularity spreads subluminally the fields
will be changing less rapidly at larger $v$ than they were previously
and the fine grid spacing required then becomes redundant. The domain
wall, and hence the need for a finer mesh associated with it, also
moves. This means that we can apply a method for removing points once
they are no longer required, allowing us to avoid unnecessary computations.
This is what is called point reduction. We have a second threshold,
smaller than the threshold for point splitting, and if the difference
in all the fields between some point and the previous one is lower
than that threshold the the point is removed from the grid.

By the combination of these two procedures the spacing in $u$ of
our grid adapts itself to maintain accuracy in rapidly changing regions
while not incurring superfluous computational cost in slowly changing
regions.

\section{Simulation Results\label{sec:Simulation-Results}}

\begin{figure}[b]
\noindent \begin{centering}
\includegraphics[height=9cm]{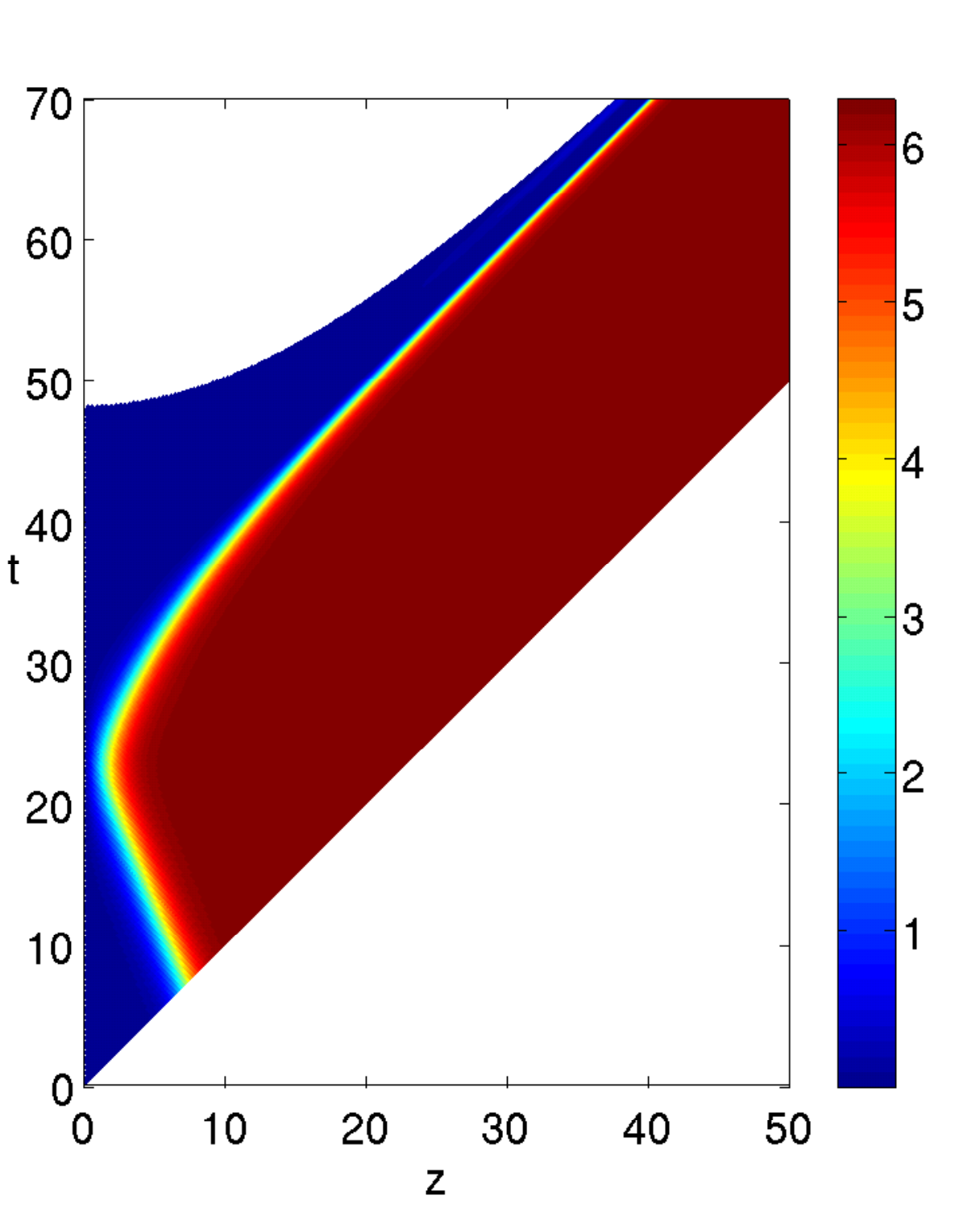}
\par\end{centering}

\caption{The scalar field during a collision in which a singularity forms.
The value of $\phi$ is shown by the colours from 0 (blue) in the
Minkowski vacuum to $2\pi$ (red) in the AdS$^{5}$ vacuum. The position
of the brane is the transition region between the two, i.e.\ the
green-yellow-orange band. We see the brane bouncing away from its
partner in the $z<0$ region just after $t=20$ before the singularity
forms at around $t=50$.\label{fig:phi}}
\end{figure}

\begin{figure}

\begin{minipage}[t]{0.45\columnwidth}%
\noindent \begin{center}
\includegraphics[height=8cm]{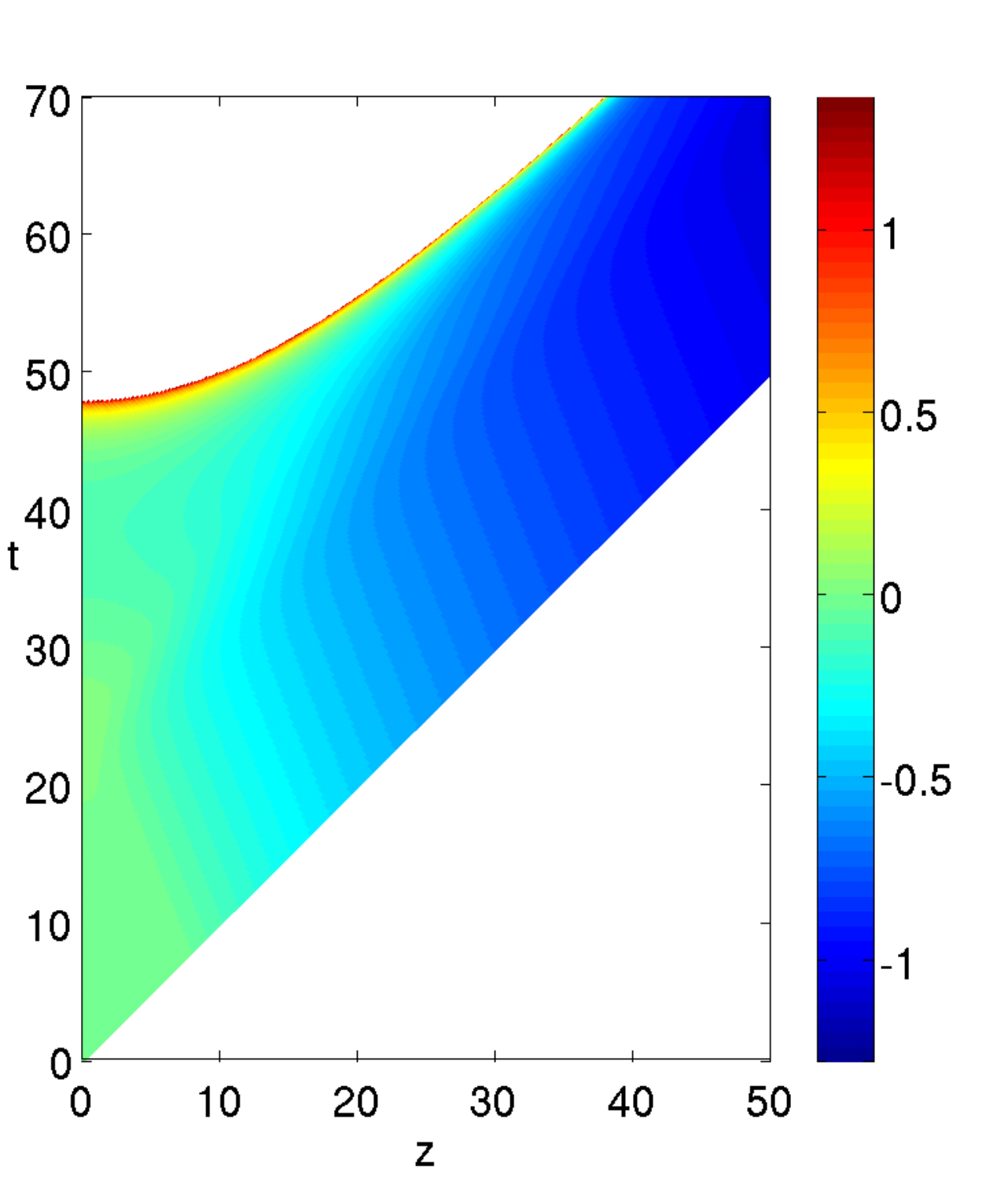}\caption{The value of the metric function $A$, as defined in Equation \eqref{eq:Dynamical_metric},
over the $t-z$ plane. $A$ is not dramatically affected by the bounce
just after $t=20$. It diverges later, driven by $B$.\label{fig:A_over_t-z}}

\par\end{center}%
\end{minipage}\hfill{}%
\begin{minipage}[t]{0.45\columnwidth}%
\noindent \begin{center}
\includegraphics[height=8cm]{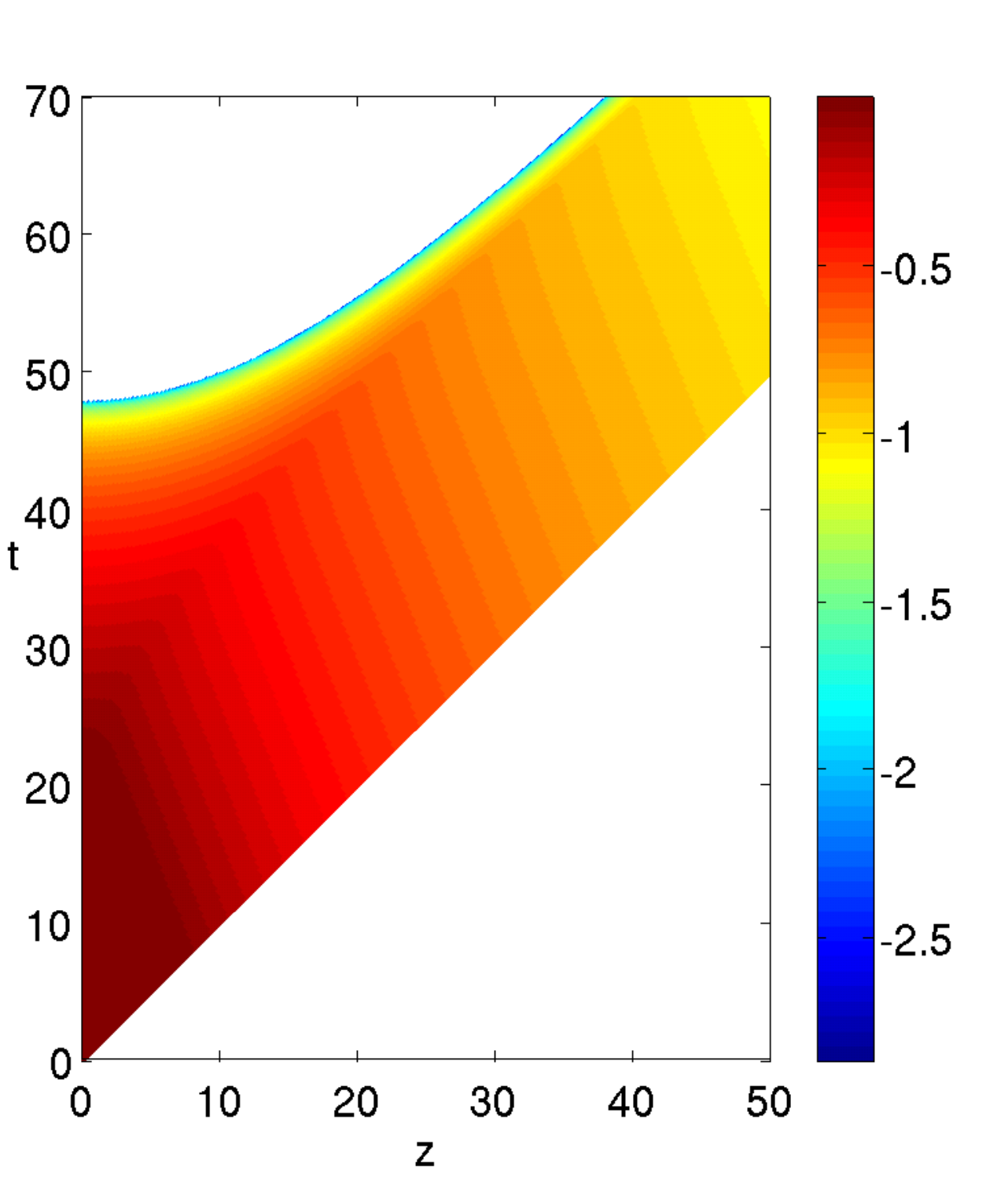}\caption{The value of the metric function B, as defined in Equation \eqref{eq:Dynamical_metric},
over the $t-z$ plane. We can see that after the branes bounce just
after $t=20$ $B$ begins to decrease, continuing until the singularity
forms around $t=50$. Hence the singularity formation is driven by
the collapse in the size of the transverse slices, that is the $x^{i}$
directions.\label{fig:B_over_t-z}}

\par\end{center}%
\end{minipage}

\end{figure}

The outcome of a simulation, using the numerical method in double-null
co-ordinates, is shown in Figure \ref{fig:phi}. In this simulation
a singularity forms. Only half of the spacetime is shown, the other
follows from reflection symmetry about $z=0$. The initial conditions
are imposed along the $u=0$ line which borders the blank triangle
in the bottom right of the plot. We see the domain wall between the
AdS$^{5}$ region with $\phi\rightarrow2\pi$ at large $z$ and the
Minkowski region with $\phi\rightarrow0$ around $z=0$. The other
domain wall is its mirror image in the $z<0$ region. The domain walls
approach each other initially before bouncing apart. After the bounce
a singularity forms in the centre. The simulation is cut off when
the curvature becomes too large near the singularity: this is the
blank area in the top left of the plot. The formation of the singularity
is driven by the metric function $B$, as defined in Equation \eqref{eq:Dynamical_metric},
which controls the $x^{i}$ components of the metric. As we see in
Figure \ref{fig:B_over_t-z}, $B$ begins to decrease after the bounce,
and diverges as the singularity is approached. This causes the other
metric function, $A$, also to diverge (Figure \ref{fig:A_over_t-z}).

\section{Horizon Structure}

Having seen that there is a singularity, we would like to check whether
there is a horizon, and examine its structure if there is one. Indeed
we may expect a horizon, since we see from Figure \ref{fig:phi} that
there is a region within which all timelike geodesics will end on
the singularity. To find a horizon we study the structure of the null
geodesics. However, given the dynamical nature of the spacetime it
is hard to study global properties of the geometry, like the event
horizon; it is more convenient to find an alternative with a local
definition. This we do in the definition by Hayward \cite{Hayward:1993wb}
of trapping horizons, which are given in terms of the expansion of
incoming and outgoing null geodesics.

Since we are using double-null co-ordinates, we have null vectors
to hand already,
\begin{align}
 & N_{+}=\partial_{u},\quad N_{-}=\partial_{v}
\end{align}
where for $z>0$ $N_{+}$ is ingoing and $N_{-}$ is outgoing. The
dual one-forms to these are
\begin{align}
 & n_{+}=-e^{2A}\dd v,\quad n_{-}=-e^{2A}\dd u
\end{align}
and we can define unit normalized null vectors
\begin{align}
 & u_{+}=e^{-2A}N_{-},\quad u_{-}=e^{-2A}N_{+}
\end{align}
so that $n_{\pm}(u_{\pm})=-1$. Then the induced metric on three-dimensional
surfaces normal to both $u_{+}$ and $u_{-}$ is
\begin{align}
h & =g+e^{-2A}n_{+}\otimes n_{-}+e^{-2A}n_{-}\otimes n_{+}
\end{align}
where $g$ is the full metric. We can use the induced metric to define
the expansions of null geodesics by
\begin{align}
\Theta_{\pm} & =\frac{1}{2}h^{ij}\m L_{\pm}h_{ij}
\end{align}
with $\m L_{\pm}$ being the Lie derivatives along $u_{\pm}$, $\m L_{\pm}\equiv\m L_{u_{\pm}}$.
\begin{figure}
\noindent \raggedright{}%
\begin{minipage}[t]{0.45\columnwidth}%
\noindent \begin{center}
\includegraphics[height=8cm]{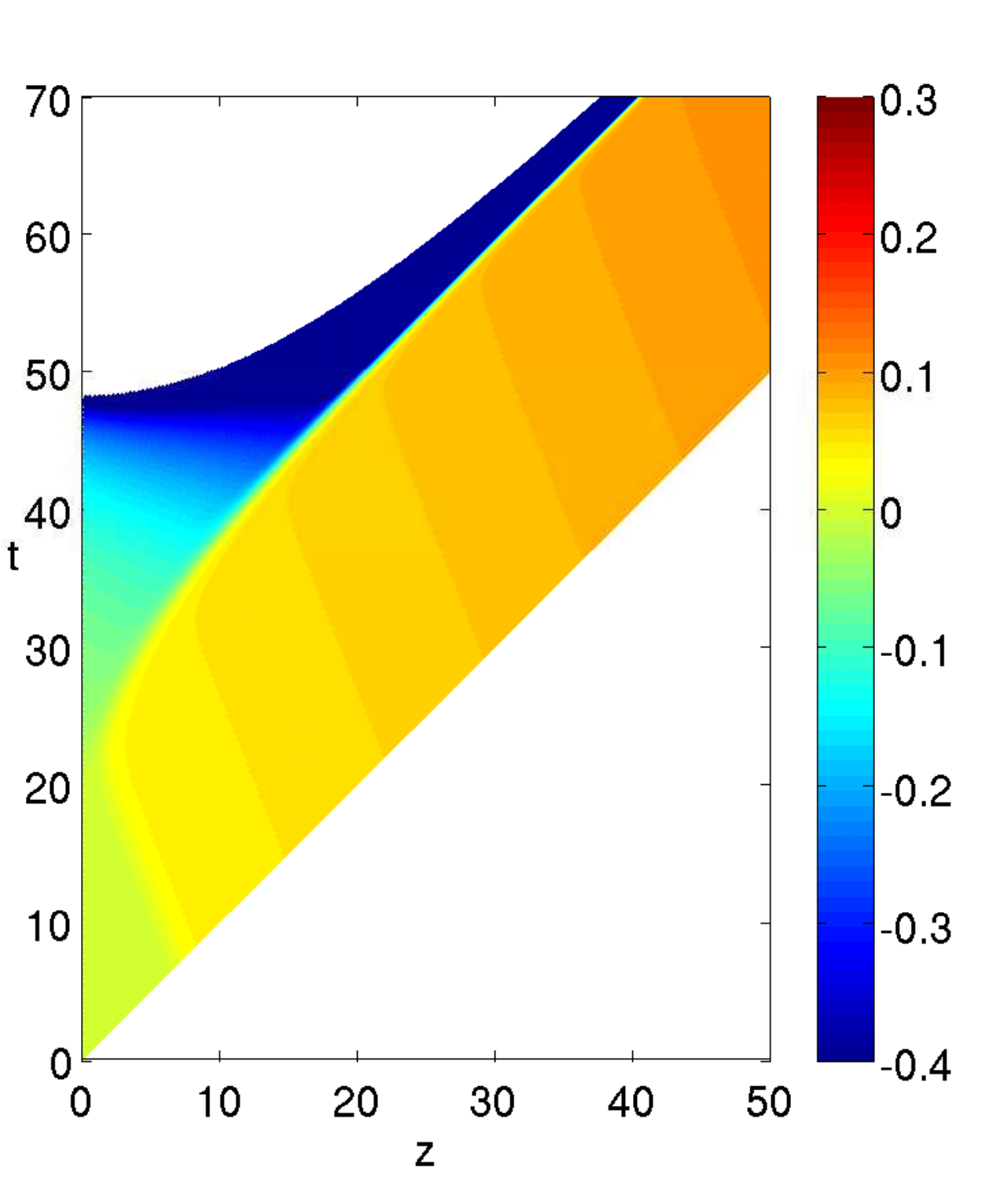}
\par\end{center}

\noindent \begin{center}
\caption{$\Theta_{-}$ over the $t-z$ plane. The trapping horizon is the line
where $\Theta_{-}=0$. $\Theta_{-}$ is negative inside the horizon
and positive outside, which shows that we have an inner horizon, since
$u_{+}$ crosses the horizon from inside to outside.\label{fig:Theta_-}}

\par\end{center}%
\end{minipage}\hfill{}%
\begin{minipage}[t]{0.45\columnwidth}%
\noindent \begin{center}
\includegraphics[height=8cm]{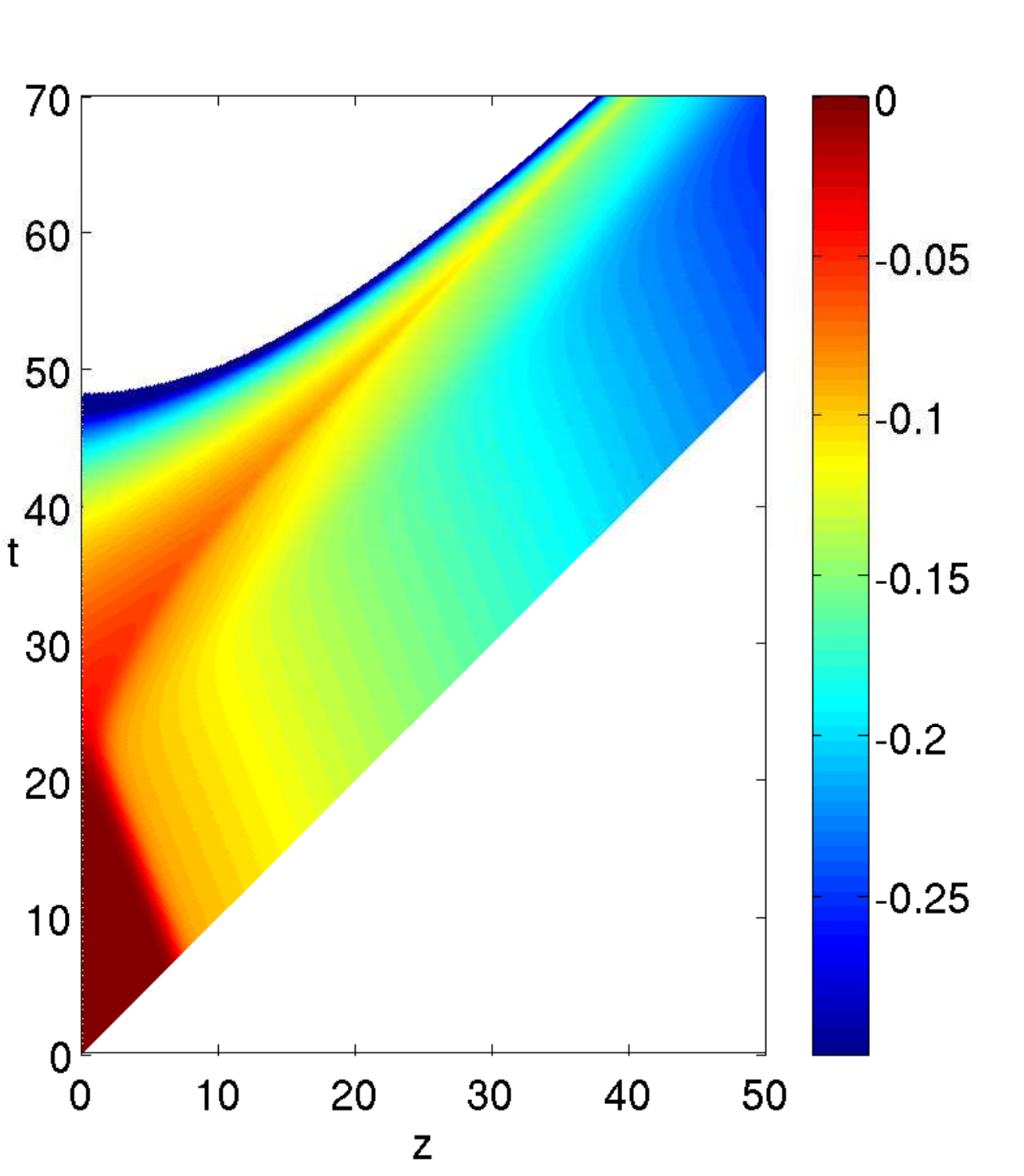}
\par\end{center}

\noindent \begin{center}
\caption{$\Theta_{+}$ over the $t-z$ plane. $\Theta_{+}$ is negative everywhere
(apart from the initial Minkowski region between the branes, where
it vanishes) and in particular on the horizon, showing that the horizon
is future.\label{fig:Theta_+}}

\par\end{center}%
\end{minipage}
\end{figure}
 A marginal surface is a three-surface on which one of the expansions
vanishes, i.e.\ it is a point in the $t-z$ plane. Finally we can
define a trapping horizon as the locus of marginal surfaces, which
is a line in the $t-z$ plane. We may classify trapping horizons firstly
as past or future and secondly as inner or outer. Consider a trapping
horizon defined by $\Theta_{-}=0$. It is future if $\Theta_{+}<0$
and past if $\Theta_{+}>0$. It is outer if $\m L_{+}\Theta_{-}<0$
and inner if $\m L_{+}\Theta_{-}>0$.

The expansions in our example are given in Figures \ref{fig:Theta_-}
and \ref{fig:Theta_+} where the values of the expansions are shown
by the colours over the $t-z$ plane. We can see from Figure \ref{fig:Theta_-}
that there is a region where $\Theta_{-}$ is negative (near the singularity)
and another where it is positive (far from the singularity) so there
is a line where $\Theta_{-}$ changes sign, which is a trapping horizon.
Figure \ref{fig:Theta_+} shows that $\Theta_{+}$ is negative along
the trapping horizon, so it is a future horizon. Evaluating $\m L_{+}\Theta_{-}$
on the horizon shows that it is positive (the horizon is slightly
steeper than a null ray) and so it is an inner horizon. Now a black
hole has a future-outer trapping horizon, suggesting again that our
singularity is not a black brane. However, it might be a big crunch
singularity which does have a future-inner trapping horizon.

By way of confirmation of this conclusion, we can measure the area
of the horizon, which for a future-inner trapping horizon should be
non-increasing. Figure \ref{fig:B_horizon}, which gives the value
of $B$ along the horizon, shows that this is indeed the case.

\section{Closing Off the Spacetime}

\begin{figure}
\centering{}%
\begin{minipage}[t]{0.45\columnwidth}%
\begin{center}
\includegraphics[width=1\columnwidth]{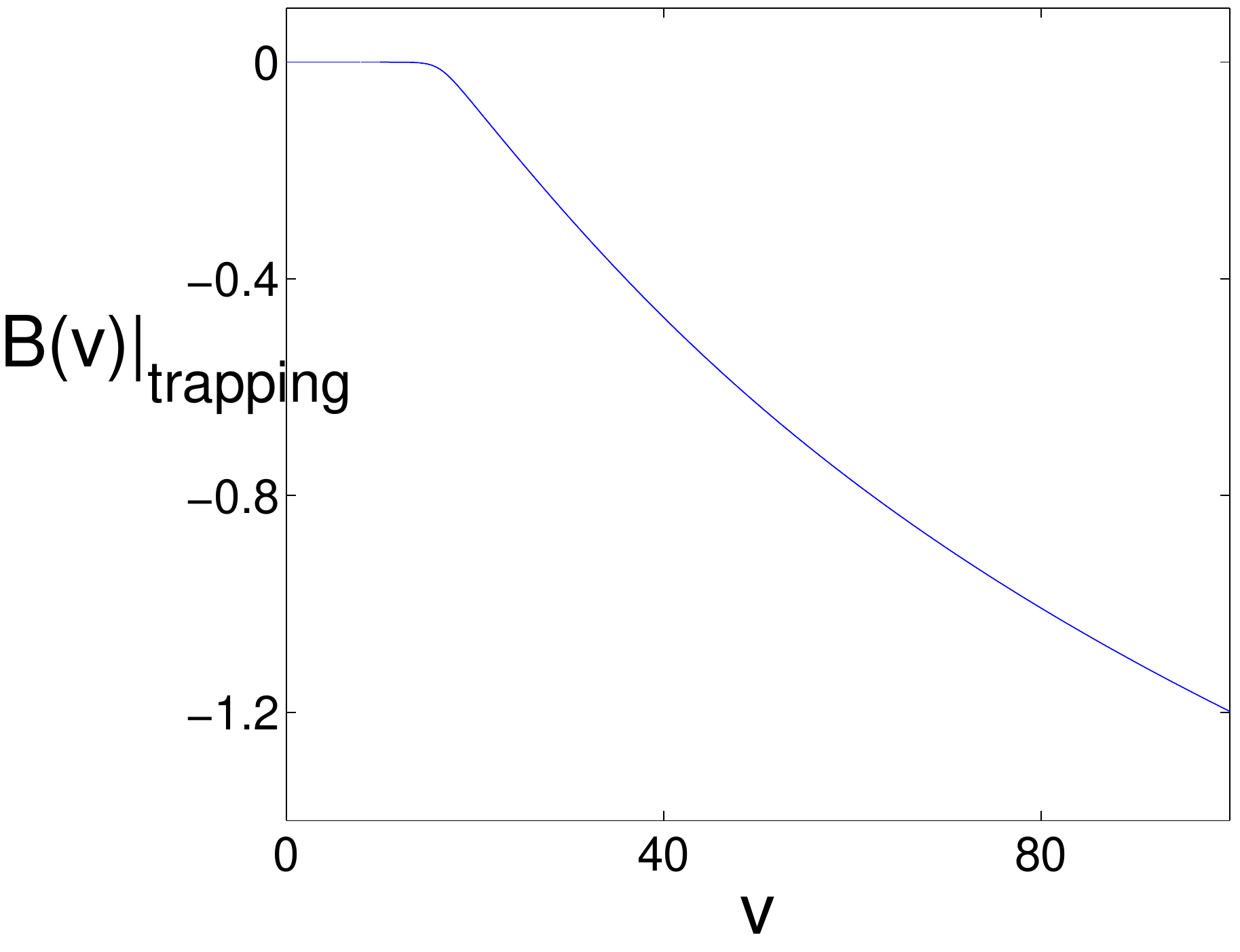}\caption{$B$ evaluated on the trapping horizon, showing that the area of the
horizon decreases monotonically.\label{fig:B_horizon}}

\par\end{center}%
\end{minipage}\hfill{}%
\begin{minipage}[t]{0.45\columnwidth}%
\begin{center}
\includegraphics[width=1\columnwidth]{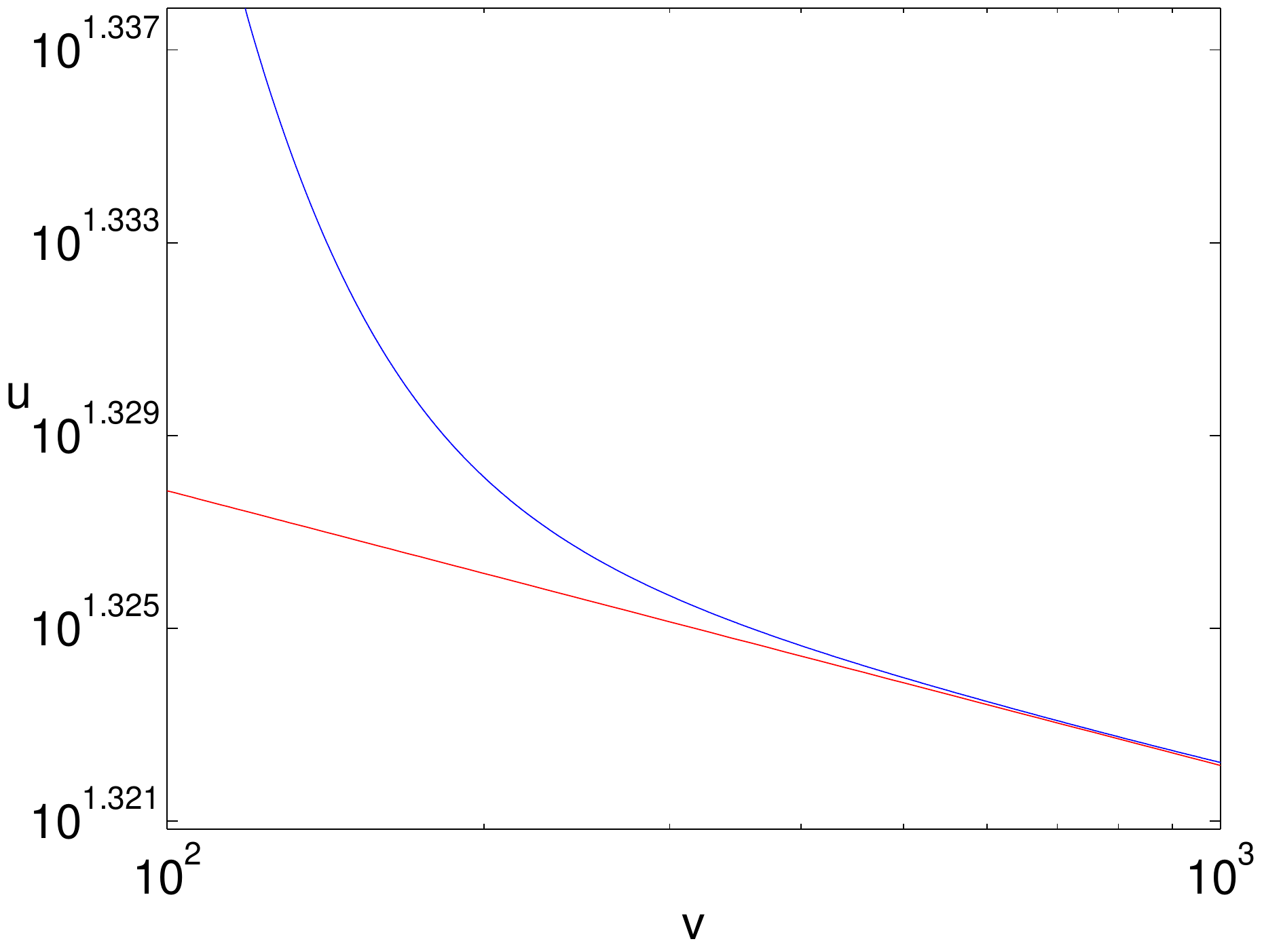}\caption{The level set for $R=-5$, in blue, and the line $u=\frac{1}{v^{5.7\times10^{-3}}}$,
in red.\label{fig:R_level_set}}

\par\end{center}%
\end{minipage}
\end{figure}

If our singularity is a big crunch then it should close off the entire
spacetime: there should be no region where an observer could survive
forever without encountering the singularity. It is challenging to
show this robustly numerically as it would require us to simulate
the spacetime out to very large values of $v$. As something of a
substitute we can look at the level surfaces of the Ricci scalar.
If they were all to reach $u=0$ eventually, then any observer would
ultimately have to cross all of them and thus reach the singularity.
The level set for $R=-5$, plotted in Figure \ref{fig:R_level_set},
shows $u$ decreasing rather slowly as $v$ increases but it is at
least consistent with crossing $u=0$ eventually and thus consistent
with a big crunch singularity.

\section{Conclusions\label{sec:Colliding_Branes_Conclusions}}

We have reconsidered the formation of curvature singularities in domain
wall\linebreak{}
collisions \cite{Takamizu:2007ks} with an eye on their global structure.
Our initial observation was that the asymptotic AdS$^{5}$ structures
of the domain walls and black branes are mutually incompatible and
so the singularities formed by domain walls cannot be black branes.
This is corroborated by looking at the dynamical geometry using numerical
simulations. These show us that the trapping horizons formed around
the singularity after the domain wall collision are of the type associated
with a big crunch rather than a black brane. Further confirmation
comes from the non-increasing area of the horizon and the trend of
the Ricci level sets across $u=0$ at large $v$, though the evidence
for the latter is tentative because of the numerical difficulties
presented by the very slow rate of decrease. As the type of singularity
seems to be dictated by the asymptotic structure, one might expect
similar qualitative behaviour as we see for the scalar domain wall
to apply also to a generic braneworld collision if they share the
same asymptotic structure.

In \cite{Chamblin:1999cj} it was predicted that the AdS$^{5}$ Cauchy
horizon would generically be replaced by a pp singularity if the AdS
region were perturbed. A pp singularity is a singularity where all
the scalar curvature invariants are finite, but the components of
the Riemann curvature diverge so that an observer in an orthonormal
frame parallelly propagated along a timelike geodesic (i.e.\ a freely
falling observer) would observe the divergence, which would manifest
in the tidal forces experienced by that observer. Here we find that
when a curvature singularity is formed it closes off the geometry
in a big crunch and no pp singularity is formed. However, when the
collision is not sufficiently violent to form a curvature singularity
there is no reason not to expect the prediction of a pp singularity
to hold.

Now that it is established that the singularity formed in the collision
of domain walls is a big crunch, further simulations would be useful
to map out the parameter space for where singularities do and do not
form. If a singularity does form then the universe on the domain wall
must end in finite time, since it will encounter the singularity.
Therefore in order to construct models of an eternally expanding universe
one would have to set the initial conditions in order to avoid singularity
formation. One could also now return to the investigation of particle
dynamics on the domain walls, as in \cite{Takamizu:2004rq,Gibbons:2006ge,Saffin:2007ja,Saffin:2007qa},
but including the gravitational dynamics.

\chapter{Braneworlds and Galileons\label{cha:Galileons}}

For modifications of gravity to be phenomenologically viable (as solutions
to the dark energy problem) they must give undetectably small corrections
to general relativity on solar system scales, but large corrections
approaching horizon scales. This forces such modifications to be non-linear,
but in non-linear theories instabilities, such as ghosts, often become
problematic. By generalizing the four dimensional effective field
theory arising from the DGP model, Nicolis et al.\ found a promising
class of models \cite{Nicolis:2008in}. These have a scalar field,
$\pi$, coupled to gravity whose Lagrangian is symmetric under the
`Galilean' transformation $\pi\rightarrow\pi+b_{\mu}x^{\mu}+c$, leading
them to christen $\pi$ the `galileon'. Ghosts can be avoided by demanding
that the equations of motion contain at most second derivatives and
this demand together with Galilean symmetry restricts the models to
a family with only five free parameters (in four dimensions): in dimension
$d$ the terms in the Lagrangian take the form $?\eta^{\mu_{1}}{}_{[\nu_{1}}?\ldots?\eta^{\mu_{m}}{}_{\nu_{m}]}?\pi\partial_{\mu_{1}}\partial^{\nu1}\pi\ldots\partial_{\mu_{m}}\partial^{\nu_{m}}\pi$
for $m\leq d$.

As the galileon theory was inspired initially by the DGP braneworld
model, it is not surprising that it can itself be embedded in a braneworld
model as the effective theory of a codimension one probe brane, as
was found by de Rham and Tolley for flat spacetime \cite{deRham:2010eu}
and generalized by Goon et al. \cite{Goon:2011qf}. This is achieved
by using more general gravity theories than the DGP model which has
just the Einstein-Hilbert term in both brane and bulk actions. The
requirement of second order equations of motion means that those gravity
theories must be what one might call Lovelock-Myers theories, with
Lovelock terms \cite{Lovelock:1971yv} both in the bulk and on the
brane supplemented by Myers-type surface terms \cite{Myers:1987yn}
on the brane. On a co-dimension one brane (which is equivalent to
a boundary) there are a series of Lovelock terms for the intrinsic
curvature (cosmological constant $\Lambda$, Einstein-Hilbert $R$,
Gauss-Bonnet $R^{2}-4R_{\mu\nu}R^{\mu\nu}+R_{\mu\nu\rho\sigma}R^{\mu\nu\rho\sigma}$,\ldots{})
and Myers terms for the extrinsic curvature, which are the surface
terms corresponding to the bulk Lovelock terms (Gibbons-Hawking $K$,
$K_{GB}=-\frac{1}{3}K^{3}+K_{\mu\nu}K^{\mu\nu}K-\frac{2}{3}K_{\mu}^{\nu}K_{\nu}^{\rho}K_{\rho}^{\mu}-2(R_{\mu\nu}-\frac{1}{2}Rg_{\mu\nu})K^{\mu\nu}$,\ldots{}).
In a given dimension these series terminate, with all the higher terms
only contributing total derivatives to the action. The galileon theory
is constructed by introducing a four dimensional probe brane into
a maximally symmetric five dimensional bulk, neglecting the backreaction
of the brane modes, i.e.\ of the galileon. The galileon is the displacement
of the brane in the fifth dimension with the Galilean symmetry thus
following from the Poincaré symmetry of the bulk: an infinitesimal
transformation is given by $x^{a}\rightarrow x^{a}+?\epsilon^{a}{}_{b}?x^{b}+v^{a}$
so if we shift the brane from $y=\pi(x)$ to $y=\pi(x)+?\epsilon^{y}{}_{\mu}?x^{\mu}+v^{y}$
there can be no physical change, and so we can see that the theory
is symmetric under Galilean transformations of $\pi$. Four of the
five parameters of the galileon field theory then come from the coefficients
of the first two Lovelock terms and the first two Myers terms. The
fifth comes from the coefficient of the tadpole term $\int^{\pi}\sqrt{g}dy$
which also turns out to be allowed.

On the other hand, one can also construct four dimensional field theories
with more than one galileon \cite{Deffayet:2010zh,Padilla:2010de}.
These share many of the attractive properties of the single galileon
theories, although as the number of galileons increases the number
of possible terms in the action, which is \cite{Padilla:2010ir} $\sum_{n=1}^{5}\frac{(m+n-1)!}{n!(m-1)!}$
for $m$ galileons, increases rather fast giving 55 free parameters
for three galileons and 125 for four galileons. Just as a single galileon
is the displacement of a codimension one brane in the transverse direction,
so should $m$ galileons be the transverse displacements of a codimension
$m$ brane.

Such a construction has been performed for a Minkowski probe brane
in a Mink\-owski bulk by Hinterbichler et al. \cite{Hinterbichler:2010xn}\ but
with an extremely restricted choice of terms in the action, leaving
only one free parameter. This can only be a braneworld embedding of
a rather small class of the possible field theory galileon models.
This choice was made on the strength of the assertion, based on the
results of \cite{Charmousis:2005ey,Charmousis:2005ez}, that the only
possible terms in the action for a four dimensional brane of even
codimension $N$ are: a cosmological constant and a term $\sqrt{-g}\left(R(g)-(K^{i})^{2}+K_{\mu\nu}^{i}K_{i}^{\mu\nu}\right)$
for $N=2$; and a cosmological constant and the Einstein-Hilbert term
of the induced metric for $N>2$. However, the actions found in \cite{Charmousis:2005ey,Charmousis:2005ez}
are those which allow consistent matching conditions for a distributional
(i.e.\ zero `thickness') brane with a tension which backreacts on
the bulk geometry. They are not relevant to the case here where we
are interested in probe branes which do not backreact. For example,
black hole theorems restricting the possible solutions and thus the
allowed sources cannot be a restriction here because there is no backreaction
and thus no source for the bulk curvature. We are allowed any curvature
terms we can write down which result in second order equations of
motion. This certainly includes all the Lovelock terms of the induced
curvature (which in four dimensions are just the cosmological constant
and the Einstein-Hilbert term). Presumably some terms in the extrinsic
curvature are also allowed but it is not clear how to construct the
analogues in codimension greater than one of the Myers terms now that
we have an extra (transverse) index on the extrinsic curvature.

Here we describe some work done in collaboration with Ian Moss, Antonio
Padilla and Paul Saffin. We describe a slightly more general class
of models, allowing for maximally symmetric spaces with arbitrary
curvature but with the same restriction as \cite{Hinterbichler:2010xn}
on the terms in the action. This restriction arises from the $SO(m)$
symmetry of the extra dimensions which follows if both brane and bulk
are maximally symmetric (and so restricts the case in \cite{Hinterbichler:2010xn}
as well). It seems that the only way to find the complete set of the
models which can be constructed in the field theory approach is to
construct less than maximally symmetric models, because of the strong
restriction placed by $SO(m)$ symmetry on the possible terms in the
action. This remains yet to be done. It appears to be a rather challenging
task since the analogues in higher codimension of the Myers surface
terms induced by bulk Lovelock invariants in codimension one are unclear:
though in the case we consider here they are incompatible with the
$SO(m)$ symmetry up to the order to which we work. Also unclear is
the role of the tadpole term in higher codimension since in this case
there is ambiguity over how one integrates up to the brane, in contrast
to codimension one. However, the formalism we have developed for constructing
the braneworld-galileon actions may be of some interest; we hope that
it might be useful for constructing more general galileon actions
than just those considered here if a scheme for breaking the $SO(m)$
symmetry can be specified.

\section{Galileons in Codimension One}

The codimension one case has two properties which make it much more
straightforward than the higher codimension case. The first is that
as mentioned above the possible terms in the action are clear in arbitrary
dimension. The second is that one can find exact expressions for all
these in terms of the galileon $\pi$, the displacement of the brane
in the transverse direction. The reason for this is that one can analytically
invert the induced metric on the brane in this case.

Consider a codimension one probe brane. Taking a Gaussian normal foliation,
the full metric of the bulk spacetime is
\begin{align}
ds^{2} & =d\rho^{2}+f(\rho)^{2}g_{\mu\nu}(x)dx^{\mu}dx^{\nu}
\end{align}
with the co-ordinates $x^{\mu}$ propagated through the leaves of
the foliation at constant $\rho$ by the trajectories normal to them.
Using the co-ordinates $x^{\mu}$ on the brane also and defining the
galileon, $\pi$, as the $\rho$-value of each point $x^{\mu}$ on
the brane, 
\begin{align}
\pi(x) & =\rho|_{\text{brane}}(x)
\end{align}
then the induced metric on the brane is
\begin{align}
\tilde{g}_{\mu\nu} & =f(\pi)^{2}g_{\mu\nu}+\partial_{\mu}\pi\partial_{\nu}\pi
\end{align}
whose inverse is
\begin{align}
\tilde{g}^{\mu\nu} & =\frac{1}{f^{2}}\left(g^{\mu\nu}-\frac{\partial^{\mu}\pi\partial^{\nu}\pi}{f^{2}\left(1+\frac{1}{f^{2}}\partial_{\xi}\pi\partial^{\xi}\pi\right)}\right)\label{eq:inverse-cod-one}
\end{align}
With these two expressions it is possible to compute exactly the determinant
and intrinsic curvature of the induced metric, and also the normal
vector and hence the extrinsic curvature.

Note however that we only have an analytic expression of the inverse,
\eqref{eq:inverse-cod-one}, because the order of the $\pi$'s does
not matter: $\left(\partial_{\mu}\pi\partial_{\xi}\pi\right)\left(\partial^{\xi}\pi\partial^{\nu}\pi\right)=\partial_{\mu}\pi\partial_{\nu}\pi\left(\partial_{\xi}\pi\partial^{\xi}\pi\right)$.
In the higher codimension case which is our focus here, this trick
fails. We now have a galileon for each transverse direction so if
the codimension is $m$ we have $\pi^{I}$ with $I=1,\ldots,m$. The
induced metric is similar
\begin{align}
\tilde{g}_{\mu\nu} & =g_{\mu\nu}+g_{IJ}\nabla_{\mu}\pi^{I}\nabla_{\nu}\pi^{J}
\end{align}
except that now $\pi^{I}$ fields come in pairs with contracted transverse
indices. Thus $\left(\nabla_{\mu}\pi_{I}\nabla_{\xi}\pi^{I}\right)\left(\nabla^{\xi}\pi_{J}\nabla^{\nu}\pi^{J}\right)\neq\nabla_{\mu}\pi_{I}\nabla_{\nu}\pi^{I}\left(\nabla_{\xi}\pi_{J}\nabla^{\xi}\pi^{J}\right)$
and so we cannot write an analytic expression for the inverse metric
like \eqref{eq:inverse-cod-one}. Since we therefore cannot use a
fully explicit, analytic formalism in the higher codimension case
we instead work perturbatively from the outset. We write the full
metric of the bulk space as a Taylor expansion in the distance from
the `zero position' of the brane (its position when all the galileon
fields $\pi^{I}$ vanish) and truncate the rest of our expressions
to fourth order in the $\pi^{I}$. This approach allows us to use
the existing machinery for calculating, for example, the Ricci scalar
in terms of metric perturbations. We now turn to the elaboration of
this perturbative formalism.

\section{Perturbative Formalism\label{sec:galileons-formalism}}

Consider a codimension $m$ probe 3-brane in an ($m+4$)-dimensional
bulk. We have then $m$ galileons $\pi^{I}$ ($I=1,\ldots,m$) which
are the displacements of the brane in the transverse directions. We
will work perturbatively, assuming from the outset that the galileons
are small. We set the co-ordinates on the bulk spacetime by choosing
a foliation adapted to the brane. Let $\Sigma_{0}$ denote a four
dimensional slice of the spacetime (the position of the unperturbed
brane with $\pi^{I}=0$) with co-ordinates $x^{\mu}$ ($\mu=0,1,2,3$).
Then we extend the co-ordinates to the neighbourhood of $\Sigma_{0}$
along the $m$ directions normal to the surface, i.e.\ $x^{a}=\left\{ x^{\mu},x^{I}\right\} $
($a=0,\ldots,3+m$) so that the co-ordinate basis vectors $\boldsymbol{e}_{\mu}\equiv\frac{\partial}{\partial x^{\mu}}$
and $\boldsymbol{e}_{I}\equiv\frac{\partial}{\partial x^{I}}$ are
orthogonal. The galileons are just the displacements of the brane,
$\Sigma$, away from $\Sigma_{0}$: the co-ordinates of points on
the brane are $x^{a}|_{brane}=\left\{ x^{\mu},\pi^{I}\left(x^{\mu}\right)\right\} $.

Let $\boldsymbol{\xi}=\xi^{I}\boldsymbol{e}_{I}$ denote the normalized
($\boldsymbol{\xi}\cdot\boldsymbol{\xi}=1$) tangent vector to some
geodesic normal to $\Sigma_{0}$ and $\sigma$ denote the proper distance
along this geodesic from $\Sigma_{0}$. Since we are working perturbatively,
we take $\sigma$ to be small and we can write quantities on the brane
as a truncated Taylor series in $\sigma$ about their values on $\Sigma_{0}$.
The (full) metric at the point a distance $\sigma$ away from $\Sigma_{0}$
along the geodesic whose tangent is $\boldsymbol{\xi}$ is
\begin{align}
g_{ab} & =\sum_{n=0}^{\infty}\frac{1}{n!}\sigma^{n}\left[\nabla_{\boldsymbol{\xi}}^{n}\left(\boldsymbol{e}_{a}\cdot\boldsymbol{e}_{b}\right)\right]_{\Sigma_{0}}\label{eq:metric expansion}
\end{align}
where of course $\boldsymbol{e}_{a}\cdot\boldsymbol{e}_{b}$ is just
$g_{ab}$ but this form is convenient since we can determine how $\nabla_{\boldsymbol{\xi}}$
acts on the basis vectors $\boldsymbol{e}_{a}$:
\begin{align}
\left.\nabla_{\boldsymbol{\xi}}\boldsymbol{e}_{\mu}\right|_{\Sigma_{0}} & =\xi^{I}?k_{I\mu}^{\nu}?\boldsymbol{e}_{\nu}+\xi^{I}?a_{I}^{J}{}_{\mu}?\boldsymbol{e}_{J}\\
\left.\nabla_{\boldsymbol{\xi}}^{n}\boldsymbol{e}_{\mu}\right|_{\Sigma_{0}} & =\nabla_{\boldsymbol{\xi}}^{n-2}\boldsymbol{R}(\boldsymbol{\xi},\boldsymbol{e}_{\mu})\boldsymbol{\xi}\\
\left.\nabla_{\boldsymbol{\xi}}\boldsymbol{e}_{I}\right|_{\Sigma_{0}} & =0\\
\left.\nabla_{\boldsymbol{\xi}}^{n}\boldsymbol{e}_{I}\right|_{\Sigma_{0}} & =\frac{\left(n-1\right)}{\left(n+1\right)}\nabla_{\boldsymbol{\xi}}^{n-2}\boldsymbol{R}(\boldsymbol{\xi},\boldsymbol{e}_{I})\boldsymbol{\xi}
\end{align}
where $?k^{I}{}_{\mu\nu}?$ and $?a_{IJ}^{\mu}?$ are the Weingarten
coefficients of the surface $\Sigma_{0}$. For proofs of these statements
see Appendix \ref{cha:galileon-appendix}. In our normal co-ordinates
they are defined quite straightforwardly: the extrinsic curvature%
\footnote{The relation of the extrinsic curvature defined like this to the usual
definition for a codimension one hypersurface is discussed in Appendix
\ref{cha:galileon-appendix}%
} is
\begin{align}
?k^{I}{}_{\mu\nu}? & =\left.-?\Gamma^{I}{}_{\mu\nu}?\right|_{\Sigma_{0}}\label{eq:extrinsic_curvature}
\end{align}
and the `twist connection' is
\begin{align}
?a_{IJ}^{\mu}? & =\left.-?\Gamma^{\mu}{}_{IJ}?\right|_{\Sigma_{0}}\label{eq:twist_connection}
\end{align}

Up to this point we are entirely agnostic as to the structure of either
the bulk spacetime or the unperturbed brane. To make further progress,
we assume that both are maximally symmetric, as this greatly simplifies
the expressions for the bulk curvature, and the extrinsic curvature
and twist connection of the unperturbed brane. This assumption would
need to be relaxed somehow to construct more general galileon theories,
but this being achieved one could presumably continue in a similar
spirit to the method described below for the maximally symmetric case.

In the maximally symmetric case we have
\begin{align}
k_{I\mu\nu} & =c_{I}g_{\mu\nu}\label{eq:k_Imunu}\\
a_{IJ\mu} & =0\\
R_{abcd} & =\kappa\left(g_{ac}g_{bd}-g_{ad}g_{bc}\right)\label{eq:R(g)}
\end{align}
where $c_{I}$ and $\kappa$ are constants. We will work at up to
fourth order in the galileon fields. To do so we need to evaluate
the terms in the expansion of the metric, \eqref{eq:metric expansion},
at the brane, $\Sigma$. There $\sigma\boldsymbol{\xi}=\pi^{I}\boldsymbol{e}_{I}$
and so, writing $\pi=c_{I}\pi^{I}$ and $\boldsymbol{\pi}\cdot\boldsymbol{\pi}=g_{IJ}\pi^{I}\pi^{J}$,
\begin{align}
\sigma\nabla_{\boldsymbol{\xi}}\boldsymbol{e}_{\mu} & =\pi\boldsymbol{e}_{\mu}\\
\sigma^{2}\nabla_{\boldsymbol{\xi}}^{2}\boldsymbol{e}_{\mu} & =\sigma^{2}\boldsymbol{R}(\boldsymbol{\xi},\boldsymbol{e}_{\mu})\boldsymbol{\xi}=\pi^{I}\pi^{J}?R^{\nu}{}_{JI\mu}?\boldsymbol{e}_{\nu}=-\kappa\boldsymbol{\pi}\cdot\boldsymbol{\pi}\boldsymbol{e}_{\mu}\\
\sigma^{3}\nabla_{\boldsymbol{\xi}}^{3}\boldsymbol{e}_{\mu} & =\sigma^{3}\nabla_{\boldsymbol{\xi}}\boldsymbol{R}(\boldsymbol{\xi},\boldsymbol{e}_{\mu})\boldsymbol{\xi}=\sigma^{3}\boldsymbol{R}(\boldsymbol{\xi},\nabla_{\boldsymbol{\xi}}\boldsymbol{e}_{\mu})\boldsymbol{\xi}=\pi^{I}\pi\pi^{J}?R^{\nu}{}_{JI\mu}?\boldsymbol{e}_{\nu}=-\kappa\pi\boldsymbol{\pi}\cdot\boldsymbol{\pi}\boldsymbol{e}_{\mu}\\
\sigma^{4}\nabla_{\boldsymbol{\xi}}^{4}\boldsymbol{e}_{\mu} & =\sigma^{4}\boldsymbol{R}(\boldsymbol{\xi},\nabla_{\boldsymbol{\xi}}^{2}\boldsymbol{e}_{\mu})\boldsymbol{\xi}=-\kappa\pi^{I}\boldsymbol{\pi}^{2}\pi^{J}?R^{\nu}{}_{JI\mu}?\boldsymbol{e}_{\nu}=\kappa^{2}\left(\boldsymbol{\pi}\cdot\boldsymbol{\pi}\right)^{2}\boldsymbol{e}_{\mu}\\
\sigma^{2}\nabla_{\boldsymbol{\xi}}^{2}\boldsymbol{e}_{I} & =\sigma^{2}\frac{1}{3}\boldsymbol{R}(\boldsymbol{\xi},\boldsymbol{e}_{I})\boldsymbol{\xi}=\frac{1}{3}\pi^{J}\pi^{K}?R^{L}{}_{KJI}?\boldsymbol{e}_{L}=\frac{1}{3}\kappa\left(\pi_{I}\pi^{J}\boldsymbol{e}_{J}-\boldsymbol{\pi}\cdot\boldsymbol{\pi}\boldsymbol{e}_{I}\right)
\end{align}
Then from \ref{eq:metric expansion}, and calling the metric at $\Sigma_{0}$
$\bar{g}_{ab}$, the metric at $\Sigma$ is
\begin{align}
g_{\mu\nu} & =\bar{g}_{\mu\nu}+2\pi\bar{g}_{\mu\nu}-\kappa\boldsymbol{\pi}\cdot\boldsymbol{\pi}\bar{g}_{\mu\nu}+\pi^{2}\bar{g}_{\mu\nu}-\frac{4}{3}\kappa\pi\boldsymbol{\pi}\cdot\boldsymbol{\pi}\bar{g}_{\mu\nu}\nonumber \\
 & \quad+\frac{1}{3}\kappa^{2}\left(\boldsymbol{\pi}\cdot\boldsymbol{\pi}\right)^{2}\bar{g}_{\mu\nu}-\frac{1}{3}\kappa\pi^{2}\boldsymbol{\pi}\cdot\boldsymbol{\pi}\bar{g}_{\mu\nu}+\m O(\pi^{5})\label{eq:g_mu-nu-at-the-brane}\\
g_{\mu I} & =0\\
g_{IJ} & =\bar{g}_{IJ}+\frac{1}{3}\kappa\left(\pi_{I}\pi_{J}-\boldsymbol{\pi}\cdot\boldsymbol{\pi}\bar{g}_{IJ}\right)+\m O(\pi^{4})
\end{align}
We use the co-ordinates $x^{\mu}$ on $\Sigma$, with $\pi^{I}=\pi^{I}(x^{\mu})$
so the induced metric on $\Sigma$ is
\begin{align}
\tilde{g}_{\mu\nu} & =g_{\mu\nu}+g_{IJ}\nabla_{\mu}\pi^{I}\nabla_{\nu}\pi^{J}\nonumber \\
 & =\Omega^{2}\left(\bar{g}_{\mu\nu}+\delta g_{\mu\nu}\right)\label{eq:induced_metric}
\end{align}
where we have
\begin{align}
\Omega^{2} & =1+2\pi+\pi^{2}-\kappa\boldsymbol{\pi}\cdot\boldsymbol{\pi}-\frac{4}{3}\kappa\pi\boldsymbol{\pi}\cdot\boldsymbol{\pi}-\frac{1}{3}\kappa\pi^{2}\boldsymbol{\pi}\cdot\boldsymbol{\pi}+\frac{1}{3}\kappa^{2}\left(\boldsymbol{\pi}\cdot\boldsymbol{\pi}\right)^{2}+\m O(\pi^{5})\\
\delta g_{\mu\nu} & =\Omega^{-2}\left(\boldsymbol{\pi}_{\mu}\cdot\boldsymbol{\pi}_{\nu}+\frac{1}{3}\kappa\left(\boldsymbol{\pi}\cdot\boldsymbol{\pi}_{\mu}\boldsymbol{\pi}\cdot\boldsymbol{\pi}_{\nu}-\boldsymbol{\pi}\cdot\boldsymbol{\pi}\boldsymbol{\pi}_{\mu}\cdot\boldsymbol{\pi}_{\nu}\right)+\m O(\pi^{6})\right)\label{eq:deltag}
\end{align}
We define the scalar $\pi=c_{I}\pi^{I}$ and the vector $\boldsymbol{\pi}=\pi^{I}\boldsymbol{e}_{I}$.
Spacetime indices on these denote derivatives: $\pi_{\mu}=\nabla_{\mu}\pi$,
$\boldsymbol{\pi}_{\mu}=\nabla_{\mu}\boldsymbol{\pi}$, $\boldsymbol{\pi}_{\mu\nu}=\nabla_{\mu}\nabla_{\nu}\boldsymbol{\pi}$.

With the full metric $g_{ab}$ and the induced metric $\tilde{g}_{\mu\nu}$
we now have the ingredients we need to evaluate the possible terms
in the action as expansions in $\pi^{I}$'s.

\section{Possible Actions}

What terms can be present in the brane action? In the codimension
one case one has the Lovelock terms in the intrinsic curvature ($\Lambda,R,R_{GB}=R^{2}-4R_{\mu\nu}R^{\mu\nu}+R_{\mu\nu\rho\sigma}R^{\mu\nu\rho\sigma},\ldots$);
the Myers surface terms involving the extrinsic curvature, which correspond
to the Lovelock terms in the bulk, ($K,K_{GB}=-\frac{1}{3}K^{3}+K_{\mu\nu}K^{\mu\nu}K-\frac{2}{3}K_{\mu}^{\nu}K_{\nu}^{\rho}K_{\rho}^{\mu}-2(R_{\mu\nu}-\frac{1}{2}Rg_{\mu\nu})K^{\mu\nu},\ldots$);
and a tadpole term $\int^{\pi}\sqrt{g}dy$. For a 3-brane in a five
dimensional bulk the non-trivial terms (i.e.\ those which are not
just total derivatives) are, labelling them as in \cite{Goon:2011qf},
\begin{align}
\m L_{(1)} & =\int^{\pi}\sqrt{g}dy\\
\m L_{(2)} & =\sqrt{-\bar{g}}\\
\m L_{(3)} & =\sqrt{-\bar{g}}K\\
\m L_{(4)} & =\sqrt{-\bar{g}}R(\tilde{g})\\
\m L_{(5)} & =\sqrt{-\bar{g}}K_{GB}
\end{align}
The question is what the analogues are in the higher codimension case.
The brane is still four dimensional so as before we have the Lovelock
terms $\m L_{(2)}$ and $\m L_{(4)}$ and the higher Lovelock terms
($R_{GB},\ldots$) are total derivatives. However, since a brane with
codimension greater than one is not a boundary the analogues of $\m L_{(3)}$
and $\m L_{(5)}$ are unclear as the extrinsic curvature now has an
extra, transverse index. However for a Minkowski brane in a Minkowski
bulk, the extrinsic curvature is at least $\m O(\pi^{3})$ (see Appendix
\ref{cha:galileon-appendix}). With $SO(m)$ symmetry the only way
to contract the transverse index on the extrinsic curvature is with
another extrinsic curvature. Therefore any such terms compatible with
$SO(m)$ symmetry are in this case $\m O(\pi^{6})$ and beyond the
order we consider here. The definition of the tadpole term $\m L_{(1)}$
also becomes ambiguous since one now has freedom in choosing the integration
contour. On the other hand, presumably one could dimensionally reduce
the codimension $m$ theory to codimension one with $\left(m-1\right)$
of the galileons becoming very massive. In that case one would have
to recover all the terms $\m L_{(1)}$ to $\m L_{(5)}$: analogues
of all these terms should therefore exist. However, such a dimensional
reduction would obviously have to break the $SO(m)$ symmetry in picking
out a single large dimension and so it is perhaps not too surprising
that when working within this restriction it is not clear how $\m L_{(1)}$,
$\m L_{(3)}$ and $\m L_{(5)}$ could appear. The general Lagrangian
respecting $SO(m)$ symmetry is, up to this order, the sum with arbitrary
coefficients of $\m L_{(2)}$ and $\m L_{(4)}$ which are worked out
in detail for the codimension greater than one case below.

\subsection{Volume Term $\m L_{(2)}$}

The volume term is
\begin{align}
\m L_{(2)} & =\sqrt{-\tilde{g}}
\end{align}

Now if $\hat{g}_{\mu\nu}=\bar{g}_{\mu\nu}+\delta g_{\mu\nu}$ then
\begin{align}
\sqrt{-\hat{g}} & =\sqrt{-\bar{g}}\left(1+\frac{1}{2}\bar{g}^{\mu\nu}\delta g_{\mu\nu}-\frac{1}{8}\left(\bar{g}^{\alpha\gamma}\bar{g}^{\beta\delta}+\bar{g}^{\alpha\delta}\bar{g}^{\beta\gamma}-\bar{g}^{\alpha\beta}\bar{g}^{\gamma\delta}\right)\delta g_{\alpha\beta}\delta g_{\gamma\delta}+\m O(\delta g^{3})\right)
\end{align}
so using \eqref{eq:deltag} and \eqref{eq:Omega^2n} we can expand
$\m L_{(2)}$ in terms of the $\pi$'s as 
\begin{align}
\m L_{(2)} & =\Omega^{4}\sqrt{-\hat{g}}\nonumber \\
 & =\sqrt{-\bar{g}}\bigg(1+4\pi+6\pi^{2}+\frac{1}{2}\boldsymbol{\pi}_{\mu}\cdot\boldsymbol{\pi}^{\mu}-2\kappa\boldsymbol{\pi}\cdot\boldsymbol{\pi}+4\pi^{3}+\pi\boldsymbol{\pi}_{\mu}\cdot\boldsymbol{\pi}^{\mu}-\frac{20}{3}\kappa\pi\boldsymbol{\pi}\cdot\boldsymbol{\pi}\nonumber \\
 & \qquad+\pi^{4}+\frac{1}{2}\pi^{2}\boldsymbol{\pi}_{\mu}\cdot\boldsymbol{\pi}^{\mu}-\frac{1}{4}\boldsymbol{\pi}_{\mu}\cdot\boldsymbol{\pi}_{\nu}\boldsymbol{\pi}^{\mu}\cdot\boldsymbol{\pi}^{\nu}+\frac{1}{8}\left(\boldsymbol{\pi}_{\mu}\cdot\boldsymbol{\pi}^{\mu}\right)^{2}-8\kappa\pi^{2}\boldsymbol{\pi}\cdot\boldsymbol{\pi}+\nonumber \\
 & \qquad\frac{1}{6}\kappa\boldsymbol{\pi}\cdot\boldsymbol{\pi}_{\mu}\boldsymbol{\pi}\cdot\boldsymbol{\pi}^{\mu}-\frac{2}{3}\kappa\boldsymbol{\pi}\cdot\boldsymbol{\pi}\boldsymbol{\pi}_{\mu}\cdot\boldsymbol{\pi}^{\mu}+\frac{5}{3}\kappa^{2}\left(\boldsymbol{\pi}\cdot\boldsymbol{\pi}\right)^{2}+\m O(\pi^{5})\bigg)\label{eq:L_2}
\end{align}

\subsection{Einstein-Hilbert Term $\m L_{(4)}$}

The Einstein-Hilbert term is

\begin{align}
\m L_{(4)} & =\sqrt{-\tilde{g}}R(\tilde{g}_{\mu\nu})
\end{align}

The Ricci scalar for a conformally transformed metric is
\begin{align}
R(\Omega^{2}\hat{g}_{\mu\nu}) & =\Omega^{-2}R(\hat{g}_{\mu\nu})-6\Omega^{-3}\hat{\nabla}^{2}\Omega
\end{align}
in four dimensions, where $\hat{\nabla}$ is the derivative covariant
with respect to $\hat{g}_{\mu\nu}$.

For small perturbations $\delta g_{\mu\nu}$ about a background metric
$\bar{g}_{\mu\nu}$, dropping total derivatives and up to second order
in the perturbation,
\begin{align}
\sqrt{-\hat{g}}R(\hat{g}) & =\sqrt{\bar{g}}\left(R(\bar{g})-G(\bar{g})^{\mu\nu}\delta g_{\mu\nu}-\frac{1}{2}\delta g_{\mu\nu}\Delta_{L}^{(\mu\nu)(\rho\sigma)}\delta g_{\rho\sigma}+\bar{g}^{\mu\nu}\mathfrak{F}_{\mu}\mathfrak{F}_{\nu}\right)\label{eq:R-expansion}
\end{align}
 where $G(\bar{g})_{\mu\nu}=R(\bar{g})_{\mu\nu}-\frac{1}{2}\bar{g}_{\mu\nu}R(\bar{g})$
is the Einstein tensor and $\mathfrak{F}$ is given by
\begin{align}
\mathfrak{F}_{\mu} & =\bar{g}^{\rho\sigma}\left(\nabla_{\sigma}\delta g_{\mu\rho}-\frac{1}{2}\nabla_{\mu}\delta g_{\rho\sigma}\right)
\end{align}
and the Lichnerowicz operator $\Delta_{L}$ is
\begin{align}
\Delta_{L}^{(\mu\nu)(\rho\sigma)} & =-g^{(\mu\nu)(\rho\sigma)}\nabla^{2}+g^{(\mu\nu)(\tau\upsilon)}\left(?{\t{R(\bar{g})}}_{\tau}^{\rho}{}_{\upsilon}^{\sigma}?+?{\t{G(\bar{g})}}_{\tau}^{\rho}?\delta_{\upsilon}^{\sigma}\right)
\end{align}
with the derivatives being covariant with respect to $\bar{g}_{\mu\nu}$
and defining $g^{(\mu\nu)(\rho\sigma)}=\frac{1}{2}\left(\bar{g}^{\mu\rho}\bar{g}^{\nu\sigma}+\bar{g}^{\mu\sigma}\bar{g}^{\nu\rho}-\bar{g}^{\mu\nu}\bar{g}^{\rho\sigma}\right)$.

As $\Sigma_{0}$ is maximally symmetric we have that $R(\bar{g})_{\mu\nu\rho\sigma}=\bar{\kappa}\left(\bar{g}_{\mu\rho}\bar{g}_{\nu\sigma}-\bar{g}_{\mu\sigma}\bar{g}_{\nu\rho}\right)$.
(Using the relations \eqref{eq:k_Imunu} and \eqref{eq:R(g)} it is
easy to see that $\bar{\kappa}$ is related to $\kappa$ and $c_{I}$
by $\bar{\kappa}=\kappa+c_{I}c^{I}$.) This implies that also $G(\bar{g})_{\mu\nu}=-3\bar{\kappa}\bar{g}_{\mu\nu}$
and $R(\bar{g})=12\bar{\kappa}$. These relations along with \eqref{eq:deltag}
allow us to expand the terms in \eqref{eq:R-expansion}, (given in
equations (\ref{eq:R-expansion-1}-\ref{eq:R-expansion-5})) and find
\begin{align}
\m L_{(4)} & =\sqrt{-\bar{g}}\bigg(12\bar{\kappa}\Omega^{2}+3\bar{\kappa}\boldsymbol{\pi}_{\mu}\cdot\boldsymbol{\pi}^{\mu}+6\pi_{\mu}\pi^{\mu}-12\kappa\pi_{\mu}\boldsymbol{\pi}\cdot\boldsymbol{\pi}^{\mu}\nonumber \\
 & \qquad\qquad+\kappa\bar{\kappa}\left(\boldsymbol{\pi}\cdot\boldsymbol{\pi}_{\mu}\boldsymbol{\pi}\cdot\boldsymbol{\pi}^{\mu}-\boldsymbol{\pi}\cdot\boldsymbol{\pi}\boldsymbol{\pi}_{\mu}\cdot\boldsymbol{\pi}^{\mu}\right)+12\kappa^{2}\boldsymbol{\pi}\cdot\boldsymbol{\pi}_{\mu}\boldsymbol{\pi}\cdot\boldsymbol{\pi}^{\mu}\nonumber \\
 & \qquad\qquad+4\bar{\kappa}\boldsymbol{\pi}_{\mu}\cdot\boldsymbol{\pi}_{\nu}\boldsymbol{\pi}^{\mu}\cdot\boldsymbol{\pi}^{\nu}-\bar{\kappa}\left(\boldsymbol{\pi}_{\mu}\cdot\boldsymbol{\pi}^{\mu}\right)^{2}-6\pi_{\mu}\pi_{\nu}\boldsymbol{\pi}^{\mu}\cdot\boldsymbol{\pi}^{\nu}\nonumber \\
 & \qquad\qquad+3\pi_{\mu}\pi^{\mu}\boldsymbol{\pi}_{\nu}\cdot\boldsymbol{\pi}^{\nu}-\boldsymbol{\pi}_{\mu}\cdot\boldsymbol{\pi}_{\nu\rho}\boldsymbol{\pi}^{\mu}\cdot\boldsymbol{\pi}^{\nu\rho}+\boldsymbol{\pi}_{\mu}\cdot?{\t{\boldsymbol{\pi}}}_{\nu}^{\nu}?\boldsymbol{\pi}^{\mu}?{\t{\boldsymbol{\pi}}}_{\rho}^{\rho}?\nonumber \\
 & \qquad\qquad-\boldsymbol{\pi}_{\mu}\cdot\boldsymbol{\pi}_{\nu\rho}\boldsymbol{\pi}^{\nu}\cdot\boldsymbol{\pi}^{\mu\rho}+\boldsymbol{\pi}^{\mu}\cdot\boldsymbol{\pi}_{\mu\rho}\boldsymbol{\pi}^{\nu}\cdot?{\t{\boldsymbol{\pi}}}_{\nu}^{\rho}?+\m O(\pi^{5})\bigg)\label{eq:L_4}
\end{align}

\subsection{Tadpole Term $\m L_{(1)}$}

Unbroken $SO(m)$ in the transverse directions restricts the candidates
for the tadpole term to just the volume of the surface, $S$, given
by the locus of the geodesics connecting $\Sigma_{0}$ and $\Sigma$,
whose co-ordinates in the transverse dimensions are $x^{I}=s\pi^{I}$,
$s\in\left[0,1\right]$. Clearly we can obtain the induced metric
on the slice at each value of $s$ just by scaling \eqref{eq:L_2}
by a factor of $s$ for each $\pi$ while the component of the metric
along the $s$ direction is given by $g_{ss}ds^{2}=g_{IJ}dx^{I}dx^{J}=ds^{2}\boldsymbol{\pi}\cdot\boldsymbol{\pi}$.
So the volume element on $S$ is $\sqrt{\tilde{g}(s)\cdot g_{ss}}=\sqrt{\boldsymbol{\pi}\cdot\boldsymbol{\pi}}\sqrt{\tilde{g}(s)}$
and
\begin{align}
\mathcal{L}_{(1)} & =\int_{0}^{1}ds\sqrt{-\tilde{g}(s)\cdot g_{ss}}\nonumber \\
 & =\sqrt{\bar{g}}\sqrt{\boldsymbol{\pi}\cdot\boldsymbol{\pi}}\bigg(1+2\pi+2\pi^{2}+\frac{1}{6}\boldsymbol{\pi}_{\mu}\cdot\boldsymbol{\pi}^{\mu}-\frac{2}{3}\kappa\boldsymbol{\pi}\cdot\boldsymbol{\pi}\nonumber \\
 & \qquad\qquad\qquad+\pi^{3}+\frac{1}{4}\pi\boldsymbol{\pi}_{\mu}\cdot\boldsymbol{\pi}^{\mu}-\frac{5}{3}\kappa\pi\boldsymbol{\pi}\cdot\boldsymbol{\pi}\bigg)+\m O(\pi^{5})\label{eq:Tadpole}
\end{align}
It seems likely that this is not in fact the correct prescription
for the tadpole term, which is probably simply absent in the $SO(m)$
symmetric case, as seems to be suggested by the field theory \cite{Padilla:2010ir}.
If \eqref{eq:Tadpole} were to be the correct form, the interpretation
of $\sqrt{\boldsymbol{\pi}\cdot\boldsymbol{\pi}}$ would present a
puzzle, or perhaps suggest new possibilities for the field theory
galileons.

If instead the $SO(m)$ symmetry were broken so that a constant vector
field, $V^{I}$, were available somehow, one could integrate along
it from $\infty$ to the brane and have something like $\m L_{(1)}=\sqrt{\bar{g}}V_{I}\pi^{I}\left(1+\ldots\right)$
which would be rather more like the field theory tadpole term $\alpha_{I}\pi^{I}$.

\section{Conclusions\label{sec:Galileons_Conclusions}}

The results presented in this chapter, \eqref{eq:L_2} and \eqref{eq:L_4},
are only slightly more general than those which have been given before.
We obtained these results by working perturbatively from the outset
so that the geometrical quantities on the brane could be constructed
straightforwardly as Taylor expansions. This contrasts to the approach
in \cite{Hinterbichler:2010xn} which was rather closer to the method
used in the codimension one case. There, with the assumption that
both brane and bulk are Minkowski, the induced curvature of the brane
was calculated directly from the induced metric and its inverse. In
this simpler case the only thing without an exact expression (and
therefore requiring a perturbative expansion) is the inverse induced
metric of the brane, which makes the direct approach feasible. However
as we can see in equations (\ref{eq:g_mu-nu-at-the-brane}-\ref{eq:induced_metric})
the metric has a trivial exact expression only for this very restricted
case with no bulk curvature and no extrinsic curvature (i.e.\ $\kappa=0$
and $c_{I}=0$) which is lost even for the rather modest relaxation
which we allow here.

We hope that techniques such as those we have introduced here may
prove to be of further use in constructing models in which the $SO(m)$
symmetry of the bulk is broken, which might thereby describe all of
the models allowed in the field theory approach.

\chapter{Conclusions}

In this thesis we have considered several topics concerning different
aspects of braneworld models, in which the structure at boundaries
of gravitational theories plays a crucial role.

We have constructed a five dimensional reduction of Ian Moss's improved
version of Heterotic M-Theory \cite{Moss:2003bk,Moss:2004ck,Moss:2005zw,Moss:2008ng},
whose action, boundary conditions and supersymmetry transformations
are summarized in Appendix \ref{cha:Summary}. Since this formulation
of the eleven dimensional theory is consistent to all orders in the
gravitational coupling, it imposes fewer restrictions on model building
than the original version due to Ho\v{r}ava and Witten \cite{Horava:1995qa,Horava:1996ma}.
This five dimensional reduction provides a starting point for the
development of the phenomenology of the improved theory. Most of the
previous phenomenological work concerning the bosonic sector of the
theory, which called only for terms of first order in $\kappa^{2/3}$,
carries over unchanged to the improved theory. However, it is now
possible to consider the effects of higher order terms consistently
and so this reduction opens up a new avenue of research in the phenomenology
of Heterotic M-Theory. This may include qualitatively new features
of the improved theory, such as the appearance of the gaugino condensate
in the boundary condition of the gravitino discovered by Ahmed and
Moss \cite{Ahmed:2008jz,Ahmed:2009ty} which produces a positive Casimir
energy.

We have examined singularity formation in brane collisions. The formation
of a singularity when two scalar field domain walls collide had been
observed before \cite{Takamizu:2007ks} but it had been assumed that
the singularity was a black brane. We observe that in fact the asymptotic
structure is not consistent with that of a black brane, and further
investigation through numerical simulations showed that the trapping
horizon formed was a future-inner horizon, not the future-outer horizon
of a black brane. This horizon structure suggests rather a big crunch
singularity; tentative confirmation of this conclusion comes from
the level sets of the Ricci scalar which, up to the limits imposed
by the finite size of our simulations, seem to be consistent with
the hypothesis that they cross all the outgoing null rays so that
the spacetime is indeed cut off by the singularity (which would be
the level set $R=\infty$). The consequence of this is that the universes
on the branes must also end in a finite time; they cannot persist
outside the horizon of the singularity forever, as the could perhaps
if it were a black brane. To further examine the implications of this
big crunch for cosmological models it would be useful to continue
the simulations to map out the parameter space (in the approach velocity
of the domain walls) where the singularity forms. Having established
the nature of the singularity one could also pursue further the investigation
of topics such as particle exchange and reheating in gravitating brane
collisions, presumably while taking care in this case to avoid singularity
formation.

We have developed a formalism for the calculation of the geometrical
quantities associated with a probe brane of arbitrary codimension.
This allows us to recover previous results for the action of multi-galileon
theories which arise as the perturbations of a flat probe brane in
a flat bulk but more importantly it provides a tool-kit for the construction
of more general multi-galileon theories. However, another ingredient
is also needed: that is, how to break the symmetry of the bulk spacetime
so that we can have preferred directions and thereby break the $SO(m)$
symmetry between the several galileon fields, which is a necessary
consequence of having a maximally symmetric bulk and brane. The specification
of a breaking which would allow the recovery of the most general galileon
theories, as found in the four dimensional effective field theory
approach, remains to be discovered.

The project to find extra dimensional models relevant to the description
of the real world is an ongoing one which still presents many challenges.
We have described here some results in such models which move them
forward a step or two towards the ultimate goal of representing the
universe in which we live and which, as discussed above, also suggest
directions for further inquiry.

\appendix

\chapter{\label{cha:Conventions}Heterotic M-Theory Conventions}

\section*{Eleven Dimensions}
\begin{itemize}
\item $\m M_{11}$ denotes the bulk spacetime with boundaries $\partial\m M_{10,1}$
(with inward-pointing normal) and $\partial\m M_{10,2}$ (with outward
pointing normal).
\item Indices:

\begin{itemize}
\item $I,J,K,\ldots=0,\ldots,10$ are the bulk spacetime indices.
\item $A,B,C,\ldots=0,\ldots,9$ are the boundary spacetime indices.
\item $\hat{I},\hat{J},\hat{K},\ldots$ and $\hat{A},\hat{B},\hat{C},\ldots$
are the corresponding tangent space indices.
\item $N$ is the direction normal to the boundaries (i.e.\ $V^{N}=g_{IJ}^{(11)}V^{I}n_{\left(11\right)}^{J}$
where $n_{\left(11\right)}^{J}$ is the unit normal vector).
\end{itemize}
\item The metric is mostly plus.
\item The volume element is $dv=\sqrt{-g^{(11)}}d^{11}x$.
\item The Riemann tensor is $?{R(\Gamma)}^{I}{}_{JKL}?=\partial_{K}?\Gamma^{I}{}_{LJ}?-\partial_{L}?\Gamma^{I}{}_{KJ}?+?\Gamma^{I}{}_{KM}??\Gamma^{M}{}_{LJ}?-?\Gamma^{I}{}_{LM}??\Gamma^{M}{}_{KJ}?$
or $R(\omega)_{\hat{I}\hat{J}KL}=\partial_{K}\omega_{L\hat{I}\hat{J}}-\partial_{L}\omega_{K\hat{I}\hat{J}}+?\omega_{K\hat{I}}^{\hat{M}}?\omega_{L\hat{M}\hat{J}}-?\omega_{L\hat{I}}^{\hat{M}}?\omega_{K\hat{M}\hat{J}}$,
the Ricci tensor is $R_{IJ}=?R^{K}{}_{IKJ}?$ and the Ricci scalar
is $g^{(11)IJ}R_{IJ}$.
\item The trace over $SO(1,9)$ indices of the curvature two-form is given
by a standard matrix trace so for example $\tr(R_{AB}R_{CD})=?R_{ABE}^{F}??R_{CDF}^{E}?$.
\item The $\Gamma_{A}$ are imaginary, $\Gamma_{A}^{*}=-\Gamma_{A}$, while
$\Gamma_{N}$ is real, $\Gamma_{N}^{*}=\Gamma_{N}$, and satisfy $\left\{ \Gamma_{I},\Gamma_{J}\right\} =2g_{IJ}^{(11)}$
and $\Gamma_{IJ}=\Gamma_{[I}\Gamma_{J]}$, etc.
\item The Dirac conjugate is $\bar{\Psi}=-i\Psi^{\dagger}\Gamma_{0}$.
\item Fermionic fields swap places under complex conjugation, i.e.\ $\left(\psi\chi\right)^{*}=\chi^{*}\psi^{*}$.
\item $\pm$ and $\mp$ refer to the signs of boundary terms, with the upper
sign referring to $\partial\m M_{10,1}$ and the lower to $\partial\m M_{10,2}$.
\end{itemize}

\section*{Five Dimensions}
\begin{itemize}
\item $\m M$ denotes the bulk spacetime with boundaries $\partial\m M_{1}$
(with inward-pointing normal) and $\partial\m M_{2}$ (with outward
pointing normal).
\item Indices:

\begin{itemize}
\item $\alpha,\beta,\gamma,\ldots=0,\ldots,4$ are the bulk spacetime indices.
\item $\mu,\nu,\rho,\ldots=0,\ldots,3$ are the boundary spacetime indices.
\item $\hat{\alpha},\hat{\beta},\hat{\gamma},\ldots$ and $\hat{\mu},\hat{\nu},\hat{\rho},\ldots$
are the corresponding tangent space indices.
\item $z$ is the direction normal to the boundaries (i.e.\ $V^{z}=g_{\alpha\beta}V^{\alpha}n^{\beta}$).
\item $I,J,K,\ldots$ are $E_{6}$ gauge indices on $\partial\m M_{1}$
and $E_{8}$ gauge indices on $\partial\m M_{2}$.
\item $A,B,C,\ldots=1,2$ are $SU(2)$ indices.
\end{itemize}
\item The metric is mostly plus.
\item The volume element is $dv=\sqrt{-g}d^{5}x$.
\item The Riemann tensor is $?{R(\Gamma)}^{\alpha}{}_{\beta\gamma\delta}?=\partial_{\gamma}?\Gamma^{\alpha}{}_{\delta\beta}?-\partial_{\delta}?\Gamma^{\alpha}{}_{\gamma\beta}?+?\Gamma^{\alpha}{}_{\gamma\epsilon}??\Gamma^{\epsilon}{}_{\delta\beta}?-?\Gamma^{\alpha}{}_{\delta\epsilon}??\Gamma^{\epsilon}{}_{\gamma\beta}?$
or $R(\omega)_{\hat{\alpha}\hat{\beta}\gamma\delta}=\partial_{\gamma}\omega_{\delta\hat{\alpha}\hat{\beta}}-\partial_{\delta}\omega_{\gamma\hat{\alpha}\hat{\beta}}+?\omega_{\gamma\hat{\alpha}}^{\hat{\epsilon}}?\omega_{\delta\hat{\epsilon}\hat{\beta}}-?\omega_{\delta\hat{\alpha}}^{\hat{\epsilon}}?\omega_{\gamma\hat{\epsilon}\hat{\beta}}$,
the Ricci tensor is $R_{\alpha\beta}=?R^{\gamma}{}_{\alpha\gamma\beta}?$
and the Ricci scalar is $g^{\alpha\beta}R_{\alpha\beta}$.
\item The alternating tensor $\epsilon_{\hat{\alpha}\hat{\beta}\hat{\gamma}\hat{\delta}}$
is normalized so that $\epsilon_{0\ldots4}=1,\epsilon^{0\ldots4}=-1$
and $\epsilon_{\mu\nu\rho\sigma}=\epsilon_{\mu\nu\rho\sigma5}$.
\item The gamma matrices satisfy $\left\{ \gamma_{\alpha},\gamma_{\beta}\right\} =2g_{\alpha\beta}$
and $\gamma_{\alpha\beta}=\gamma_{[\alpha}\gamma_{\beta]}$, etc.
The $\gamma_{\mu}$ are real, $\gamma_{\mu}^{*}=\gamma_{\mu}$ and
$\gamma_{5}\equiv\gamma_{z}$ is imaginary $\gamma_{5}^{*}=-\gamma_{5}$.
\item The Dirac conjugate is $\bar{\psi}=-i\psi^{\dagger}\gamma_{0}$.
\item Fermionic fields swap places under complex conjugation, i.e.\ $\left(\psi\chi\right)^{*}=\chi^{*}\psi^{*}$.
\item $\pm$ and $\mp$ refer to the signs of boundary terms, with the upper
sign referring to $\partial\m M_{1}$ and the lower to $\partial\m M_{2}$.
\end{itemize}

\section*{Calabi-Yau Manifold}
\begin{itemize}
\item $X$ denotes the Calabi-Yau space over which the reduction is performed,
so $\m M_{11}=\m M_{5}\times X$, $\partial\m M_{10,1}=\partial\m M_{1}\times X$
and $\partial\m M_{10,2}=\partial\m M_{2}\times X$.
\item Indices:

\begin{itemize}
\item $\underline{a},\underline{b},\underline{c},\ldots=\underline{1},\ldots,\underline{6}$
are real indices.
\item $a,b,c,\ldots=1,2,3$ are holomorphic indices, related to the real
indices by $V^{1}=V_{\underline{1}}+iV_{\underline{2}}$, $V_{1}=\frac{1}{2}\left(V_{\underline{1}}-iV_{\underline{2}}\right)$,
etc.
\item $\bar{a},\bar{b},\bar{c},\ldots=\bar{1},\bar{2},\bar{3}$ are anti-holomorphic
indices on the Calabi-Yau space, related to the real indices by $V^{\bar{1}}=V_{\underline{1}}-iV_{\underline{2}}$,
$V_{\bar{1}}=\frac{1}{2}\left(V_{\underline{1}}+iV_{\underline{2}}\right)$,
etc.
\end{itemize}
\item The metric is Euclidean.
\item The volume element is $dv=\sqrt{g^{CY}}d^{6}x$.
\item The reference volume of $X$ after scaling out the geometric moduli
is $v=\int_{X}dv$.
\item The alternating tensor $\epsilon_{abc}$ is normalized so that $\epsilon_{123}=1,\epsilon_{abc}\epsilon^{abc}=48$.
\item The gamma matrices satisfy $\left\{ \gamma_{a},\gamma_{\bar{b}}\right\} =2g_{a\bar{b}}$
and $\gamma_{a\bar{b}}=\gamma_{[a}\gamma_{\bar{b}]}$, etc. The $\gamma_{a}$
are imaginary $\left(\gamma_{a}\right)^{*}=-\gamma_{\bar{a}}$ and
$\gamma_{7}^{*}=-\gamma_{7}$.
\item The Hermitian conjugate of the covariantly constant spinor $u^{A}$
is defined to be $\left(u^{A}\right)^{\dagger}=\bar{u}_{A}$ and then
the Dirac conjugate is given by $\overline{\left(u^{A}\right)}=\bar{u}_{A}\gamma_{7}$\end{itemize}

\chapter{Calabi-Yau Geometry\label{cha:Calabi-Yau_Geometry}}

Useful identities for spinors and gamma matrices on the Calabi-Yau
follow from the fact that there is a unique constant spinor, $u_{A}$;
there is no constant vector, so $\bar{u}^{A}\gamma_{a}u_{A}=0$; there
is a unique holomorphic three-form, so $\bar{u}^{A}\gamma_{abc}u_{A}\propto\Omega_{abc}=i\epsilon_{abc}$;
and the Dirac algebra $\left\{ \gamma_{a},\gamma_{\bar{b}}\right\} =2g_{a\bar{b}}$.
We choose the spinor basis so that $\gamma^{a}u_{1}=0$, and as $u_{1}^{*}=u_{2}$
it follows that
\begin{align}
\gamma^{a}u_{1} & =0\\
\gamma^{\bar{a}}u_{2} & =0\\
\gamma^{a\bar{b}}u_{A} & =g^{a\bar{b}}?\tau^{B}{}_{A}?u_{B}\\
\gamma_{abc}u_{1} & =i\epsilon_{abc}u_{2}\\
\gamma_{\bar{a}\bar{b}\bar{c}}u_{2} & =i\epsilon_{\bar{a}\bar{b}\bar{c}}u_{1}
\end{align}
where we have the matrix $?\tau^{A}{}_{B}?=?{\left(\begin{array}{cc}
1 & 0\\
0 & -1
\end{array}\right)}^{A}{}_{B}?$. Tensors formed from arbitrary products of gamma matrices using the
constant spinor, such as those that appear in all the two fermion
terms, are given by some combination of $g_{a\bar{b}}$, $\epsilon_{abc}$
and $\epsilon_{\bar{a}\bar{b}\bar{c}}$ by using these identities
and the Dirac algebra.

The following results concerning the geometry of the (1,1) moduli
of the Calabi-Yau space are based on \cite{Candelas:1990pi}.

The components of the complex structure $\omega$ and the metric $g$
are related by
\begin{align}
\omega_{a\bar{b}} & =ig_{a\bar{b}}
\end{align}
and the reference volume of the Calabi-Yau (the unit in which $V$
is measured) is
\begin{align}
v & =\frac{1}{6}\int\omega\wedge\omega\wedge\omega\label{eq:v}
\end{align}
Given a basis $\omega_{ia\bar{b}}$ of the cohomology group $H^{1,1}$
the complex structure is given by a set of real moduli fields $b^{i}$
\begin{align}
\omega_{a\bar{b}} & =b^{i}\omega_{ia\bar{b}}\label{eq:Kahler_form}
\end{align}
The metric on the moduli space is
\begin{align}
G_{ij} & =\frac{1}{2v}\int\omega_{i}\wedge*\omega_{j}\label{eq:G_ij}
\end{align}
The dual basis of $H^{2,2}$, $\nu^{i}$, is defined by
\begin{align}
\int\omega_{i}\wedge\nu^{j} & =v\delta_{i}^{j}
\end{align}
and we can see by comparing this to \eqref{eq:G_ij} that $\nu_{a\bar{b}c\bar{d}}^{i}=\frac{1}{2}G^{ij}\left(*\omega\right)_{a\bar{b}c\bar{d}}$

The intersection numbers are
\begin{align}
\m K_{ijk} & =\int\omega_{i}\wedge\omega_{j}\wedge\omega_{k}\label{eq:K_ijk}
\end{align}
and it is useful to define
\begin{align}
\m K & =\m K_{ijk}b^{i}b^{j}b^{k}=6v
\end{align}
From \eqref{eq:v} and \eqref{eq:G_ij} we may find that
\begin{align}
*\omega_{i} & =b_{i}\omega\wedge\omega-\omega\wedge\omega_{i}\label{eq:star_omega_i}\\
G_{ij} & =\frac{1}{2}\omega_{ia\bar{b}}?\omega_{j}^{a\bar{b}}?\label{eq:G_ij2}
\end{align}
and using (\ref{eq:G_ij}-\ref{eq:star_omega_i})
\begin{align}
G_{ij} & =-\frac{1}{2}\frac{\partial\ln\m K}{\partial b^{i}\partial b^{j}}\label{eq:G_ij3}
\end{align}
The contractions of $\m K_{ijk}$ are
\begin{align}
\m K^{-1}\m K_{ijk}b^{k} & =\frac{2}{3}b_{i}b_{j}-\frac{1}{3}G_{ij}\\
\m K^{-1}\m K_{ijk}b^{j}b^{k} & =\frac{2}{3}b_{i}
\end{align}
and
\begin{align}
b_{i}b^{i} & =\frac{3}{2}
\end{align}
From \eqref{eq:Kahler_form} and \eqref{eq:G_ij2} we may see that
the connection coefficients
\begin{align}
\Gamma_{ijk} & =-\frac{i}{2}?\omega_{ia}^{b}??\omega_{jb}^{c}??\omega_{kc}^{a}?
\end{align}
can be used to define the covariant derivative on fields with $i,j,k,\ldots$
indices
\begin{align}
\m D_{\alpha}C^{i} & =\nabla_{\alpha}C^{i}+\partial_{\alpha}b^{k}?\Gamma^{i}{}_{jk}?C^{j}
\end{align}
which is compatible with the metric $G_{ij}$, $\m D_{\alpha}G_{ij}=0$.

The Levi-Civita components follow from \eqref{eq:G_ij3}
\begin{align}
\Gamma_{i\left(jk\right)}= & -\frac{3}{2}\m K^{-1}\m K_{ijk}-3b_{(i}G_{jk)}+2b_{i}b_{j}b_{k}
\end{align}
with a corresponding curvature tensor
\begin{align}
R_{ijkl} & =\frac{9}{4}\m K^{-2}?{\m K}_{il}^{m}?\m K_{kjm}-\frac{9}{4}\m K^{-2}?{\m K}_{ik}^{m}?\m K_{ljm}+\frac{1}{2}G_{il}G_{kj}-\frac{1}{2}G_{ik}G_{lj}
\end{align}

It is useful to decompose the tangent space to the moduli into a direction
parallel to $b^{i}$ and the perpendicular directions, which we may
do using the projection tensor
\begin{align}
{\delta^{\perp}}_{j}^{i} & =\delta_{j}^{i}-\frac{2}{3}b^{i}b_{j}
\end{align}
and, extending the notation to $G$ and $\m K$, 
\begin{align}
G_{ij}^{\perp} & =G_{ij}-\frac{2}{3}b_{i}b_{j}\\
G^{\perp ij} & =G^{ij}-\frac{2}{3}b^{i}b^{j}\\
\m K^{-1}\m K_{ijk}^{\perp} & =\m K^{-1}\m K_{ijk}+\frac{2}{3}b_{(i}G_{jk)}-\frac{20}{27}b_{i}b_{j}b_{k}
\end{align}

Cyclic contractions of the cohomology generators appear frequently
in the calculations of the reduction. A collection of useful identities
for these are
\begin{align}
?\omega_{ia}^{a}? & =2ib_{i}\\
?\omega_{ia}^{b}??\omega_{jb}^{a}? & =-2G_{ij}\\
?\omega_{ia}^{b}??\omega_{jb}^{c}??\omega_{kc}^{a}? & =2i\Gamma_{ijk}\\
?\omega_{ia}^{[d}??\omega_{jb}^{e}??\omega_{kc}^{f]}? & =-\frac{i}{48}\epsilon_{abc}\epsilon^{def}\m K^{-1}\m K_{ijk}\\
\frac{1}{2}\left(?\omega_{ia}^{b}??\omega_{jb}^{c}??\omega_{kc}^{a}?+?\omega_{ja}^{b}??\omega_{ib}^{c}??\omega_{kc}^{a}?\right) & =-3i\m K^{-1}\m K_{ijk}-6ib_{(i}G_{jk)}+4ib_{i}b_{j}b_{k}\\
\frac{1}{2}\left(?\omega_{ia}^{b}??\omega_{jb}^{c}??\omega_{kc}^{d}??\omega_{ld}^{a}?+?\omega_{ka}^{b}??\omega_{jb}^{c}??\omega_{ic}^{d}??\omega_{ld}^{a}?\right) & =-\frac{9}{2}\m K^{-2}\m K_{ikm}?{\m K}_{jl}^{m}?+G_{il}G_{jk}+G_{ij}G_{kl}
\end{align}

\chapter{$E_{8}$ Gauge Theory\label{cha:E8-Gauge-Theory}}

Our reference for the gauge theory was \cite{Cvitanovic:2008zz}.
We include here the particulars needed above for ease of reference
and to make clear the normalizations used.

We need to split the (imaginary) generators of $E_{8}$, $Y^{A}$,
into: $S^{i}$ which generate $SU(3)$; $X^{I}$ which generate $E_{6}$;
$T_{\hat{a}p}$ and $T^{\hat{a}p}=\left(T_{\hat{a}p}\right)^{\dagger}$
which generate the cosets. This corresponds to $\mathbf{248}=\left(\mathbf{8},\mathbf{1}\right)\oplus\left(\mathbf{1},\mathbf{78}\right)\oplus\left(\mathbf{3},\mathbf{27}\right)\oplus\left(\mathbf{\bar{3}},\overline{\mathbf{27}}\right)$.
So the $S^{i}$ and $X^{I}$ form adjoint representations of $SU\left(3\right)$
and $E_{6}$ respectively, i.e.
\begin{eqnarray}
 & \left[S^{i},S^{j}\right]=i?f^{ij}{}_{k}?S^{k};\;?{\left(S^{i}\right)}^{j}{}_{k}?=-i?f^{ij}{}_{k}?\\
 & \left[X^{I},X^{J}\right]=i?F^{IJ}{}_{K}?X^{K};\;?{\left(X^{I}\right)}^{J}{}_{K}?=-i?F^{IJ}{}_{K}?
\end{eqnarray}
where $?f^{ij}{}_{k}?$ and $?F^{IJ}{}_{K}?$ are the (real) structure
constants of $SU(3)$ and $E_{6}$. The $T_{ap}$ transform as the
fundamental both of $SU(3)$ and of $E_{6}$
\begin{eqnarray}
 & \left[S^{i},T_{\hat{a}p}\right]=-?\lambda^{i}{}_{\hat{a}}^{\hat{b}}?T_{\hat{b}p};\;\left[S^{i},T^{\hat{a}p}\right]=?\lambda^{i\hat{a}}{}_{\hat{b}}?T^{\hat{b}p}\\
 & \left[X^{I},T_{\hat{a}p}\right]=-?\Lambda^{I}{}_{p}^{q}?T_{\hat{a}q};\;\left[X^{I},T^{\hat{a}p}\right]=?\Lambda^{Ip}{}_{q}?T^{\hat{a}q}
\end{eqnarray}
where $?\lambda^{i\hat{a}}{}_{\hat{b}}?$ and $?\Lambda^{Ip}{}_{q}?$
are the fundamental generators of $SU(3)$ and $E_{6}$ respectively,
and of course the $SU(3)$ and $E_{6}$ subgroups commute
\begin{equation}
\left[S^{i},X^{I}\right]=0
\end{equation}
We normalize the antisymmetric invariant tensor of $SU(3)$ by
\begin{equation}
\hat{\epsilon}_{123}=1\text{, or }\hat{\epsilon}_{\hat{a}\hat{b}\hat{c}}\hat{\epsilon}^{\hat{a}\hat{b}\hat{c}}=6
\end{equation}
the symmetric invariant tensor of $E_{6}$ by
\begin{equation}
d_{prs}d^{qrs}=\delta_{p}^{q}
\end{equation}
the generators by
\begin{align}
\mathrm{Tr}\left(S^{i}S^{j}\right) & =30\delta^{ij}\\
\mathrm{Tr}\left(X^{I}X^{J}\right) & =30\delta^{IJ}\\
\mathrm{Tr}\left(T^{\hat{a}p}T_{\hat{b}q}\right) & =30\delta_{\hat{b}}^{\hat{a}}\delta_{q}^{p}
\end{align}
and the fundamental generators by
\begin{align}
?{\left(\lambda^{i}\right)}^{\hat{a}}{}_{\hat{b}}??{\left(\lambda^{j}\right)}^{\hat{b}}{}_{\hat{a}}? & =\frac{1}{2}\delta^{ij}\\
?{\left(\Lambda^{I}\right)}^{p}{}_{q}??{\left(\Lambda^{J}\right)}^{q}{}_{p}? & =3\delta^{IJ}
\end{align}
where $\mathrm{Tr}$ is the trace in the adjoint of $E_{8}$. Then
by considering the Jacobi identities the algebra of the $T^{ap}$
and $T_{ap}$ is fixed to be
\begin{align}
\left[T_{\hat{a}p},T_{\hat{b}q}\right] & =\sqrt{5}\hat{\epsilon}_{\hat{a}\hat{b}\hat{c}}d_{pqr}T^{\hat{c}r}\\
\left[T^{\hat{a}p},T^{\hat{b}q}\right] & =-\sqrt{5}\hat{\epsilon}^{\hat{a}\hat{b}\hat{c}}d^{pqr}T_{\hat{c}r}\\
\left[T_{\hat{a}p,}T^{\hat{b}q}\right] & =-\delta_{\hat{a}}^{\hat{b}}?\Lambda^{I}{}_{p}^{q}?X^{I}-\delta_{p}^{q}?\lambda^{i}{}_{\hat{a}}^{\hat{b}}?S^{i}
\end{align}
Whence we can also find that
\begin{align}
\mathrm{Tr}\left(T_{\hat{a}p}T_{\hat{b}q}T_{\hat{c}r}\right) & =15\sqrt{5}\hat{\epsilon}_{\hat{a}\hat{b}\hat{c}}d_{pqr}
\end{align}
By considering the singlet and adjoint projectors, we can find the
contractions of the fundamental generators,
\begin{align}
?{\left(\lambda^{i}\right)}^{\hat{a}}{}_{\hat{c}}??{\left(\lambda^{i}\right)}^{\hat{b}}{}_{\hat{d}}? & =\frac{1}{2}\left(\delta_{\hat{d}}^{\hat{a}}\delta_{\hat{c}}^{\hat{b}}-\frac{1}{3}\delta_{\hat{c}}^{\hat{a}}\delta_{\hat{d}}^{\hat{b}}\right)\\
?{\left(\Lambda^{I}\right)}^{p}{}_{q}??{\left(\Lambda^{I}\right)}^{r}{}_{s}? & =\frac{1}{6}\left(\delta_{q}^{p}\delta_{s}^{r}+3\delta_{s}^{p}\delta_{q}^{r}-30d_{qst}d^{prt}\right)\label{eq:E6-adjoint-projector}
\end{align}
In using these results it is important to note that in the reduction
we have everywhere $\tr$ rather than $\mathrm{Tr}$, which is defined
for $E_{8}$ (since there is no fundamental with lower dimension than
the adjoint for $E_{8}$) as $\tr=\frac{1}{30}\mathrm{Tr}$ so
\begin{align}
\tr\left(S^{i}S^{j}\right) & =\delta^{ij}\\
\tr\left(X^{I}X^{J}\right) & =\delta^{IJ}\\
\tr\left(T^{\hat{a}p}T_{\hat{b}q}\right) & =\delta_{\hat{b}}^{\hat{a}}\delta_{q}^{p}\\
\tr\left(T_{\hat{a}p}T_{\hat{b}q}T_{\hat{c}r}\right) & =\frac{\sqrt{5}}{2}\hat{\epsilon}_{\hat{a}\hat{b}\hat{c}}d_{pqr}
\end{align}
and that the $SU\left(3\right)$ indices used here, $\hat{a},\hat{b},\hat{c},\ldots$
are raised and lowered by complex conjugation, so they are not the
same as the (anti-)holomorphic indices of the Calabi-Yau, $a,b,c,\ldots$
and we must use a dreibein to convert between them. This is particularly
relevant with respect to the $\epsilon$ tensors since $\hat{\epsilon}_{\hat{a}\hat{b}\hat{c}}\hat{\epsilon}^{\hat{a}\hat{b}\hat{c}}=6$
while $\epsilon_{abc}\epsilon^{abc}=48$ and so it is necessary to
keep track of which is which: we have
\begin{align}
\hat{\epsilon}_{abc} & =\hat{\epsilon}_{\hat{a}\hat{b}\hat{c}}?e_{a}^{\hat{a}}??e_{b}^{\hat{b}}??e_{c}^{\hat{c}}?=\frac{1}{2\sqrt{2}}\epsilon_{abc}
\end{align}

\chapter{\label{cha:Quaternionic_Appendix}Gauging a Quaternionic Isometry}

The material in this appendix is not in any way new, but is given
for ease of reference and clarity of conventions.

Suppose we have a quaternionic manifold with coordinates $q^{x}$.
The holonomy group is $SU(2)\otimes SU(2)$ so we have a vierbein
which converts vector indices \linebreak{}
$x,y,z,\ldots=1,\ldots,4$ into pairs of $SU(2)$ indices $A,B,C,\ldots=1,2$
which we can write as
\begin{align}
?f_{x}^{A}{}_{B}? & =?{\left(\begin{array}{cc}
\bar{v}_{x} & u_{x}\\
-\bar{u}_{x} & v_{x}
\end{array}\right)}^{A}{}_{B}?\label{eq:f}
\end{align}
$SU(2)$ indices are raised and lowered with $\epsilon^{AB}$ and
$\epsilon_{AB}$ ($X_{A}=X^{B}\epsilon_{BA}$, $X^{A}=\epsilon^{AB}X_{B}$)
and the $x,y,z$ metric is $g_{xy}=\epsilon_{AC}\epsilon^{BD}?f_{x}^{A}{}_{B}??f_{y}^{C}{}_{D}?$.

Metric compatibility $\nabla_{x\vphantom{B}}^{\vphantom{A}}?f_{y}^{A}{}_{B}?=0\Rightarrow\dd?f^{A}{}_{B}?+?{\t{\tilde{\omega}}}_{\left(L\right)}^{A}{}_{C}?\wedge?f^{C}{}_{B}?+?f^{A}{}_{C}?\wedge?{\t{\tilde{\omega}}}_{\left(R\right)}^{C}{}_{B}?=0$
tells us that the connections are
\begin{align}
?{\t{\tilde{\omega}}}_{\left(L\right)}^{A}{}_{B}? & =?{\left(\begin{array}{cc}
\frac{1}{4}\left(v-\bar{v}\right) & -u\\
\bar{u} & -\frac{1}{4}\left(v-\bar{v}\right)
\end{array}\right)}^{A}{}_{B}?\\
?{\t{\tilde{\omega}}}_{\left(R\right)}^{A}{}_{B}? & =?{\left(\begin{array}{cc}
\frac{3}{4}\left(v-\bar{v}\right) & 0\\
0 & -\frac{3}{4}\left(v-\bar{v}\right)
\end{array}\right)}^{A}{}_{B}?
\end{align}

Since the manifold is quaternionic it has three complex structures
$?{\left(J^{u}\right)}^{x}{}_{y}?$ (where $u,v,w=1,2,3$) satisfying
the quaternion algebra $J^{u}J^{v}=-\delta^{uv}+\epsilon^{uvw}J^{w}$.
There are correspondingly three Kähler forms satisfying $?{\left(K^{u}\right)}_{x}^{z}?\left(K^{v}\right)_{zy}=-\delta^{uv}g_{xy}+\epsilon^{uvw}\left(K^{w}\right)_{xy}$.
A suitable choice of $K^{u}$ is given by
\begin{align}
K^{1} & =-i\left(u\wedge\bar{v}-\bar{u}\wedge v\right)\nonumber \\
K^{2} & =-\left(u\wedge\bar{v}+\bar{u}\wedge v\right)\nonumber \\
K^{3} & =-i\left(u\wedge\bar{u}-v\wedge\bar{v}\right)
\end{align}
which we may write as $?K_{A}^{B}?=\frac{i}{2}?{\left(\sigma^{u}\right)}_{A}^{B}?K^{u}$,
where $?{\left(\sigma^{u}\right)}_{A}^{B}?$ are the standard Pauli
matrices, so that
\begin{align}
?K_{A}^{B}? & =?{\left(\begin{array}{cc}
\frac{1}{2}\left(u\wedge\bar{u}-v\wedge\bar{v}\right) & -\bar{u}\wedge v\\
u\wedge\bar{v} & -\frac{1}{2}\left(u\wedge\bar{u}-v\wedge\bar{v}\right)
\end{array}\right)}_{A}^{B}?
\end{align}

Now if the manifold has some isometries, labelled by indices $i,j,k,\ldots$,
generated by Killing vectors $k_{i}^{x}$ then we may gauge them by
adding a prepotential term to the connections
\begin{align}
?\omega_{\left(L\right)}^{A}{}_{B}? & =?{\t{\tilde{\omega}}}_{\left(L\right)}^{A}{}_{B}?-\frac{1}{2}g\m A^{i}\nabla_{x}k_{i}^{y}?f_{y}^{A}{}_{C}??f^{x}{}_{B}^{C}?\label{eq:omega_L}\\
?\omega_{\left(R\right)}^{A}{}_{B}? & =?{\t{\tilde{\omega}}}_{\left(R\right)}^{A}{}_{B}?-\frac{1}{2}g\m A^{i}\nabla_{x}k_{i}^{y}?f^{x}{}_{C}^{A}??f_{y}^{C}{}_{B}?\label{eq:omega_R}
\end{align}
where $-\frac{1}{2}\nabla_{x}k_{i}^{y}?f_{y}^{A}{}_{C}??f^{x}{}_{B}^{C}?=?{\m P}_{i}^{A}{}_{B}?$
is the prepotential defined by $i_{k_{i}}?K^{A}{}_{B}?=\dd?{\m P}_{i}^{A}{}_{B}?+?{\left[\omega_{\left(L\right)},\m P_{i}\right]}^{A}{}_{B}?$
($i_{k_{i}}$ stands for the inner product $i_{k_{i}}?K^{A}{}_{B}?=\left(k_{i}\right)^{x}\left(?K^{A}{}_{B}?\right)_{xy}\dd q^{y}$).

\chapter{Summary of the Reduced Heterotic M-Theory\label{cha:Summary}}

\section{Action}

The bulk action is
\begin{align}
S\left(\m M\right)=\; & \frac{1}{2\kappa_{5}^{2}}\int_{\m M}dv\Bigg(R-\frac{1}{2}V^{-2}\partial_{\alpha}V\partial^{\alpha}V-\frac{1}{2}\m D_{\alpha}\sigma\m D^{\alpha}\sigma-2V^{-1}\partial_{\alpha}\bar{\xi}\partial^{\alpha}\xi\nonumber \\
 & \qquad\qquad\quad-G_{ij}^{\perp}\partial_{\alpha}b^{i}\partial^{\alpha}b^{j}-\frac{1}{4}\m F_{i\alpha\beta}\m F^{i\alpha\beta}-\frac{1}{2}V^{-2}\alpha_{i}\alpha^{i}\nonumber \\
 & \qquad\qquad\quad+\bar{\psi}_{A\alpha}\gamma^{\alpha\beta\gamma}\m D_{\beta}\psi_{\gamma}^{A}+\bar{\zeta}_{A}\gamma^{\beta}\m D_{\beta}\zeta^{A}+G_{ij}^{\perp}\bar{\lambda}_{A}^{i}\gamma^{\beta}\m D_{\beta}\lambda^{jA}\nonumber \\
 & \qquad\qquad\quad-\sqrt{2}i\bar{\zeta}_{A}\gamma^{\alpha}\gamma^{\beta}\psi_{\alpha}^{B}?f_{\beta}^{A}{}_{B}?+\frac{i}{2}G_{ij}\bar{\lambda}_{A}^{i}\gamma^{\alpha}\gamma^{\beta}\psi_{\alpha}^{A}\partial_{\beta}b^{j}\nonumber \\
 & \qquad\qquad\quad-\frac{\sqrt{2}i}{8}\left(\bar{\psi}_{A\gamma}\gamma^{\alpha}\gamma^{\gamma\delta}\gamma^{\beta}\psi_{\delta}^{A}-\bar{\zeta}_{A}\gamma^{\alpha\beta}\zeta^{A}-\frac{1}{3}G_{jk}^{\perp}\bar{\lambda}_{A}^{j}\gamma^{\alpha\beta}\lambda^{Ak}\right)b_{i}\m F_{\alpha\beta}^{i}\nonumber \\
 & \qquad\qquad\quad-\frac{\sqrt{2}}{4}G_{ij}^{\perp}\bar{\lambda}_{A}^{i}\gamma^{\gamma}\gamma^{\alpha\beta}\psi_{\gamma}^{A}\m F_{\alpha\beta}^{j}+\frac{3\sqrt{2}i}{8}\m K^{-1}\m K_{ijk}^{\perp}\bar{\lambda}_{A}^{i}\gamma^{\alpha\beta}\lambda^{Aj}\m F_{\alpha\beta}^{k}\nonumber \\
 & \qquad\qquad\quad-\frac{\sqrt{2}}{4}V^{-1}\alpha?\tau^{A}{}_{B}?\bar{\psi}_{A\alpha}\gamma^{\alpha\beta}\psi_{\beta}^{B}+iV^{-1}\alpha?\tau^{A}{}_{B}?\bar{\zeta}_{A}\gamma^{\alpha}\psi_{\alpha}^{B}\nonumber \\
 & \qquad\qquad\quad-\frac{\sqrt{2}i}{2}V^{-1}\alpha_{i}^{\perp}?\tau^{A}{}_{B}?\bar{\lambda}_{A}^{i}\gamma^{\alpha}\psi_{\alpha}^{B}-2V^{-1}\alpha_{i}^{\perp}?\tau^{A}{}_{B}?\bar{\zeta}_{A}\lambda^{Bi}\nonumber \\
 & \qquad\qquad\quad+\frac{3\sqrt{2}}{4}V^{-1}\alpha?\tau^{A}{}_{B}?\bar{\zeta}_{A}\zeta^{B}\nonumber \\
 & \qquad\qquad\quad+\frac{3\sqrt{2}}{4}V^{-1}\left(\m K^{-1}?{\m K}^{\perp i}{}_{jk}?+\frac{1}{9}b^{i}G_{jk}^{\perp}\right)\alpha_{i}?\tau^{A}{}_{B}?\bar{\lambda}_{A}^{j}\lambda^{kB}\Bigg)
\end{align}

The action on $\partial\m M_{1}$ is
\begin{align}
S\left(\partial\m M_{1}\right) & =\frac{1}{2\kappa_{5}^{2}}\int_{\partial\m M_{1}}dv\left(-2K-\sqrt{2}V^{-1}\alpha\vphantom{\frac{\partial\bar{W}}{\partial\bC jp}}\right.\nonumber \\
 & \qquad\qquad\qquad\quad+\frac{1}{2}?\tau^{A}{}_{B}?\bar{\psi}_{A\mu}\gamma^{\mu\nu}\psi_{\nu}^{B}+\frac{1}{2}?\tau^{A}{}_{B}?\bar{\zeta}_{A}\zeta^{B}+\frac{1}{2}G_{ij}^{\perp}?\tau^{A}{}_{B}?\bar{\lambda}_{A}^{i}\lambda^{Bj}+\m T\Bigg)\nonumber \\
 & +\frac{1}{g^{2}}\int_{\partial\m M_{1}}dv\left(-\frac{1}{4}VF_{\mu\nu}^{I}F^{I\mu\nu}-2G_{ij}\m D_{\mu}C^{ip}\m D^{\mu}\bC jp\vphantom{\frac{\partial\bar{W}}{\partial\bC jp}}\right.\nonumber \\
 & \qquad\qquad\qquad\quad-\frac{1}{2}V^{-1}G^{ij}\frac{\partial W}{\partial C^{ip}}\frac{\partial\bar{W}}{\partial\bC jp}-2V^{-1}\bar{C}^{i}\Lambda^{I}C_{i}\bar{C}^{j}\Lambda^{I}C_{j}\nonumber \\
 & \qquad\qquad\qquad\quad+\frac{1}{2}\bar{\chi}_{A}^{I}\gamma^{\mu}\m D_{\mu}\chi^{IA}+\frac{1}{2}G_{ij}\tr\left(?{\t{\bar{\eta}}}_{A}^{i}?\gamma^{\mu}\m D_{\mu}\eta^{Aj}\right)\nonumber \\
 & \qquad\qquad\qquad\quad+\frac{3\sqrt{10}}{2}V^{-\frac{1}{2}}\m K^{-1}\m K_{ijk}\left(d_{pqr}C^{ip}?{\t{\bar{\eta}}}_{R}^{jq}??\eta_{L}^{kr}?+d^{pqr}\bC ip?{\t{\bar{\eta}}}_{L}^{j}{}_{q}??\eta_{R}^{k}{}_{r}?\right)\nonumber \\
 & \qquad\qquad\qquad\quad+2G_{ij}V^{-\frac{1}{2}}\left(\bar{\chi}_{L}^{I}?\eta_{R}^{i}{}_{q}?C^{jp}?\Lambda^{Iq}{}_{p}?-\bar{\chi}_{R}^{I}?\eta_{L}^{ip}?\bC jq?\Lambda^{Iq}{}_{p}?\right)\nonumber \\
 & \qquad\qquad\qquad\quad+\bar{\psi}_{A\mu}j^{A\mu}+\bar{\zeta}_{A}j^{A}+G_{ij}^{\perp}\bar{\lambda}_{A}^{i}j^{Aj}+\Theta^{\mu}\m D_{\mu}\sigma\Bigg)
\end{align}
Where $\m T=-\bar{\psi}_{A\mu}\gamma^{\mu}\left(\psi_{z}^{A}-\frac{\sqrt{2}i}{3}\gamma_{z}\zeta^{A}\right)+\frac{7\sqrt{2}i}{3}\bar{\zeta}_{A}\psi_{z}^{A}$,
the superpotential is
\begin{align}
W & =3\sqrt{10}\m K^{-1}\m K_{ijk}d_{pqr}C^{ip}C^{jq}C^{kr}
\end{align}
and the currents which couple to bulk fields are
\begin{align}
j^{A\mu}=\; & -\frac{1}{4}V^{\frac{1}{2}}\gamma^{\nu\rho}\gamma^{\mu}F_{\nu\rho}^{I}\chi^{IA}\nonumber \\
 & -\gamma^{\nu}\gamma^{\mu}P_{R}?\eta^{A}{}_{ip}?\m D_{\nu}C^{ip}+\gamma^{\nu}\gamma^{\mu}P_{L}?\eta^{A}{}_{i}^{p}?\m D_{\nu}\bC ip\nonumber \\
 & +\frac{3\sqrt{10}}{2}V^{-\frac{1}{2}}\gamma^{\mu}\m K^{-1}\m K_{ijk}\left(d_{pqr}P_{L}\eta^{Aip}C^{jq}C^{kr}+d^{pqr}P_{R}?\eta^{Ai}{}_{p}?\bC jq\bC kr\right)\nonumber \\
j^{A}=\; & -\frac{\sqrt{2}i}{4}V^{\frac{1}{2}}\gamma^{\mu\nu}F_{\mu\nu}^{I}\chi^{IA}\nonumber \\
 & +\frac{2\sqrt{2}i}{9}b_{i}b_{j}\gamma^{\mu}\left(P_{R}?\eta^{Aj}{}_{p}?\m D_{\mu}C^{ip}-P_{L}\eta^{Aip}\m D_{\mu}\bC jp\right)\nonumber \\
 & -\frac{\sqrt{2}i}{3}V^{-\frac{1}{2}}\left(G_{ij}+\frac{4}{3}b_{i}b_{j}\right)?\tau^{A}{}_{B}?\chi^{IB}\bar{C}^{j}\Lambda^{I}C^{i}\nonumber \\
 & -3\sqrt{5}iV^{-\frac{1}{2}}\m K^{-1}\m K_{ijk}\left(d_{pqr}C^{ip}C^{jq}P_{L}\eta^{Akr}+d^{pqr}\bC ip\bC jqP_{R}?\eta^{Ak}{}_{r}?\right)\nonumber \\
j^{Ai}=\; & i\left(?\Gamma^{\perp i}{}_{jk}?-\frac{4}{3}{\delta^{\perp}}_{(j}^{i}b_{k)}^{\vphantom{i}}\right)\gamma^{\mu}\left(P_{R}?\eta^{Ak}{}_{p}?\m D_{\mu}C^{jp}-P_{L}\eta^{Ajp}\m D_{\mu}\bC kp\right)\nonumber \\
 & -2iV^{-\frac{1}{2}}\left(?\Gamma^{\perp i}{}_{jk}?-\frac{4}{3}{\delta^{\perp}}_{(j}^{i}b_{k)}^{\vphantom{i}}\right)?\tau^{A}{}_{B}?\chi^{IB}\bar{C}^{k}\Lambda^{I}C^{j}\nonumber \\
\Theta^{\mu}=\; & \frac{1}{12}\epsilon^{\mu\nu\rho\sigma}\left(\omega_{\nu\rho\sigma}^{Y}-\frac{1}{4}V^{-1}\bar{\chi}_{A}^{I}\gamma_{\nu\rho\sigma}\chi^{IA}-\frac{1}{4}V^{-1}\tr\left(\bar{\eta}_{iA}\gamma_{\nu\rho\sigma}\eta^{iA}\right)\right)\label{eq:M1-currents}
\end{align}

The action on $\partial\m M_{2}$ is
\begin{align*}
S\left(\partial\m M_{2}\right)=\; & \frac{1}{2\kappa_{5}^{2}}\int_{\partial\m M_{2}}dv\left(2K+\sqrt{2}V^{-1}\alpha\vphantom{\frac{\partial\bar{W}}{\partial\bC jp}}\right.\\
 & \qquad\qquad\qquad-\frac{1}{2}?\tau^{A}{}_{B}?\bar{\psi}_{A\mu}\gamma^{\mu\nu}\psi_{\nu}^{B}-\frac{1}{2}?\tau^{A}{}_{B}?\bar{\zeta}_{A}\zeta^{B}-\frac{1}{2}G_{ij}^{\perp}?\tau^{A}{}_{B}?\bar{\lambda}_{A}^{i}\lambda^{Bj}+\m T\Bigg)\\
 & +\frac{1}{g^{2}}\int_{\partial\m M_{2}}dv\left(-\frac{1}{4}VF_{\mu\nu}^{I}F^{I\mu\nu}+\frac{1}{2}\bar{\chi}_{A}^{I}\gamma^{\mu}D_{\mu}\chi^{IA}\vphantom{\frac{\partial\bar{W}}{\partial\bC jp}}\right.\\
 & \qquad\qquad\qquad\quad+\bar{\psi}_{A\mu}j^{A\mu}+\bar{\zeta}_{A}j^{A}+\Theta^{\mu}\m D_{\mu}\sigma\Bigg)
\end{align*}
where $\m T=\bar{\psi}_{A\mu}\gamma^{\mu}\left(\psi_{z}^{A}-\frac{\sqrt{2}i}{3}\gamma_{z}\zeta^{A}\right)-\frac{7\sqrt{2}i}{3}\bar{\zeta}_{A}\psi_{z}^{A}$
and the currents on $\partial\m M_{2}$ are:
\begin{align}
j^{A\mu} & =-\frac{1}{4}V^{\frac{1}{2}}F_{\nu\rho}^{I}\gamma^{\nu\rho}\gamma^{\mu}\chi^{IA}\nonumber \\
j^{A} & =-\frac{\sqrt{2}i}{2}V^{\frac{1}{2}}F_{\mu\nu}^{I}\gamma^{\mu\nu}\chi^{IA}\nonumber \\
\Theta^{\mu} & =\frac{1}{12}\epsilon^{\mu\nu\rho\sigma}\frac{\kappa_{5}^{2}}{g^{2}}\left(\omega_{\nu\rho\sigma}^{Y}-\frac{1}{4}V^{-1}\bar{\chi}_{A}^{I}\gamma_{\nu\rho\sigma}\chi^{IA}\right)\label{eq:M2-currents}
\end{align}

\section{Boundary Conditions\label{sec:Summary-Boundary-Conditions}}

On $\partial\m M_{1}$:
\begin{align}
\xi & =\frac{\kappa_{5}^{2}}{g^{2}}\left(\sqrt{10}\m K^{-1}\m K_{ijk}d_{pqr}C^{ip}C^{jq}C^{kr}+\frac{1}{2}V^{\frac{1}{2}}\bar{\chi}_{1}^{I}\chi^{I2}\right)\\
\bar{\xi} & =\frac{\kappa_{5}^{2}}{g^{2}}\left(\sqrt{10}\m K^{-1}\m K_{ijk}d^{pqr}\bC ip\bC jq\bC kr-\frac{1}{2}V^{\frac{1}{2}}\bar{\chi}_{2}^{I}\chi^{I1}\right)\\
\m A_{\mu}^{i} & =-\frac{\sqrt{2}\kappa_{5}^{2}}{g^{2}}\bigg(i?\Gamma^{i}{}_{jk}?\left(\bC kp\m D_{\mu}C^{jp}-C^{jp}\m D_{\mu}\bC kp\right)-\frac{i}{4}b^{i}?\tau^{A}{}_{B}?\bar{\chi}_{A}^{I}\gamma_{\mu}\chi^{IB}\nonumber \\
 & \qquad\quad-\frac{i}{4}\left(?\Gamma^{i}{}_{jk}?-b^{i}G_{jk}\right)\left(?{\t{\bar{\eta}}}_{L}^{jp}?\gamma_{\mu}?\eta_{R}^{k}{}_{p}?-?{\t{\bar{\eta}}}_{L}^{k}{}_{p}?\gamma_{\mu}?\eta_{L}^{jp}?\right)\bigg)\\
\m D_{5}\sigma & =-\frac{1}{12g^{2}}V^{2}\partial_{\mu}\left(\omega_{\mu\nu\rho}^{Y}-\frac{1}{4}V^{-1}\bar{\chi}_{A}^{I}\gamma_{\mu\nu\rho}\chi^{IA}-\frac{1}{4}V^{-1}\tr\left(\bar{\eta}_{iA}\gamma_{\mu\nu\rho}\eta^{iA}\right)\right)\nonumber \\
 & \quad-\frac{i}{4}V?\tau^{A}{}_{B}?\bar{\psi}_{A\alpha}\gamma^{\alpha\beta}\gamma_{5}\psi_{\beta}^{B}-\frac{3i}{4}V?\tau^{A}{}_{B}?\bar{\zeta}_{A}\gamma_{5}\zeta^{B}+\frac{i}{4}V?\tau^{A}{}_{B}?G_{ij}^{\perp}\bar{\lambda}_{A}^{i}\gamma_{5}\lambda^{Bj}\nonumber \\
 & \quad+\frac{\sqrt{2}}{2}V?\tau^{A}{}_{B}?\bar{\zeta}_{A}\gamma^{\alpha}\gamma_{5}\psi_{\alpha}^{B}\\
K^{\mu\nu}-Kg^{\mu\nu} & =-\kappa_{5}^{2}T_{YM}^{\mu\nu}-\sqrt{2}\alpha V^{-1}g^{\mu\nu}+\frac{1}{2}?\tau^{A}{}_{B}?\bar{\psi}_{A}^{(\mu}\gamma^{\nu)\rho}\psi_{\rho}^{B}\nonumber \\
 & \quad+\frac{1}{4}?\tau^{A}{}_{B}?\left(\bar{\psi}_{A\rho}\gamma^{\rho\sigma}\psi_{\sigma}^{B}+\bar{\zeta}_{A}\zeta^{B}+G_{ij}^{\perp}\bar{\lambda}_{A}^{i}\lambda^{Bj}\right)g^{\mu\nu}\\
\partial_{5}V & =-\sqrt{2}\alpha-\frac{\kappa_{5}^{2}}{g^{2}}V^{2}\frac{\partial\m L_{YM}}{\partial V}-\frac{\sqrt{2}i}{2}V\bar{\zeta}_{A}\gamma^{\alpha}\gamma_{5}\psi_{\alpha}^{A}\\
?P_{-}^{A}{}_{B}?\psi_{\mu}^{B} & =-\frac{1}{3}\left(\gamma_{\mu\nu}-3g_{\mu\nu}\right)?\tau^{A}{}_{B}?j^{B\nu}\\
?P_{+}^{A}{}_{B}?\zeta^{B} & =-?\tau^{A}{}_{B}?j^{B}\\
?P_{+}^{A}{}_{B}?\lambda^{\perp iB} & =-?\tau^{A}{}_{B}?j^{iB}
\end{align}
where
\begin{align}
T_{YM}^{\mu\nu} & =2\frac{\delta\m L_{YM}}{\delta g_{\mu\nu}}-g^{\mu\nu}\m L_{YM}
\end{align}
and the currents $j^{B\nu}$, $j^{B}$ and $j^{iB}$ are given by
\eqref{eq:M1-currents}.

On $\partial\m M_{2}$: 
\begin{align}
\xi & =\frac{\kappa_{5}^{2}}{2g^{2}}V^{\frac{1}{2}}\bar{\chi}_{2}^{I}\chi^{I1}\\
\bar{\xi} & =-\frac{\kappa_{5}^{2}}{2g^{2}}V^{\frac{1}{2}}\bar{\chi}_{1}^{I}\chi^{I2}\\
\m A_{\mu}^{i} & =-\frac{\sqrt{2}i\kappa_{5}^{2}}{4g^{2}}b^{i}?\tau^{A}{}_{B}?\bar{\chi}_{A}^{I}\gamma_{\mu}\chi^{IB}\\
\m D_{5}\sigma & =\frac{1}{12g^{2}}\partial_{\mu}\left(\omega_{\mu\nu\rho}^{Y}-\frac{1}{4}V^{-1}\bar{\chi}_{A}^{I}\gamma_{\mu\nu\rho}\chi^{IA}\right)\nonumber \\
 & \quad-\frac{i}{4}V?\tau^{A}{}_{B}?\bar{\psi}_{A\alpha}\gamma^{\alpha\beta}\gamma_{5}\psi_{\beta}^{B}-\frac{3i}{4}V?\tau^{A}{}_{B}?\bar{\zeta}_{A}\gamma_{5}\zeta^{B}+\frac{i}{4}V?\tau^{A}{}_{B}?G_{ij}^{\perp}\bar{\lambda}_{A}^{i}\gamma_{5}\lambda^{Bj}\nonumber \\
 & \quad+\frac{\sqrt{2}}{2}V?\tau^{A}{}_{B}?\bar{\zeta}_{A}\gamma^{\alpha}\gamma_{5}\psi_{\alpha}^{B}\\
K^{\mu\nu}-Kg^{\mu\nu} & =\kappa_{5}^{2}T_{YM}^{\mu\nu}-\sqrt{2}\alpha V^{-1}g^{\mu\nu}+\frac{1}{2}?\tau^{A}{}_{B}?\bar{\psi}_{A}^{(\mu}\gamma^{\nu)\rho}\psi_{\rho}^{B}\nonumber \\
 & \quad+\frac{1}{4}?\tau^{A}{}_{B}?\left(\bar{\psi}_{A\rho}\gamma^{\rho\sigma}\psi_{\sigma}^{B}+\bar{\zeta}_{A}\zeta^{B}+G_{ij}^{\perp}\bar{\lambda}_{A}^{i}\lambda^{Bj}\right)g^{\mu\nu}\\
\partial_{5}V & =-\sqrt{2}\alpha+\frac{\kappa_{5}^{2}}{g^{2}}V^{2}\frac{\partial\m L_{YM}}{\partial V}-\frac{\sqrt{2}i}{2}V\bar{\zeta}_{A}\gamma^{\alpha}\gamma_{5}\psi_{\alpha}^{A}\\
?P_{-}^{A}{}_{B}?\psi_{\mu}^{B} & =\frac{1}{3}\left(\gamma_{\mu\nu}-3g_{\mu\nu}\right)?\tau^{A}{}_{B}?j^{B\nu}\\
?P_{+}^{A}{}_{B}?\zeta^{B} & =?\tau^{A}{}_{B}?j^{B}\\
?P_{+}^{A}{}_{B}?\lambda^{\perp iB} & =?\tau^{A}{}_{B}?j^{iB}
\end{align}
with the currents $j^{B\nu}$, $j^{B}$ and $j^{iB}$ given by \eqref{eq:M2-currents}.

\section{Supersymmetry\label{sec:Summary-Supersymmetry}}

The supersymmetry transformations of the bulk fields are:
\begin{align}
\delta?e^{\hat{\alpha}}{}_{\alpha}? & =-\frac{1}{2}\bar{s}_{A}\gamma^{\hat{\alpha}}\psi_{\alpha}^{A}\\
\delta\psi_{\alpha}^{A} & =\m D_{\alpha}s^{A}-\frac{\sqrt{2}i}{12}\left(?\gamma_{\alpha}^{\beta\gamma}?-4\delta_{\alpha}^{\beta}\gamma^{\gamma}\right)b_{i}\m F_{\beta\gamma}^{i}s^{A}-\frac{\sqrt{2}}{6}V^{-1}\alpha?\tau^{A}{}_{B}?\gamma_{\alpha}s^{B}\\
\delta\zeta^{A} & =-\frac{\sqrt{2}i}{2}\gamma^{\alpha}s^{B}?f_{xB}^{A}?D_{\alpha}q^{x}\\
\delta V & =-\frac{i}{\sqrt{2}}?\tau^{A}{}_{B}?\bar{s}_{A}\zeta^{B}\\
\delta\sigma & =\frac{\sqrt{2}}{2}?{\left(\begin{array}{cc}
V & -V^{\frac{1}{2}}\xi\\
-V^{\frac{1}{2}}\bar{\xi} & -V
\end{array}\right)}^{A}{}_{B}?\bar{s}_{A}\zeta^{B}\\
\delta\xi & =-\frac{\sqrt{2}i}{2}V^{\frac{1}{2}}\bar{s}_{2}\zeta^{1}\\
\delta\bar{\xi} & =\frac{\sqrt{2}i}{2}V^{\frac{1}{2}}\bar{s}_{1}\zeta^{2}\\
\bigg(\text{i.e.\,}\delta q^{x} & =-\frac{\sqrt{2}i}{2}?f^{xA}{}_{B}?\bar{s}_{A}\zeta^{B}\bigg)\\
\delta\m A_{\alpha}^{i} & =-\frac{\sqrt{2}i}{2}b^{i}?\tau^{A}{}_{B}?\bar{s}_{A}\psi_{\alpha}^{B}+\frac{\sqrt{2}}{2}?\tau^{A}{}_{B}?\bar{s}_{A}\gamma_{\alpha}\lambda^{\perp Bi}\\
\delta\lambda^{\perp iA} & =-\frac{i}{2}\partial_{\alpha}b^{i}\gamma^{\alpha}s^{A}+\frac{\sqrt{2}}{4}\m F_{\alpha\beta}^{i\perp}\gamma^{\alpha\beta}s^{A}-\frac{\sqrt{2}i}{4}V^{-1}G^{\perp ij}\alpha_{j}?\tau^{A}{}_{B}?s^{B}\\
\delta b^{i} & =-\frac{i}{2}?\tau^{A}{}_{B}?\bar{s}_{A}\lambda^{\perp iB}
\end{align}
Those of the $E_{6}$ gauge fields on $\partial\m M_{1}$ are
\begin{align}
\delta A_{\mu}^{I} & =\frac{1}{2}V^{-\frac{1}{2}}\bar{s}_{A}\gamma_{\mu}\chi^{A}\\
\delta\chi^{IA} & =\frac{1}{4}V^{\frac{1}{2}}\gamma^{\mu\nu}F_{\mu\nu}^{I}s^{A}+V^{-\frac{1}{2}}G_{ij}\bar{C}^{i}\Lambda^{I}C^{j}?\tau^{A}{}_{B}?s^{B}\\
\delta?\eta_{L}^{ip}? & =-\m D_{\mu}C^{ip}\gamma^{\mu}s^{2}-\frac{3\sqrt{10}}{2}V^{-\frac{1}{2}}\m K^{-1}?{\m K}^{i}{}_{jk}?d^{pqr}\bC jq\bC krs^{1}\\
\delta?\eta_{R}^{i}{}_{p}? & =\m D_{\mu}\bC ip\gamma^{\mu}s^{1}-\frac{3\sqrt{10}}{2}V^{-\frac{1}{2}}\m K^{-1}?{\m K}^{i}{}_{jk}?d_{pqr}C^{jq}C^{kr}\\
\delta C^{ip} & =-\frac{1}{2}\bar{s}_{2}?\eta_{L}^{ip}?-\frac{i}{4}?\Gamma^{i}{}_{jk}?C^{jp}\bar{s}_{A}\lambda^{\perp Ak}\\
\delta\bC ip & =-\frac{1}{2}\bar{s}_{1}?\eta_{R}^{i}{}_{p}?-\frac{i}{4}?\Gamma^{i}{}_{kj}?\bC jp\bar{s}_{A}\lambda^{\perp Ak}
\end{align}
and of the $E_{8}$ fields on $\partial\m M_{2}$ are
\begin{align}
\delta A_{\mu}^{I} & =\frac{1}{2}V^{-\frac{1}{2}}\bar{s}_{A}\gamma_{\mu}\chi^{A}\\
\delta\chi^{IA} & =\frac{1}{4}V^{\frac{1}{2}}\gamma^{\mu\nu}F_{\mu\nu}^{I}s^{A}
\end{align}

\chapter{Galileons: Further Details\label{cha:galileon-appendix}}

\section*{Derivatives of the co-ordinate vectors}

\subsection*{Tangential}

$\nabla_{\boldsymbol{\xi}}\boldsymbol{e}_{\mu}=\xi^{I}\nabla_{I}\boldsymbol{e}_{\mu}=\xi^{I}?\Gamma^{b}{}_{I\mu}?\boldsymbol{e}_{b}$
and due to the normal co-ordinates $\Gamma_{I\mu\nu}=-\Gamma_{\mu I\nu}$
and $\Gamma_{\mu IJ}=-\Gamma_{I\mu J}$ so using the definitions \eqref{eq:extrinsic_curvature}
and \eqref{eq:twist_connection} we can see that
\begin{align}
\left.\nabla_{\boldsymbol{\xi}}\boldsymbol{e}_{\mu}\right|_{\Sigma_{0}} & =\xi^{I}?k_{I\mu}^{\nu}?\boldsymbol{e}_{\nu}+\xi^{I}?a_{I}^{J}{}_{\mu}?\boldsymbol{e}_{J}
\end{align}

$\boldsymbol{\xi}$ and $\boldsymbol{e}_{\mu}$ are orthogonal, $\left[\boldsymbol{\xi},\boldsymbol{e}_{\mu}\right]=0$,
so $\nabla_{\boldsymbol{\xi}}\boldsymbol{e}_{\mu}=\nabla_{\mu}\boldsymbol{\xi}$
and $\nabla_{\boldsymbol{\xi}}\nabla_{\boldsymbol{\xi}}\boldsymbol{e}_{\mu}=\nabla_{\boldsymbol{\xi}}\nabla_{\mu}\boldsymbol{\xi}$.
Then since $\boldsymbol{\xi}$ is the tangent of a geodesic we have
$\nabla_{\boldsymbol{\xi}}\boldsymbol{\xi}=0$ and so $\nabla_{\boldsymbol{\xi}}\nabla_{\mu}\boldsymbol{\xi}=\left[\nabla_{\boldsymbol{\xi}},\nabla_{\mu}\right]\boldsymbol{\xi}=\boldsymbol{R}(\boldsymbol{\xi},\boldsymbol{e}_{\mu})\boldsymbol{\xi}$
(recall that $\boldsymbol{R}(\boldsymbol{A},\boldsymbol{B})\boldsymbol{C}=?R^{a}{}_{bcd}?A^{c}B^{d}C^{b}\boldsymbol{e}_{a}$).
Applying $\nabla_{\boldsymbol{\xi}}^{n-2}$ to this expression we
find
\begin{align}
\left.\nabla_{\boldsymbol{\xi}}^{n}\boldsymbol{e}_{\mu}\right|_{\Sigma_{0}} & =\nabla_{\boldsymbol{\xi}}^{n-2}\boldsymbol{R}(\boldsymbol{\xi},\boldsymbol{e}_{\mu})\boldsymbol{\xi}
\end{align}

\subsection*{Transverse}

$\sigma$ is the proper distance along the geodesic whose tangent
is $\boldsymbol{\xi}$ from $\Sigma_{0}$, so $\nabla_{\boldsymbol{\xi}}\sigma=1$,
$\partial_{I}\sigma=\xi_{I}$ and points along the geodesic have co-ordinates
$x^{I}=\sigma\xi^{I}$. Thus we can evaluate the Lie bracket to be
$\left[\boldsymbol{\xi},\sigma\boldsymbol{e}_{I}\right]=\partial_{\boldsymbol{\xi}}\sigma\boldsymbol{e}_{I}-\sigma\partial_{I}\left(\frac{1}{\sigma}x^{J}\right)\boldsymbol{e}_{J}=\xi_{I}\boldsymbol{\xi}$
and so $\nabla_{\boldsymbol{\xi}}\left(\sigma\boldsymbol{e}_{I}\right)=\sigma\nabla_{I}\boldsymbol{\xi}+\xi_{I}\boldsymbol{\xi}$.
Combining this with the geodesic equation $\nabla_{\boldsymbol{\xi}}\boldsymbol{\xi}=0$
and the fact that $\boldsymbol{\xi}$ is normalized, $\nabla_{\boldsymbol{\xi}}\xi^{I}=0$,
we find that 
\begin{equation}
\nabla_{\boldsymbol{\xi}}^{2}\left(\sigma\boldsymbol{e}_{I}\right)=\nabla_{I}\boldsymbol{\xi}+\sigma\nabla_{\boldsymbol{\xi}}\nabla_{I}\boldsymbol{\xi}\label{eq:d2sigmae_I}
\end{equation}
On the other hand, factoring out $\sigma$, we have $\sigma\left[\boldsymbol{\xi},\boldsymbol{e}_{I}\right]=\left(\xi_{I}\boldsymbol{\xi}-\boldsymbol{e}_{I}\right)$,
so using $\nabla_{\boldsymbol{\xi}}\boldsymbol{\xi}=0$ again 
\begin{equation}
\boldsymbol{R}(\boldsymbol{\xi},\boldsymbol{e}_{I})\boldsymbol{\xi}\equiv\nabla_{\boldsymbol{\xi}}\nabla_{I}\boldsymbol{\xi}-\nabla_{I}\nabla_{\boldsymbol{\xi}}\boldsymbol{\xi}-\nabla_{\left[\boldsymbol{\xi},\boldsymbol{e}_{I}\right]}\boldsymbol{\xi}=\nabla_{\boldsymbol{\xi}}\nabla_{I}\boldsymbol{\xi}+\frac{1}{\sigma}\nabla_{I}\boldsymbol{\xi}\label{eq:R(xi,e_I)xi}
\end{equation}
Then comparing \eqref{eq:d2sigmae_I} and \eqref{eq:R(xi,e_I)xi}
we can see that 
\begin{align}
\nabla_{\boldsymbol{\xi}}^{2}\left(\sigma\boldsymbol{e}_{I}\right) & =\sigma\boldsymbol{R}(\boldsymbol{\xi},\boldsymbol{e}_{I})\boldsymbol{\xi}
\end{align}

Using this result, since $\nabla_{\boldsymbol{\xi}}^{2}\left(\sigma\boldsymbol{e}_{I}\right)=2\nabla_{\boldsymbol{\xi}}\boldsymbol{e}_{I}+\sigma\nabla_{\boldsymbol{\xi}}^{2}\boldsymbol{e}_{I}$
then if we evaluate on $\Sigma_{0}$, where $\sigma=0$, then we find
\begin{align}
\left.\nabla_{\boldsymbol{\xi}}\boldsymbol{e}_{I}\right|_{\Sigma_{0}} & =0
\end{align}
while applying $\nabla_{\boldsymbol{\xi}}^{n}$ gives
\begin{equation}
\nabla_{\boldsymbol{\xi}}^{n}\left(\nabla_{\boldsymbol{\xi}}^{2}\left(\sigma\boldsymbol{e}_{I}\right)\right)=\nabla_{\boldsymbol{\xi}}^{n}\left(2\nabla_{\boldsymbol{\xi}}\boldsymbol{e}_{I}+\sigma\nabla_{\boldsymbol{\xi}}^{2}\boldsymbol{e}_{I}\right)=\left(2+n\right)\nabla_{\boldsymbol{\xi}}^{n+1}\boldsymbol{e}_{I}+\sigma\nabla_{\boldsymbol{\xi}}^{n+2}\boldsymbol{e}_{I}
\end{equation}
and
\begin{equation}
\nabla_{\boldsymbol{\xi}}^{n}\left(\sigma\boldsymbol{R}(\boldsymbol{\xi},\boldsymbol{e}_{I})\boldsymbol{\xi}\right)=n\nabla_{\boldsymbol{\xi}}^{n-1}\boldsymbol{R}(\boldsymbol{\xi},\boldsymbol{e}_{I})\boldsymbol{\xi}+\sigma\nabla_{\boldsymbol{\xi}}^{n}\boldsymbol{R}(\boldsymbol{\xi},\boldsymbol{e}_{I})\boldsymbol{\xi}
\end{equation}
so evaluating on $\Sigma_{0}$ once again
\begin{align}
\left.\nabla_{\boldsymbol{\xi}}^{n}\boldsymbol{e}_{I}\right|_{\Sigma_{0}} & =\frac{\left(n-1\right)}{\left(n+1\right)}\nabla_{\boldsymbol{\xi}}^{n-2}\boldsymbol{R}(\boldsymbol{\xi},\boldsymbol{e}_{I})\boldsymbol{\xi}
\end{align}

\section*{Relation of $?k^{I}{}_{\mu\nu}?$ to codimension one}

Here we define the extrinsic curvature as
\begin{align}
?k^{I}{}_{\mu\nu}? & =\left.-?\Gamma^{I}{}_{\mu\nu}?\right|_{\Sigma_{0}}\label{eq:k_normal_coordinates}
\end{align}
The definition of extrinsic curvature which does not depend on choosing
normal co-ordinates is 
\begin{equation}
?k^{I}{}_{\mu\nu}?=-?h_{\mu}^{a}??h_{\nu}^{b}??\sigma^{I}{}_{c}?\nabla_{a}?h_{b}^{c}?
\end{equation}
where $h_{\mu\nu}=\boldsymbol{e}_{\mu}\cdot\boldsymbol{e}_{\nu}$
and $\sigma_{IJ}=\boldsymbol{e}_{I}\cdot\boldsymbol{e}_{J}$ are the
parts of the metric transverse and normal respectively to the surface
whose extrinsic curvature we are calculating. These definitions are
equivalent since $\nabla_{a}?g_{b}^{c}?=0\Rightarrow\nabla_{a}?h_{b}^{c}?=-\nabla_{a}?\sigma_{b}^{c}?$
so $?k^{I}{}_{\mu\nu}?=?h_{\mu}^{a}??h_{\nu}^{b}??\sigma^{I}{}_{c}?\nabla_{a}?\sigma_{b}^{c}?$
while $?h_{\nu}^{b}??\sigma_{b}^{d}?=0$ and $?h_{\nu}^{b}??\sigma^{I}{}_{c}?\partial_{a}?\sigma_{b}^{c}?=-?h_{\nu}^{b}??\sigma_{b}^{c}?\partial_{a}?\sigma^{I}{}_{c}?=0$
which leaves just the one connection term on the right hand side,
corresponding to \ref{eq:k_normal_coordinates}. In co-dimension one,
$h_{ab}=g_{ab}-n_{a}n_{b}$, with $n_{a}$ being the normal to the
surface, and $?\sigma^{I}{}_{c}?=\delta_{c}^{I}$ while $I$ only
takes one value so $?\sigma^{I}{}_{c}?\nabla_{a}?h_{b}^{c}?=-\nabla_{a}\left(n_{b}n^{I}\right)$
but $n^{I}=n^{5}=1$ and so we recover 
\begin{equation}
k_{\mu\nu}=?h_{\mu}^{a}??h_{\nu}^{b}?\nabla_{a}n_{b}
\end{equation}
 which is the familiar definition.

\section*{$\Omega^{2n}$}

Since expansions for various powers of $\Omega$ are employed, it
is useful to note that for any $n$
\begin{align}
\Omega^{2n} & =1+2n\pi+n(2n-1)\pi^{2}-n\kappa\boldsymbol{\pi}\cdot\boldsymbol{\pi}+\frac{2}{3}n(n-1)(2n-1)\pi^{3}\nonumber \\
 & \quad-\frac{2}{3}n\left(3n-1\right)\kappa\pi\boldsymbol{\pi}\cdot\boldsymbol{\pi}+\frac{1}{6}n(n-1)(4n^{2}-14n+15)\pi^{4}\nonumber \\
 & \quad-\frac{1}{3}n(3n^{2}+2n-4)\kappa\pi^{2}\boldsymbol{\pi}\cdot\boldsymbol{\pi}+\frac{1}{6}n(3n-1)\kappa^{2}\left(\boldsymbol{\pi}\cdot\boldsymbol{\pi}\right)^{2}+\m O(\pi^{5})\label{eq:Omega^2n}
\end{align}

\section*{Definition of $\pi$, $\boldsymbol{\pi}$,\ldots{}}

As defined at the end of Section~\eqref{sec:galileons-formalism}
the symbols $\pi$ and $\boldsymbol{\pi}$ with various indices are
\[
\pi=c_{I}\pi^{I};\;\pi_{\mu}=\nabla_{\mu}\left(c_{I}\pi^{I}\right);\;\text{etc.}
\]
\[
\boldsymbol{\pi}=\pi^{I}\boldsymbol{e}_{I};\;\boldsymbol{\pi}_{\mu}=\nabla_{\mu}\left(\pi^{I}\boldsymbol{e}_{I}\right);\;\boldsymbol{\pi}_{\mu\nu}=\nabla_{\mu}\nabla_{\nu}\left(\boldsymbol{\pi}^{I}\boldsymbol{e}_{I}\right);\;\text{etc.}
\]
and the $\boldsymbol{\pi}$'s always appear in scalar products, for
example $\boldsymbol{\pi}\cdot\boldsymbol{\pi}=\pi^{I}\pi^{J}\boldsymbol{e}_{I}\cdot\boldsymbol{e}_{J}=\bar{g}_{IJ}\pi^{I}\pi^{J}$.

\section*{Extrinsic Curvature is $\m O(\pi^{3})$}

The extrinsic curvature is
\begin{align}
?k^{I}{}_{\mu\nu}? & =-?h_{\mu}^{a}??h_{\nu}^{b}??\sigma^{I}{}_{c}?\nabla_{a}?h_{b}^{c}?
\end{align}
for the brane $h_{ab}=g_{ab}+\nabla_{a}\pi_{I}\nabla_{b}\pi^{I}$
and $\sigma_{ab}=g_{ab}-\nabla_{\mu}\pi_{a}\nabla^{\mu}\pi_{b}$.
The leading term in the extrinsic curvature is therefore
\begin{align}
?k^{I}{}_{\mu\nu}? & =-\nabla_{\mu}\left(\nabla_{\nu}\pi_{J}\nabla^{I}\pi^{J}\right)+\ldots
\end{align}
but since $\partial_{I}\pi^{J}=0$ and for a Minkowski brane in a
Minkowski bulk $\partial_{I}\bar{g}_{ab}=0$, $\nabla_{I}\pi^{J}=\m O(\pi^{2})$
and $?k^{I}{}_{\mu\nu}?=\m O(\pi^{3})$, as claimed.

\section*{Expansion of terms in $\m L_{(4)}$}

Using \eqref{eq:deltag}, and remembering that we will keep only the
terms necessary to find $\sqrt{-\tilde{g}}R(\tilde{g})$ up to fourth
order in $\pi$, we can write
\begin{align}
\mathfrak{F}_{\mu} & =\nabla_{\mu}\pi_{I}\nabla^{2}\pi^{I}\equiv\boldsymbol{\pi}_{\mu}\cdot?{\t{\boldsymbol{\pi}}}_{\nu}^{\nu}?+\m O(\pi^{3})\label{eq:R-expansion-1}
\end{align}
Dropping the total derivative allows us to write
\begin{align}
-\frac{1}{2}g^{(\mu\nu)(\rho\sigma)}\delta g_{\mu\nu}\nabla^{2}\delta g_{\rho\sigma} & =-\frac{1}{2}\nabla_{\tau}\left(g^{(\mu\nu)(\rho\sigma)}\delta g_{\mu\nu}\nabla^{\tau}\delta g_{\rho\sigma}\right)+\frac{1}{2}g^{(\mu\nu)(\rho\sigma)}\nabla_{\tau}\delta g_{\mu\nu}\nabla^{\tau}\delta g_{\rho\sigma}\nonumber \\
 & =\frac{1}{2}g^{(\mu\nu)(\rho\sigma)}\nabla_{\tau}\left(\boldsymbol{\pi}_{\mu}\cdot\boldsymbol{\pi}_{\nu}\right)\nabla^{\tau}\left(\boldsymbol{\pi}_{\rho}\cdot\boldsymbol{\pi}_{\sigma}\right)+\m O(\pi^{5})\nonumber \\
 & =\left(\boldsymbol{\pi}_{\nu}\cdot\boldsymbol{\pi}_{\mu\rho}\right)\left(\boldsymbol{\pi}^{\nu}\cdot\boldsymbol{\pi}^{\mu\rho}\right)+\left(\boldsymbol{\pi}_{\nu}\cdot\boldsymbol{\pi}_{\mu\rho}\right)\left(\boldsymbol{\pi}^{\mu}\cdot\boldsymbol{\pi}^{\nu\rho}\right)\nonumber \\
 & \quad-\left(\boldsymbol{\pi}^{\mu}\cdot\boldsymbol{\pi}_{\mu\rho}\right)\left(\boldsymbol{\pi}^{\nu}\cdot?{\t{\boldsymbol{\pi}}}_{\nu}^{\rho}?\right)+\m O(\pi^{5})\label{eq:R-expansion-2}
\end{align}
Since $R(\bar{g})_{\mu\nu\rho\sigma}=\bar{\kappa}\left(\bar{g}_{\mu\rho}\bar{g}_{\rho\sigma}-\bar{g}_{\mu\sigma}\bar{g}_{\nu\rho}\right)$
and $G(\bar{g})_{\mu\nu}=-3\bar{\kappa}\bar{g}_{\mu\nu}$ for the
maximally symmetric slice $\Sigma_{0}$ 
\begin{align}
G(\bar{g})^{\mu\nu}\delta g_{\mu\nu} & =-3\bar{\kappa}\bigg(\boldsymbol{\pi}_{\mu}\cdot\boldsymbol{\pi}^{\mu}+2\pi\boldsymbol{\pi}_{\mu}\cdot\boldsymbol{\pi}^{\mu}-\pi^{2}\boldsymbol{\pi}_{\mu}\cdot\boldsymbol{\pi}^{\mu}\nonumber \\
 & \qquad\qquad+\frac{2}{3}\kappa\boldsymbol{\pi}\cdot\boldsymbol{\pi}\boldsymbol{\pi}_{\mu}\cdot\boldsymbol{\pi}^{\mu}+\frac{1}{3}\kappa\boldsymbol{\pi}\cdot\boldsymbol{\pi}_{\mu}\boldsymbol{\pi}\cdot\boldsymbol{\pi}^{\mu}\bigg)+\m O(\pi^{5})\label{eq:R-expansion-3}
\end{align}
and
\begin{align}
 & -\frac{1}{2}\delta g_{\mu\nu}g^{(\mu\nu)(\tau\upsilon)}\left(?{\t{R(\bar{g})}}_{\tau}^{\rho}{}_{\upsilon}^{\sigma}?+?{\t{G(\bar{g})}}_{\tau}^{\rho}?\delta_{\upsilon}^{\sigma}\right)\delta g_{\rho\sigma}\label{eq:R-expansion-4}\\
 & =4\bar{\kappa}\left(\boldsymbol{\pi}_{\mu}\cdot\boldsymbol{\pi}_{\nu}\right)\left(\boldsymbol{\pi}^{\mu}\cdot\boldsymbol{\pi}^{\nu}\right)-\bar{\kappa}\left(\boldsymbol{\pi}_{\mu}\cdot\boldsymbol{\pi}^{\mu}\right)^{2}+\m O(\pi^{5})
\end{align}
Finally since
\begin{align}
\sqrt{-\hat{g}}\hat{g}^{\mu\nu} & =\sqrt{-\bar{g}}\left(\bar{g}^{\mu\nu}-\delta g^{\mu\nu}+\frac{1}{2}\bar{g}^{\mu\nu}\delta?g_{\rho}^{\rho}?+\m O(\delta g^{2})\right)\nonumber \\
 & =\sqrt{-\bar{g}}\left(\bar{g}^{\mu\nu}-\boldsymbol{\pi}^{\mu}\cdot\boldsymbol{\pi}^{\nu}+\frac{1}{2}\bar{g}^{\mu\nu}\boldsymbol{\pi}_{\rho}\cdot\boldsymbol{\pi}^{\rho}+\m O(\pi^{4})\right)
\end{align}
we find that dropping the total derivative
\begin{align}
-6\Omega^{4}\sqrt{\hat{g}}\Omega^{-3}\hat{\nabla}^{2}\Omega & =-6\sqrt{-\hat{g}}\hat{\nabla}_{\mu}\left(\Omega\hat{\nabla}^{\mu}\Omega\right)+6\sqrt{\hat{g}}\hat{g}^{\mu\nu}\partial_{\mu}\Omega\partial_{\nu}\Omega\nonumber \\
 & =6\sqrt{\hat{g}}\hat{g}^{\mu\nu}\partial_{\mu}\Omega\partial_{\nu}\Omega\nonumber \\
 & =6\sqrt{-\bar{g}}\bigg(\pi_{\mu}\pi^{\mu}-2\kappa\pi_{\mu}\boldsymbol{\pi}\cdot\boldsymbol{\pi}^{\mu}+2\kappa^{2}\boldsymbol{\pi}\cdot\boldsymbol{\pi}_{\mu}\boldsymbol{\pi}\cdot\boldsymbol{\pi}^{\mu}\nonumber \\
 & \qquad\qquad-\pi_{\mu}\pi_{\nu}\boldsymbol{\pi}^{\mu}\cdot\boldsymbol{\pi}^{\nu}+\frac{1}{2}\pi_{\mu}\pi^{\mu}\boldsymbol{\pi}_{\nu}\cdot\boldsymbol{\pi}^{\nu}+\m O(\pi^{5})\bigg)\label{eq:R-expansion-5}
\end{align}

\clearpage{}

\bibliographystyle{utphys}
\phantomsection\addcontentsline{toc}{chapter}{\bibname}\bibliography{References,References_brane-collisions,References_Galileons}

\end{document}